\documentclass[a4paper,fleqn]{cas-sc} 

\usepackage[authoryear]{natbib}

\usepackage{graphicx}	
\usepackage{amsmath}	
\usepackage{setspace} 
\usepackage{lineno} 
\usepackage{pdflscape}
\usepackage{subcaption}
\usepackage{gensymb}
\usepackage{multirow}

\def\tsc#1{\csdef{#1}{\textsc{\lowercase{#1}}\xspace}}
\tsc{WGM}
\tsc{QE}
\tsc{EP}
\tsc{PMS}
\tsc{BEC}
\tsc{DE}

\begin{document}
\let\WriteBookmarks\relax
\def\floatpagepagefraction{1}
\def\textpagefraction{.001}

\shorttitle{Spectro-photometry of Phobos simulants II.}

\shortauthors{A. Wargnier et~al.}

\title [mode = title]{Spectro-photometry of Phobos simulants}
\title[mode=sub]{II. Effects of porosity}

\author[1,2]{Antonin Wargnier}[type=editor,
                        auid=000,bioid=1,
                        orcid=0000-0001-7511-2910]

\cormark[1]

\ead{antonin.wargnier@obspm.fr}

\credit{Conceptualization, Data curation, Formal analysis, Investigation, Methodology, Resources, Writing - original draft}

\affiliation[1]{organization={LESIA, Observatoire de Paris, Université PSL, CNRS, Sorbonne Université, Université de Paris-Cité},
    addressline={5 place Jules Janssen}, 
    city={Meudon},
    postcode={92195}, 
    country={France}}
\affiliation[2]{organization={LATMOS, CNRS, Université Versailles St-Quentin, Université Paris-Saclay, Sorbonne Université},
    addressline={11 Bvd d'Alembert}, 
    city={Guyancourt},
    postcode={F-78280}, 
    country={France}}

\author[3]{Olivier Poch}
\credit{Data curation, Investigation, Methodology, Resources, Validation, Writing - review \& editing}
\affiliation[3]{organization={Université Grenoble Alpes, CNRS, Institut de Planétologie et d'Astrophysique de Grenoble (IPAG), UMR 5274},
    city={Grenoble},
    postcode={F-38041}, 
    country={France}}

\author[1,4]{Giovanni Poggiali}
\credit{Conceptualization, Resources, Writing - review \& editing}
\affiliation[4]{organization={INAF-Astrophysical Observatory of Arcetri},
    addressline={largo E. Fermi n.5}, 
    city={Firenze},
    postcode={I-50125}, 
    country={Italy}}
    
\author[2,1]{Thomas Gautier}
\credit{Conceptualization, Funding acquisition, Resources, Supervision, Validation, Writing - review \& editing}
    
\author[1]{Alain Doressoundiram}
\credit{Conceptualization, Project administration, Writing - review \& editing}

\author[3]{Pierre Beck}
\credit{Resources, Writing - review \& editing}

\author[5]{Tomoki Nakamura}
\credit{Conceptualization, Writing - review \& editing}
\affiliation[5]{organization={Department of Earth Science, Tohoku University},
    state={Sendai},
    country={Japan}}

\author[6]{Hideaki Miyamoto}
\credit{Resources, Writing - review \& editing}
\affiliation[6]{organization={Department of Systems Innovation, University of Tokyo},
    addressline={7-3-1 Hongo, Bunkyo-ku}, 
    city={Tokyo},
    country={Japan}}
    
\author[7]{Shingo Kameda}
\credit{Resources, Writing - review \& editing}
\affiliation[7]{organization={Rikkyo University},
    city={Tokyo},
    country={Japan}}
    
\author[8]{Nathalie Ruscassier}
\credit{Investigation, Writing - review \& editing}

\author[8]{Arnaud Buch}
\credit{Resources, Writing - review \& editing}
\affiliation[8]{organization={Laboratoire Génie des Procédés et Matériaux, CentraleSupélec, Université Paris-Saclay},
    city={Gif-sur-Yvette},
    country={France}}
    
\author[9]{Pedro H. Hasselmann}
\credit{Formal analysis, Writing - review \& editing}
\affiliation[9]{organization={INAF - Osservatorio di Roma},
   country={Italy}}

\author[1]{Robin Sultana}
\credit{Conceptualization, Writing - review \& editing}

\author[3]{Eric Quirico}
\credit{Resources, Writing - review \& editing}

\author[1,10]{Sonia Fornasier}
\credit{Writing - review \& editing}
\affiliation[10]{organization={Institut Universitaire de France (IUF)},
addressline={1 rue Descartes}, 
    city={Paris Cedex 05},
    postcode={F-75231}, 
    country={France}}

\author[1]{Maria Antonietta Barucci}
\credit{Writing - review \& editing}

\cortext[cor1]{Corresponding author}

\begin{abstract}
Surface porosity has been found to be an important property for small bodies. Some asteroids and comets can exhibit an extremely high surface porosity in the first millimeter layer. This layer may be produced by various processes and maintained by the lack of an atmosphere. However, the influence of porosity on the spectro-photometric properties of small body surfaces is not yet fully understood. Investigating the spectroscopic effect of porosity is necessary because it is one of the major issues when trying to interpret remote-sensing data.\\
In this study, we looked into the effect of porosity on the spectro-photometric properties of Phobos regolith spectroscopic simulants. Macro- and micro-porosity were created by mixing the simulants with ultra-pure water, producing ice-dust particles, and then sublimating the water. The sublimation of the water ice enabled the production of porous powdered simulants with significant micro- and macro-porosity associated with macro-roughness. The reflectance spectroscopic properties in the visible and near-infrared (0.5-4.2 µm) show no strong variations between the porous and compact samples. However, one simulant exhibits a bluing of the slope after increasing porosity, providing possible insights into the differences between the blue and red units observed on Phobos. In the mid-infrared range, a contrast increase of the 10-µm emissivity plateau due to silicates is observed.\\
Photometry reveals a modification in the phase reddening behavior between the compact powder and the sublimation residue for both simulants. However, the observed behavior is different between the simulants, suggesting that the phase reddening may be dependent on the composition of the simulants. The phase curve also appears to be modified by the addition of porosity, with a higher contribution of forward scattering observed for the sublimation residue. The derivation of the Hapke parameters indicates an increase in roughness for the porous sample, but no significant modification of the opposition effect.\\
This study aims to provide new insights into the understanding of porosity by using two Phobos simulants in the context of the upcoming JAXA/Martian Moons eXploration mission.
\end{abstract}


\begin{highlights}
\item We observed the emergence of a 10 µm-plateau for porous samples, absent or fainter for compact samples
\item The increase of porosity and sub-centimetric roughness modify roughness Hapke parameter and phase function parameters
\item The increase of porosity and roughness modify the phase reddening
\item The increase of porosity induces spectral slope bluing, which may explain the difference between the blue and red units on Phobos
\end{highlights}


\maketitle


\section{Introduction}
The porosity of planetary surfaces is an important aspect influencing physical, spectroscopic (e.g., \citealt{Vernazza_2012}) and photometric (e.g., \citealt{Fornasier_2015, Hapke_2016}) properties. Multi-scale porosity in planetary surfaces have been found on different bodies such as Mars (e.g., \citealt{Clark_1977, Allen_1998}), the Moon (e.g., \citealt{Ohtake_2010, Hapke_2016, Szabo_2022}), comets (e.g., \citealt{AHearn_2005, Basilevsky_2006, Thomas_2008, Skorov_2011, Patzold_2016, Hasselmann_2017, Patzold_2019, Groussin_2019}), and asteroids (e.g., \citealt{Britt_2002, Consolmagno_2008, Vernazza_2012}). The origins of the surface porosity on these bodies cannot always be deciphered but could be due to different composition and/or processes that occur at the surface. For example, on airless bodies such as the Moon, the porosity of the regolith is probably due to several cumulative effects linked to space weathering such as impact gardening, micrometeorite bombardment, and volcanic activity \citep{McKay_1991, Hapke_2016, Badyukov_2020}. On the other hand, the martian regolith exhibits a porosity, which may be due to impact gardening \citep{Hartmann_2001} but also aeolian erosion driven by the presence of an atmosphere \citep{Sullivan_2007, Warner_2017}. Small bodies such as asteroids and comets also display interesting porous properties. The study of asteroid surfaces has revealed a huge range of porosity values, influenced by different processes such as impacts (loss of ice, ejecta blankets, etc) and thermal cycling creating porous fine-grained regolith \citep{Sugita_2019, Molaro_2020, Cambioni_2021}. The small bodies, in particular cometary nuclei, tend to exhibits the highest porosity \citep{Mohlmann_2002}. The cometary nuclei are composed of volatile-rich materials and therefore often show highly porous structures resulting from surface erosion driven by sublimation of the ice (e.g., \citealt{Britt_2004, Keller_2015}). \par
It has long been known since \cite{Hapke_Wells_1981} that the porosity of a surface affects the measured reflectance. Although for a long time, no radiative transfer model (e.g., \citealt{Hapke_1981, Lumme_1981, Mishchenko_1999}) allowed to take into account the porosity of planetary bodies, the filling factor (i.e., the volume fraction occupied by grains) was then progressively added -- to model the porosity -- in radiative transfer models of planetary surfaces (e.g. \citealt{Mishchenko_1992, Shkuratov_1999, Hapke_1999}), and \cite{Hapke_2008} provided a new version of his model that included a porosity correction factor. In the last decade, laboratory experiments have provided new insights into the spectroscopic effect of porosity (e.g., \citealt{Vernazza_2012, Poch_2016_2, Poch_2016, Young_2019, Schroder_2021, Sultana_2021, Sultana_2023, Martin_2022, Martin_2023}).\par
The martian moon Phobos is a relatively small body with unknown origins. Its physical, spectroscopic, and photometric properties suggest a resemblance to an asteroid. A possible feature observed at a wavelength of approximately 10 µm has been identified in the MIR spectra of Phobos \citep{Giuranna_2011, Glotch_2018}. The emissivity feature is linked to silicates and appears with high porosity, as evidenced by both asteroid observations \citep{Emery_2006, Vernazza_2012} and laboratory measurements with KBr (potassium bromide) as a proxy for macro-porosity \citep{Vernazza_2012, Martin_2022, Martin_2023, Sultana_2023}. However, the insufficient S/N in the mid-infrared spectroscopic observations from the Planetary Fourier Spectrometer (PFS) and Thermal Emission Spectrometer (TES) observations \citep{Giuranna_2011, Glotch_2018} does not allow to give a definitive unambiguous conclusion about the presence of this feature. In addition to the 10 µm feature in the spectrum, the very low value of thermal inertia observed on Phobos \citep{Michel_2022} could indicate a high porosity as already observed on several other asteroids \citep{Vernazza_2012}.\par
We have shown in the previous article \citep{Wargnier_2024a} the results on the spectroscopic and photometric characterization of two Phobos simulants, OPPS \citep{Wargnier_2024a} and UTPS \citep{Miyamoto_2021}. These simulants have been developed respectively to reproduce the Phobos spectra from the visible to the mid-infrared (0.5 - 18 µm) and to obtain a mineralogy similar to that of the Tagish Lake meteorite, a commonly proposed analog for D-type asteroids and Phobos \citep{Pang_1978, Pajola_2013}. This paper builds upon our previous work to continue the investigation on the spectrophotometry of Phobos simulants to better understand the Phobos surface in preparation for the future Martian Moons eXploration mission (MMX, \citealt{Kuramoto_2022}) observations, with a specific focus here on the spectro-photometric effects of the porosity as a function of the geometry of  observation; an important parameter as it affects the spectral slope and reflectance. This study aimed to determine the variation of phase reddening and the phase curve for porous surfaces. The findings of this work attempt to improve our understanding of the martian moons and more generally of the porosity of small bodies and planetary surface.

\section{Materials and methods}
\subsection{Samples: preparation, characterization and measurements}
To study the spectrophotometric effects of the porosity, we used two different Phobos simulants: the OPPS described in \cite{Wargnier_2024a} and the UTPS-TB \citep{Miyamoto_2021}. The composition of the two simulants is shown in Table \ref{tab:sim_compo}. For more information on these simulants (e.g., characterization of the endmembers,...), the reader is referred to the corresponding papers. \par
As in \cite{Wargnier_2024a}, reflectance spectra were acquired using three instruments: the SHADOWS (spot size 5.2 mm) and SHINE (spot size 7.5 mm) spectrogonio-radiometer at the Institut de Planétologie et d'Astrophysique de Grenoble (IPAG, France) for the visible and near-infrared (VNIR) spectra, and a FTIR Bruker Vertex 70V equipped with the A513/QA accessory for bidirectional measurements in the mid-infrared (MIR). \par
Spectra were acquired from 0.5 to 3.6 µm for our porous Phobos simulants under more than 50 geometries from 5 to 110$\degree$ by varying incidence and emission angles in the principal plane. Measurements in the principal plane are the most favorable and result in the lowest uncertainties according to \cite{Schmidt_2015} and \cite{Schmidt_2019}. Compared to \cite{Wargnier_2024a}, some geometries at large phase angles were not accessible due to technical limitations induced by the porosity experiment. Two main photometric properties are considered: the phase reddening and the phase curve. To ensure comparability with our previous study on compact Phobos simulant \citep{Wargnier_2024a}, we also apply the Hapke IMSA model \citep{Hapke_2012a} to our datasets. A Markov Chain Monte Carlo (MCMC) was run to determine the a-posteriori probability density function (PDF), considering an initial uniform PDF for the six Hapke free parameters. The median of the a-posteriori PDF provides the best fit values, and the associated uncertainties were estimated based on the 16-50-84 percentiles. More details about method of inversion and estimation of the uncertainties of the Hapke parameters are given in our previous paper \citep{Wargnier_2024a}.\par
Endmembers of the OPPS were ground separately, sieved to a size smaller than 50 µm, and then intimately mixed together using an agate mortar. UTPS was initially present at a block formed by baking of the powder ($\sim$100°C) with particle size ranging from $\sim$30 µm to 2 mm \citep{Miyamoto_2021}. UTPS simulant grains were then directly extract and sieved to a size smaller than 50 µm. This grain size was chosen to ensure that the simulant grains could be contained within the ice particles (67 $\pm$ 31 µm, \citealt{Poch_2016_2, Poch_2016}) for the sublimation experiment.

\begin{table*}[]
    \centering
    \caption{Mineral composition of the two Phobos simulants. Values for UTPS-TB are from \cite{Miyamoto_2021} and values for OPPS from \cite{Wargnier_2024a}. For this work, we prepared OPPS with slightly smaller grain size (25-50 µm) for saponite, olivine, and DECS-19. We crushed, ground, and sieved the UTPS-TB directly to 25-50 µm grain size.}
    \label{tab:sim_compo}
    \begin{tabular}{cccc|cccc}
        \hline
        \multicolumn{4}{c}{\textbf{UTPS-TB}} & \multicolumn{4}{c}{\textbf{OPPS}}\\
        \hline
        Mineral & Grain size & UTPS-TB & UTPS-TB & Mineral & Grain size & OPPS & OPPS\\
                & & (wt.\%) & (vol.\%) & & & (wt.\%) & (vol.\%)\\
        \hline
         Serpentine & 25 - 50 µm & 60.5 & 66.8 & Saponite & 25 - 50 µm & 40.8 & 40\\ 
         Forsterite & 25 - 50 µm & 7.3 & 6.4 & Forsterite & 25 - 50 µm & 29.3 & 20\\ 
         Magnetite & 25 - 50 µm & 7.7 & 4.3 & DECS-19 coal & 25 - 50 µm & 15.5 & 20\\ 
         Fe-Ni sulfide & 25 - 50 µm & 9.2 & 5.45 & Anthracite & $<$ 1 µm & 14.4 & 20\\ 
         Carbonate & 25 - 50 µm & 10.3 & 10.6 & & & \\ 
         Carbon & $<$ 1 µm & $\sim$5 & $\sim$6.5 & & \\ 
         \hline
    \end{tabular}
\end{table*}

\subsection{Setup of the sublimation experiment} \label{sect:porosity_setup}
We created porous samples by sublimation of water ice mixed with grains of the Phobos simulants. The water ice particles were created with the Setup for Production of Icy Analogues-B (SPIPA-B). A detailed description of this setup can be found in \cite{Pommerol_2019}. The aim of this experiment is to create an icy intra-mixture, with all grains of the endmembers of the simulants captured inside ice particles. We followed the protocol described in \cite{Poch_2016} and \cite{Pommerol_2019} to produce the icy-dust particles. We mixed 200 mg of the simulant with 20 ml of ultra-pure water to obtain a suspension of 1 wt.\% of the simulant, homogenised using a sonotrode coupled with an ultrasonic generator. We put the mixture on a magnetic stirrer to keep the mixture homogeneous. The choice of the 1\% was made for technical reasons as with higher concentration, grains clog the tubes and the sonotrode used for nebulization. Moreover, others works \citep{Poch_2016, Schroder_2021} have shown that this concentration enables the creation of a hyperporous sample. The liquid suspension is moved with a peristaltic pump from a beaker to the ultrasonic sonotrode. The sonotrode is equipped with a nebulizer, which creates water droplets containing the simulant grains. The sonotrode is placed above a bowl of liquid nitrogen. The droplets fall into the liquid nitrogen and freeze instantly, producing ice particles in which the simulant grains are trapped. Resulting water ice particles from SPIPA-B shown a diameter of 67 $\pm$ 31 µm \citep{Poch_2016_2, Poch_2016}. The bowl of nitrogen is placed in a freezer at -25°C. We then wait for the nitrogen to evaporate, recovering only the icy particles containing the simulant grains. Tools for manipulating the ice particles and a sieve are also placed in the freezer and immersed in liquid nitrogen for the duration of the experiment. A 400 µm-sieve is used to deposit a homogeneous layer of simulant particles and avoid large ice particles or aggregates that may have formed. In the freezer, we then fill the sample holder -- previously cooled with liquid nitrogen -- with a quantity of between half and the entire sample holder, depending on the composition of the sample. The sample holder used in this study is a plastic one 3D-printed specifically for these experiments, with a diameter of around 2.5 cm and a depth of 2 cm, covered using black aluminium tape. When the sample is filled with the water ice particles, we remove it from the freezer, place it in a transport box filled with liquid nitrogen, and transport it to the CarboNIR chamber \citep{Grisolle_2013}. \par
\begin{figure}
\centering
\resizebox{0.5\hsize}{!}{\includegraphics{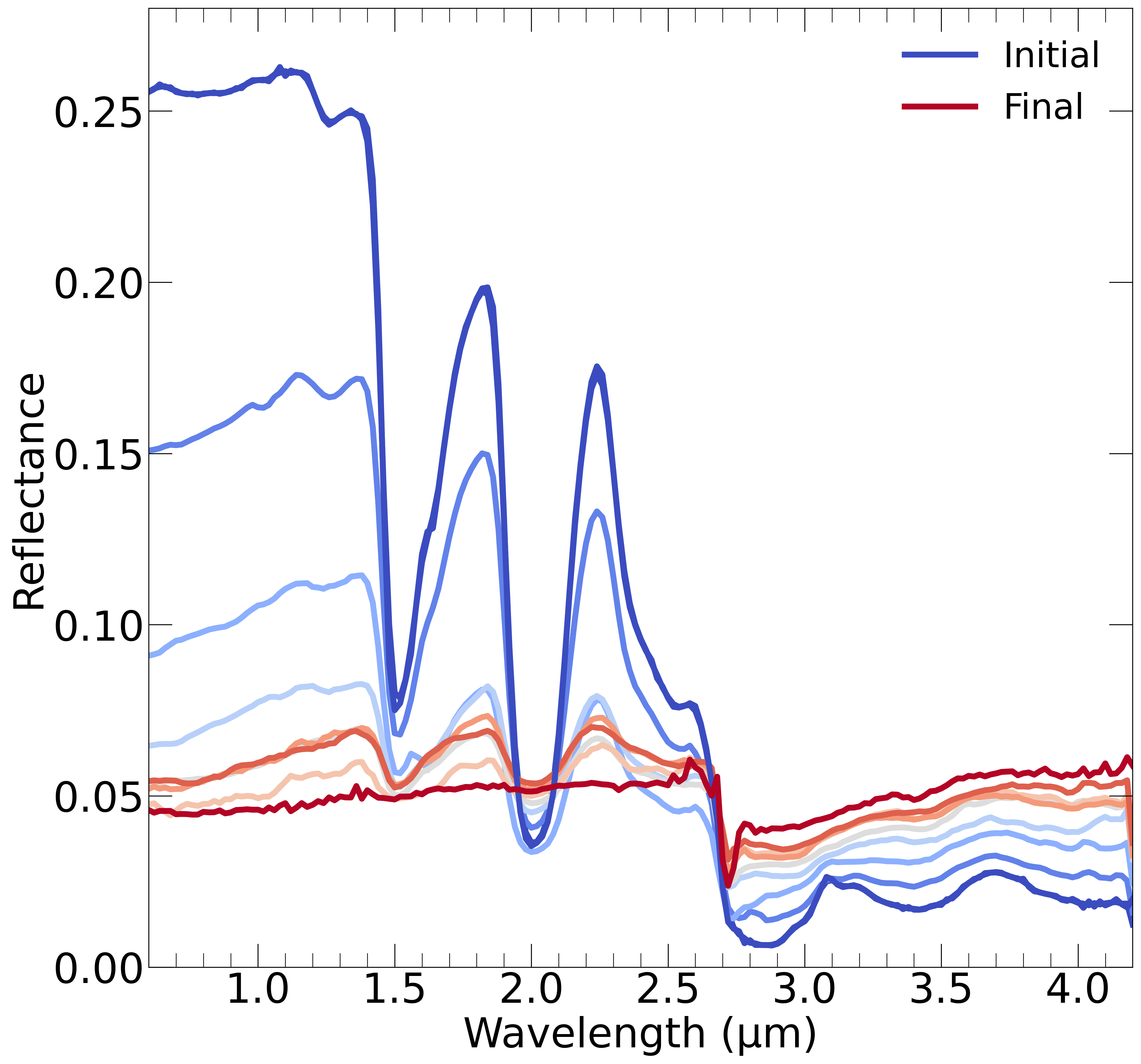}}
\caption{Evolution of the spectrum of the mixture of water ice and UTPS simulant during the sublimation process. Measurements were made under vacuum at 110 K using the SHINE spectrometer at i = 0$\degree$ and e = 30$\degree$. Spectra are acquired every 50 min. Dark blue spectrum is the spectrum at the beginning of the experiment. Note the huge water ice absorption bands at 1.5 µm, 2.0 µm, and 3.1 µm for the spectrum. The final spectrum after 12 h of sublimation is shown in dark red. Water ice signatures have completely disappeared. Phyllosilicates signatures at 2.7 µm and carbonates at 3.4 µm are now visible as expected by the spectrum of the compact UTPS simulant (\citealt{Wargnier_2024a} and Fig. \ref{fig:beforeafter_subl_MIX98andUTPS}).}
\label{fig:evolution_sublimation}
\end{figure}

The CarboNIR chamber is a cryogenic vacuum chamber on the SHINE spectrogognio-radiometer. The cell is closed at the top by a sapphire window, enabling measurements to be taken during the experiment. The sample made up of the intra-mixture of ice particles and simulant particles is placed in CarboNIR under a high vacuum at a pressure of 10$^{-6}$ mbar and a temperature $>$ 110 K, in order to avoid too rapid sublimation, which would lead to ejection of grains and sample destruction, due to water vapor flows. During the sublimation process, we measured spectra with SHINE, in the nominal geometric configuration (incidence i=$0\degree$, emission e=$30\degree$, azimuth $\phi$=$0\degree$) to assess the progress of the experiment. Spectrum acquisition takes 50 min per spectrum, and a new spectrum is automatically started at the end of the previous one. After more than 12 hours of sublimation in CarboNIR, the spectra showed no further change (Fig. \ref{fig:evolution_sublimation}), and the cell was gradually ($\sim$2 K/min) warmed up to 270 K. Before removing the sapphire window and recovering the sublimation residue, nitrogen gas is slowly injected into the chamber and the pressure is raised to atmospheric pressure. The sample is then quickly placed for 30 min in a desiccator to warm it up and avoid the adsorption of surrounding water vapor. In the following, the porous sample recovered after sublimation will be referred to as the "sublimation residue". Note that the experimental procedure used here does not intend to replicate the exact physical processes leading to porosity on Phobos but is instead a useful proxy to increase micro-porosity in our lab samples and study its spectro-photometric effect.\par
This experiment was carried out three times on two different simulants. The first trial with the UTPS simulant produced very little sublimation residue and had to be repeated to obtain an acceptable quantity (i.e., to produce an optically thick sample). With both experiments, we obtained 5.8 mg of sublimation residue. It appears that the mixture composition have important effect on the experiment. The micro-porous structure is first created during the freezing of the water droplet and finally revealed after the sublimation of the water ice. During the freezing step, some endmembers (e.g., saponite) tends to produce filaments that binds others grains (e.g., olivine) via a combination of inter-particles forces between the grains (Van der Waals interactions and electrostatic forces, \citealt{Saunders_1986}). Such mixture leads to the formation of an important internal cohesion of the sample combined with a high porosity \citep{Poch_2016_2}. The composition of UTPS, without expanding phyllosilicates, is not optimal for this kind of experiment, but it does provide a porous sublimation residue. The best endmembers for this experiment are expansive phyllosilicates such as smectites. For this reason, we produced a Phobos simulant mixture (OPPS) with a high saponite content (40 vol.\%). When mixed with water, saponite tends to disintegrate into nanometric particles. These particles then form veins and filaments that bind the grains of the other endmembers of the mixture together \citep{Poch_2016_2, Sultana_2021, Schroder_2021}. This results in the creation of a foam texture of the Phobos simulant after sublimation, in comparison with the "compact" sample. We were able to produce 5.4 mg of the OPPS sublimation residue. In this work, a "compact" sample means a conventionally prepared sample, with a powder deposited in the sample holder and its surface smoothed with a spatula. \par
From the mass of the simulant put into the sample holder before sublimation, the density of the simulant, and the volume occupied in the sample holder, we estimated the bulk porosity to be larger than 63\% for UTPS and larger than 80\% for OPPS sublimation residues. In comparison, the compact UTPS exhibits a bulk porosity of 44\% and the bulk porosity of OPPS is about 49\%. The bulk porosity is not necessarily the dominant parameter on the radiative transfer. However, such high bulk porosity could also imply a highly rough surface (as seen in Fig. \ref{fig:optical_mic_img_ressub}) that may induce a modification of the spectro-photometric properties of the surface. The microporosity introduced into our samples at the nm/µm scale has more significant effects on the optical and spectral properties of the simulants. Microporosity cannot be calculated for our samples with the measurements presented in this work, but we believe it to be high in relation to the texture of our residues, particularly for the OPPS sublimation residue, according to SEM images and previous studies in the literature \citep{Poch_2016_2, Poch_2016, Pommerol_2019, Schroder_2021, Sultana_2021, Sultana_2023}. \newline
It is noteworthy that our porous samples correspond to porous aggregates, as shown in Fig. \ref{fig:optical_mic_img_ressub}. The presence of such large aggregates implies that, in addition to macro- and micro-porosity, our sublimation residues exhibit a high macro-roughness associated with micro-roughness. The macro-roughness is at the millimeter scale for the OPPS and at $\sim$200 µm scale for the UTPS-TB. In the following, for the sake of clarity, we will use the terms 'porous samples', 'sublimation residue' or 'porous aggregates', but it should be noted that this work sheds light on the effects of both porosity and roughness. \par
We investigated the effects of porosity on a spectrum in terms of spectral slope, reflectance and band depth (Sect. \ref{sec:porosity_results}). Moreover, we also studied the effects of observation geometry for the same three parameters (Sect. \ref{sec:porosity_geometry}). In particular, we used the Hapke model to observe whether the parameters can provide important information about the state of a surface (or part of a surface) by comparing with the values of the Hapke parameters obtained in \cite{Wargnier_2024a}, for compact simulants. Mid-infrared spectra of porous simulant were also obtained and discussed in detail in the following (Sect. \ref{sec:porosity_results}).

\section{Results}
\begin{table*}
	\centering
	\caption{Spectral parameters of the Phobos simulants in the visible and near-infrared and positions of the mid-infrared features Christiansen feature (CF), Reststrahlen band (RB), and transparency feature (TF). Uncertainties on the spectral slope and reflectance measurements for the mixtures are due to SHADOWS uncertainties. Spectra are obtained with a phase angle of 30° (i=0°, e=30°). The spectral slope is given in \%/100nm. Values for MIR features position are given with uncertainties of 0.05 µm. In this table, because Phobos has a unique RB, only the position of the first RB of the simulants is given; attributions of the other RBs are visible in Fig. \ref{fig:comparison_MIR}. The CF is absent after addition of porosity and replaced by a 10 µm-plateau}
	\label{tab:mix_params}
    \resizebox{\textwidth}{!}{
	\begin{tabular}{ccccccccc} 
		\hline
		\textbf{Mixtures} & \multicolumn{2}{c}{\textbf{Spectral Slope}} & \multicolumn{2}{c}{\textbf{Reflectance}} & \multicolumn{3}{c}{\textbf{MIR features position (µm)}} & \textbf{References}\\
        & 0.72-0.9 µm & 1.5-2.4 µm & 0.6 µm & 1.8 µm & CF(s)/10µm-plateau & RB & TF & \\
		\hline
		\textbf{"Compact" samples} & & & & & & & & \\ 
        OPPS & 2.55 $\pm$ 0.01 & 2.42 $\pm$ 0.06 & 0.034 $\pm$ 0.001 & 0.045 $\pm$ 0.001 & 8.41/9.20 & 9.36 & -- & \cite{Wargnier_2024a}\\ 
		UTPS-TB  & 1.53 $\pm$ 0.01 & 1.09 $\pm$ 0.04 & 0.035 $\pm$ 0.001 & 0.046 $\pm$ 0.001 & 8.56/9.18 & 10.16 & 11.87 & \cite{Wargnier_2024a}\\
        \hline
        \textbf{Porous samples} & & & & & & & & \\ 
        OPPS & -0.82 $\pm$ 0.01 & 1.38 $\pm$ 0.04 & 0.041 $\pm$ 0.001 & 0.044 $\pm$ 0.001 & [9.2,10.4] & $\sim$10.69 & -- & This work\\
        UTPS-TB & 1.83 $\pm$ 0.01 & 0.62 $\pm$ 0.02 & 0.040 $\pm$ 0.001 & 0.050 $\pm$ 0.001 & [8.5,10.6] & $\sim$11.22 & -- & This work\\
        \hline
	\end{tabular}
    }
\end{table*}

\subsection{Influence of porosity on spectra} \label{sec:porosity_results}
\begin{figure*}
\resizebox{\hsize}{!}{\includegraphics{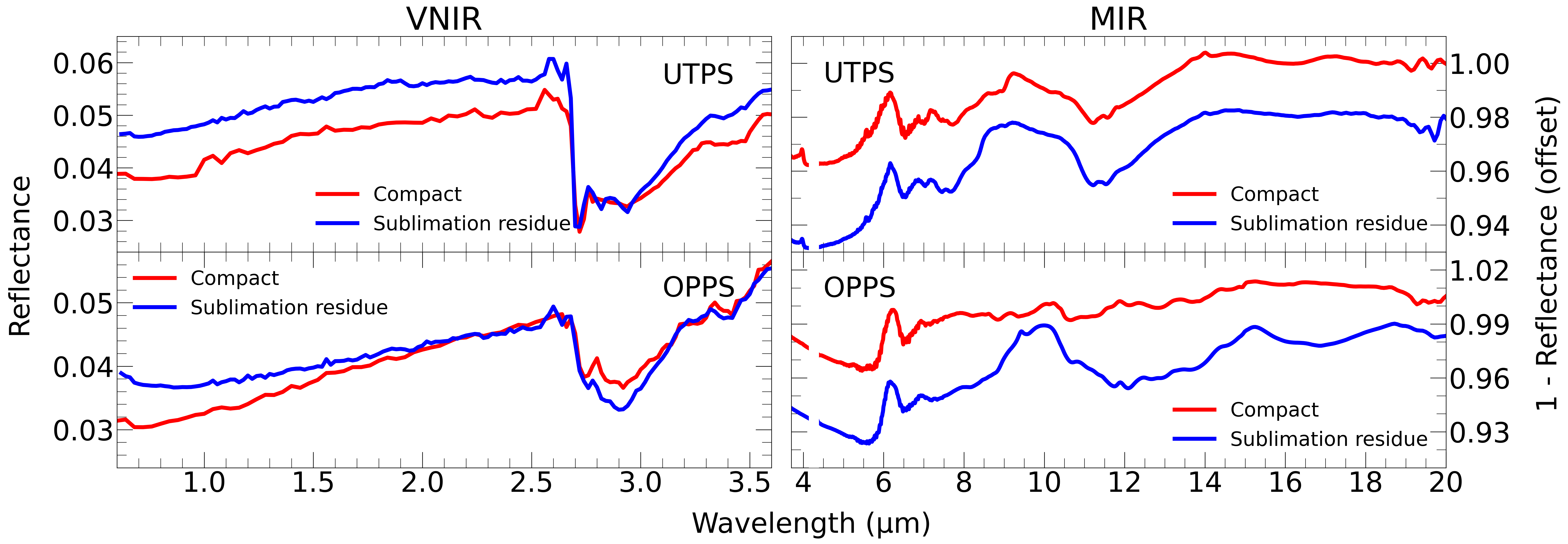}}
\caption{Comparison of the VNIR and MIR spectra of UTPS and OPPS after the sublimation experiment. The left panel includes the VNIR spectra of the two simulants and MIR spectra are plotted on the right panel. The red line represents the compact spectrum and the blue line shows the spectrum of the porous sublimation residue. Spectra are measured at atmospheric pressure and ambient temperature.}
\label{fig:beforeafter_subl_MIX98andUTPS}
\end{figure*}

We produced porous samples of the two different Phobos simulants. Spectroscopic comparison between the compact sample and the porous sublimation residue is presented in Fig. \ref{fig:beforeafter_subl_MIX98andUTPS} for both simulants. In the VNIR, the UTPS sublimation residue appears to be quite similar to the compact sample. The spectral slope is almost identical, with slightly higher reflectance (+12.5\%) and slightly deeper band at 2.7 µm (+5\%) for the sublimation residue (Table \ref{tab:mix_params}). This is because of the sublimation residue's porous texture, which tends to trap atmospheric water, resulting in an increased depth of the band around 2.8-3.0 µm. The difference in reflectance between the OPPS compact sample and the OPPS sublimation residue behaves similarly to the UTPS simulant (i.e., increase of the level of reflectance). The increase in the depth of the molecular water band is more significant here (+6\%). As previously discussed (Sect. \ref{sect:porosity_setup}), the OPPS sublimation residue has a high porosity, resulting in significant water absorption. However, it is worth noting that the spectral slope in the near-infrared is significantly modified when compared with the UTPS simulant (Table \ref{tab:mix_params}). A bluing of the simulant spectrum is observed in the case of the porous sample. \\
The mid-infrared spectra exhibit different behaviors between the two samples. In both cases, however, differences between the compact and sublimation residue samples are visible. For the OPPS, the spectral contrast is typically more enhanced in the sublimation residue. The various features, including the Christiansen feature (CF) and Reststrahlen band (RB) are nearly imperceptible in the compact simulant spectrum, while the porous sample clearly displays these bands. It is also interesting to note the appearance of silicate signatures at 15 and 19 µm. The spectrum of the UTPS sublimation residue is more similar to the UTPS compact one. Nevertheless, the CF feature presents a wider shape for the porous sublimation residue. \par
\begin{figure*}
\resizebox{\hsize}{!}{\includegraphics{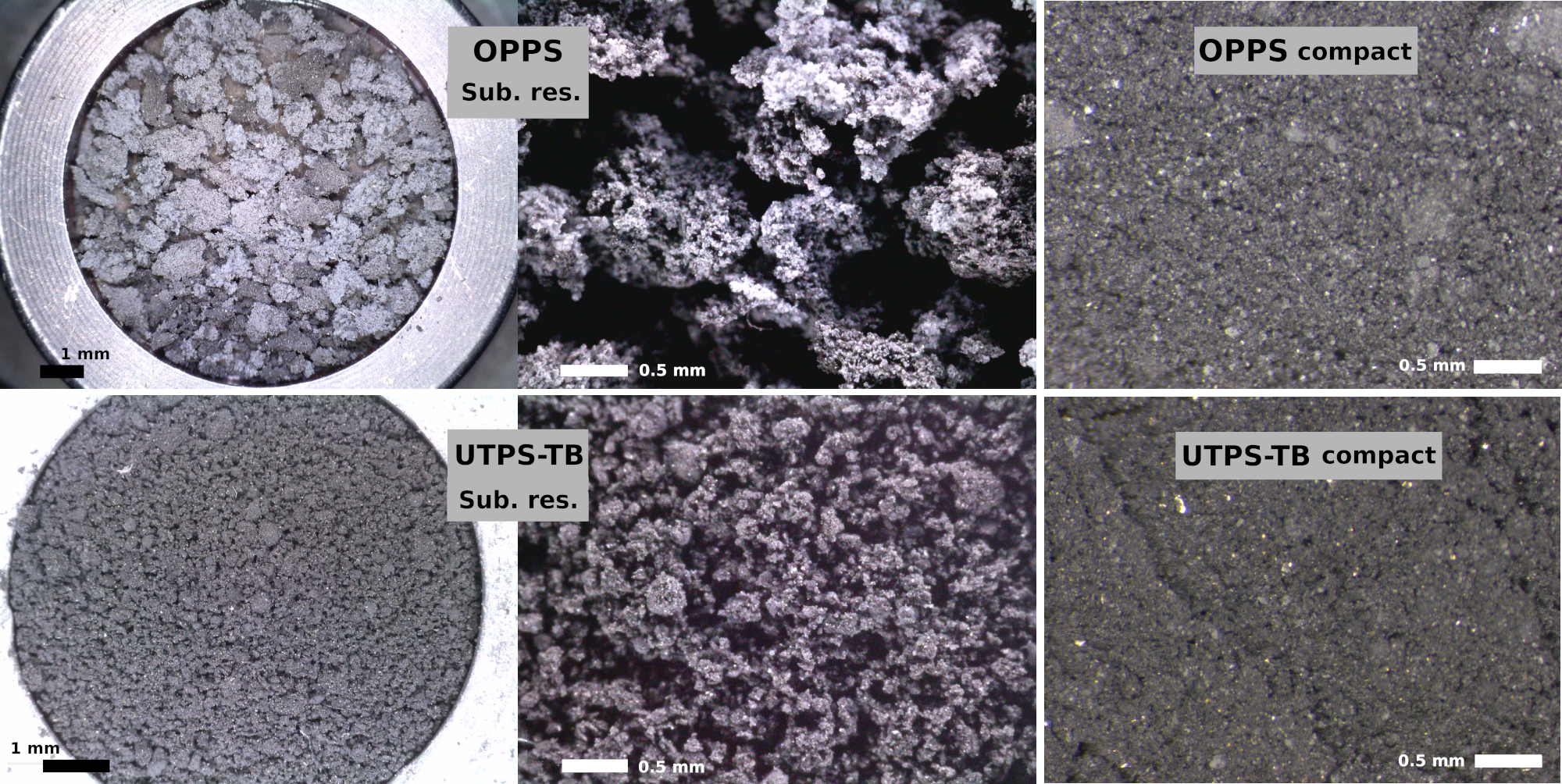}}
\caption{Optical microscope images of the sublimation residues of the OPPS (top) and the UTPS-TB simulants (bottom). Images reveal porous aggregates in both simulants following the water ice sublimation experiment. While the UTPS simulant exhibits some aggregates ($<$500 µm), it is significantly more visible in the OPPS, which has a foam-like texture with up to millimeter size porous aggregates. The aggregates in both simulants possess a microporous texture. It is also interesting to note that the size difference of the aggregates between the simulants implies differences in the macroscopic roughness of the surface. The rough and porous surface obtained can be compared to the relatively smooth surface of the powder before the water ice sublimation experiment (right).}
\label{fig:optical_mic_img_ressub}
\end{figure*}
\begin{figure*}
\resizebox{\hsize}{!}{\includegraphics{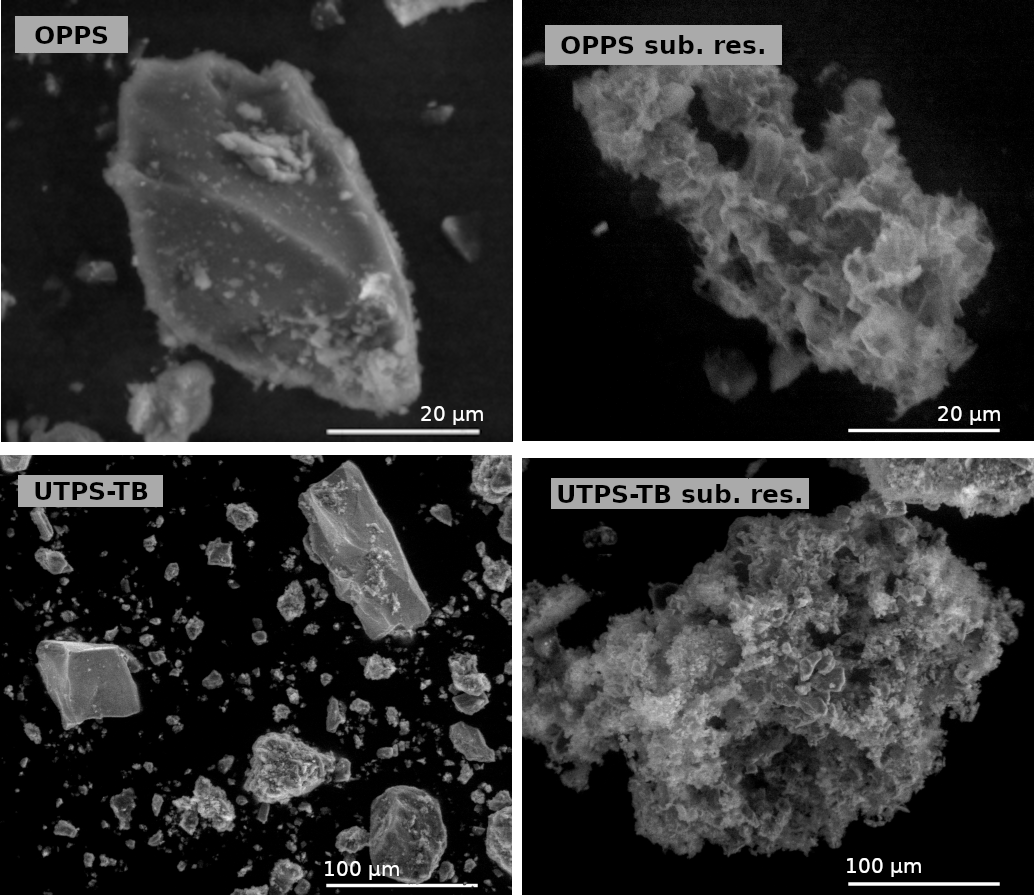}}
\caption{SEM images of Phobos simulant (UTPS and OPPS) before (left) and after (right) the water ice mixture and sublimation experiment. The compact samples (left) exhibit grains of varying sizes, with occasional instances of smaller grains being coated onto larger ones, as well as aggregates of grains. The sublimation residues (right) are primarily composed of large aggregates, which display a fluffy texture (high microporosity) created during the freezing of the drop of dusty water. It is also worth noting that the scale of the two images for the UTPS is identical, hence the porous aggregates are significantly larger compared to the aggregates or grains of the UTPS compact simulant.}
\label{fig:comparison_subres_UTPS}
\end{figure*}

The obtained results suggest that porosity can significantly affect spectra, whether in the VNIR or the MIR. The effects of porosity are particularly notable in cases with high levels of porosity, like those present in OPPS where bulk porosity reached almost 80\% (compared to 63\% for the UTPS simulant). Note also that the two porous simulants produced exhibit a high microporosity, in the form of fractal aggregates (Figs. \ref{fig:optical_mic_img_ressub} and \ref{fig:comparison_subres_UTPS}). In the visible and near-infrared (VNIR) range, previous studies \citep{Poch_2016, Schroder_2021, Sultana_2023} have investigated the impact of porosity induced by water ice sublimation. Our results are consistent with these studies, showing an increase of reflectance of $\sim$20\%, and a bluing of the spectrum of $\sim$50\% in the 1.5-2.4 µm wavelength range.\newline
It is also worth noting that the sublimation residuals of both simulants exhibit broad CF bands, which are substantially stronger in the case of OPPS. These mid-infrared results align with those previously reported in the literature, with the mixing with KBr to simulate porosity \citep{Vernazza_2012, Martin_2022, Martin_2023}. The broadening of the CF is linked to the emergence of the 10-µm plateau, as described in \cite{Vernazza_2012}, \cite{Martin_2022}, and \cite{Sultana_2023}. The spectral contrast of this 10-µm plateau is correlated with the degree of porosity (e.g., \citealt{Martin_2022}). The presence of a significant plateau is therefore associated with high porosity powders containing silicates and has been observed in emissivity spectra of Jupiter Trojans \citep{Emery_2006} or main belt asteroids like Ceres and Lutetia \citep{Vernazza_2012}.

\subsection{Influence of porosity on band depth}
As porosity can affect the visible and near-infrared spectrum, we investigated whether modifying the surface texture could affect the band depth. We compared the absorption bands of organics and hydrated minerals in the sublimation residue and compact samples of OPPS and UTPS taken at the same geometry. Looking first at the 3.42 µm feature, we found no variation between the compact and porous samples. The compact OPPS displays a C-H band of 7.7 $\pm$ 0.1 \%, and the sublimation residue's band depth is identical with 7.7 $\pm$ 0.1 \%. The 2.7 µm O-H absorption band appears to slightly increase in depth as porosity increases. The compact OPPS exhibits a band depth of 21.7 $\pm$ 0.1 \% at 2.7 µm, whereas the sublimation residue indicates a depth of 22.5 $\pm$ 0.1 \%. The sublimation residue of the UTPS, on the other hand, has a band depth of 48.8 $\pm$ 0.1 \%, and the compact sample shows a band depth of 44.6 $\pm$ 0.01 \%.

\subsection{Influence of porosity on photometry} \label{sec:porosity_geometry}
\subsubsection{Phase reddening}
Phase reddening is a phenomenon observed on the majority of small bodies, whereby the spectral slope increases with increasing phase angle. Phase reddening is likely related to a combination of different effects: (1) a higher contribution of multiple scattering at large phase angle and (2) the influence of microscopic surface roughness \citep{Beck_2012, Schroder_2014, Fornasier_2020}. Hence, it might be expected that surface texture would affect phase reddening and can help to constrain the physical properties of the surface, in terms of roughness/porosity and grain size. \par
\begin{figure*}
     \centering
     \begin{subfigure}[b]{0.49\textwidth}
         \centering
         \resizebox{\hsize}{!}{\includegraphics[width=\textwidth]{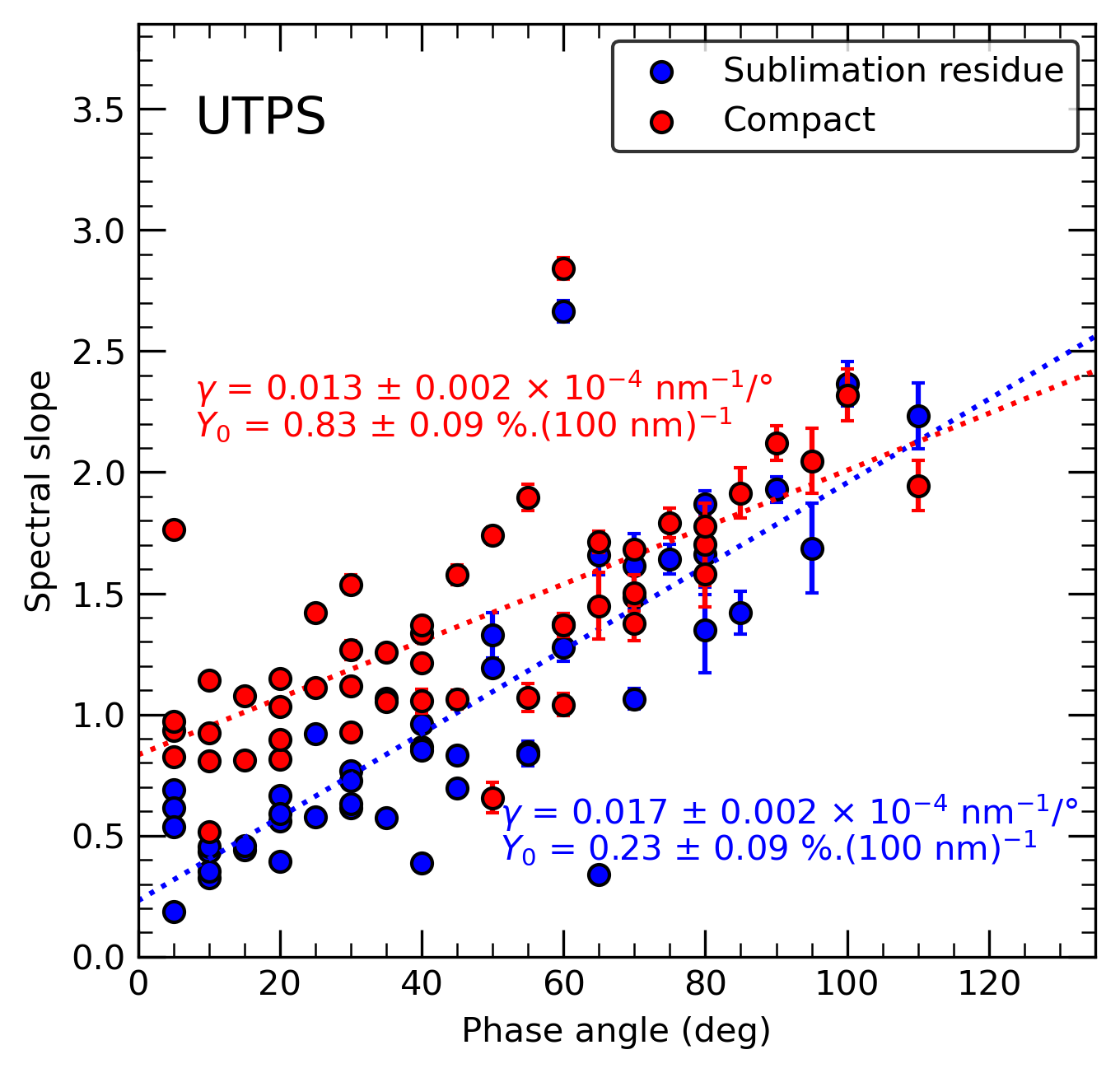}}
         \caption{UTPS}
         \label{fig:UTPSres_geometry}
     \end{subfigure}
     \hfill
     \begin{subfigure}[b]{0.49\textwidth}
         \centering
         \resizebox{\hsize}{!}{\includegraphics[width=\textwidth]{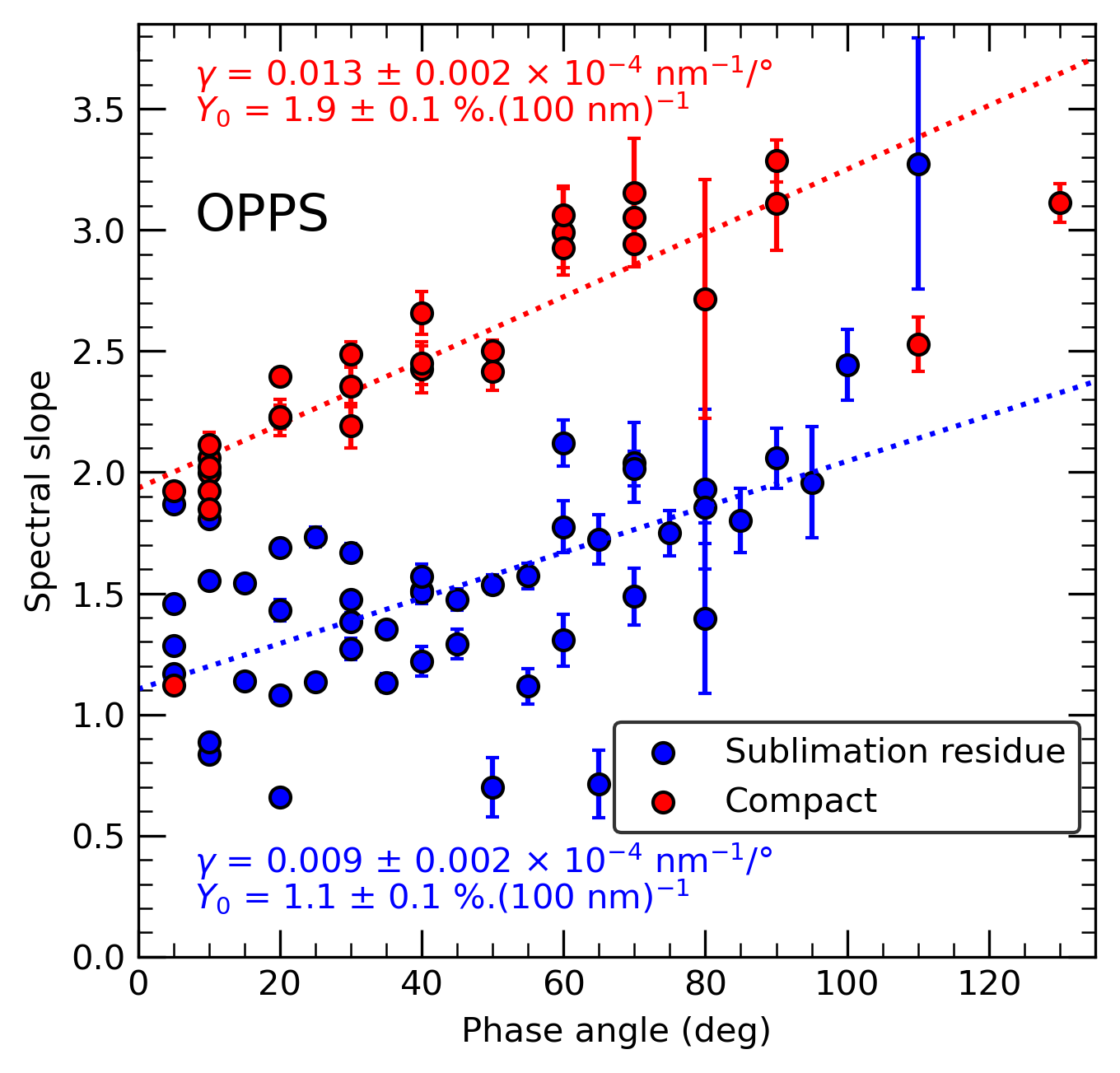}}
         \caption{OPPS}
         \label{fig:MIX98res_geometry}
     \end{subfigure}
     \caption{Evolution of the phase reddening of the (a) UTPS and (b) OPPS simulant for two different textures of the surface: compact and porous (sublimation residue). The phase reddening coefficients are derived assuming a linear relation. The $\gamma$ coefficient corresponds to the phase reddening slope and Y$_0$ is the zero-point. The UTPS phase reddening shows two outliers at 60$\degree$ of phase angle for the compact sample and the sublimation residue. These points correspond to specular reflection with i = 30$\degree$ and e = 30$\degree$ and were removed from the linear fit because of the peculiar behavior at this geometry.}
     \label{fig:geometry_residu}
\end{figure*}

We computed phase reddening parameters (the phase reddening coefficient, $\gamma$ and the spectral slope at zero phase angle, $Y_{0}$) for the two Phobos simulants, with a spectral slope calculation between 1.5 and 2.4 µm. The results are shown in Fig. \ref{fig:geometry_residu}. For the OPPS, we found $\gamma$ = 0.009 $\pm$ 0.002 10$^{-4}$ nm$^{-1}$/$\degree$ and $Y_{0}$ = 1.1 $\pm$ 0.1 \%.(100 nm)$^{-1}$. The UTPS exhibits $\gamma$ = 0.017 $\pm$ 0.002 10$^{-4}$ nm$^{-1}$/$\degree$ and $Y_{0}$ = 0.23 $\pm$ 0.09 \%.(100 nm)$^{-1}$. The parameters of the compact Phobos simulants were computed in \cite{Wargnier_2024a}. The phase reddening appears to be modified between a compact and a porous surface. Whereas the UTPS exhibits a higher $\gamma$ (i.e., a larger phase reddening) for the porous sample, the OPPS sublimation residue shows a decrease of the phase reddening compared to the compact sample. This discrepancy may be attributed to the obtained porosity, in particular the OPPS are in the form of large porous aggregates exhibiting significant macro- and micro-porosity, associated to a high surface roughness. \cite{Beck_2012} and \cite{Schroder_2014} suggested that phase reddening might be attributable to micro-roughness. \par 
In this study, we were unable to quantitatively characterize the microscopic roughness and it is also challenging to attribute the observed effect to a specific parameter that we modify in the experiment (macro- or micro-porosity or roughness). However, the fluffy texture, which is more pronounced in the OPPS, implies a higher micro-roughness. Therefore, we anticipated that the phase reddening coefficient would be higher for this sample, which was not the case. Hence, micro-roughness is likely to play an important role in the phase reddening process, although it cannot be attributed alone to this parameter. New experiments are required to gain a better understanding of this crucial effect.

\subsubsection{Phase curve}
We solved the inversion problem using the Monte Carlo Markov Chain (MCMC) technique (e.g., \citealt{Fernando_2015}) for the Hapke IMSA model \citep{Hapke_2012a}, therefore retrieving the six Hapke parameters $\omega$, b, c, B$_{sh,0}$, h$_{sh}$, and $\bar{\theta}$ from the bi-directional reflectance distribution function of the two porous sublimation residue of the Phobos simulants. The porosity factor K is determined using the relation given by \cite{Helfenstein_2011}, in order to avoid too many free parameters for the inversion.\par
The sublimation residues exhibits a higher reflectance at 600 nm, but their single-scattering albedo are slightly smaller compared to the compact sample (e.g., 0.44$^{+0.03}_{-0.06}$ for the compact UTPS and 0.36$^{+0.08}_{-0.11}$ for the UTPS sublimation residue). This is in agreement with the work of \cite{Hapke_1999} and \cite{Shepard_2007} that also found a decrease of the single-scattering albedo with increasing porosity and/or surface roughness. The parameters of interest are b and c due to their different evolution across different surface types. The two parameters represents the scattering behavior of surface particles: the light is preferentially backscattered for positive c values and forward scattered for negative c values, and the scattering lobe is modified by the value of b. The sublimation residues exhibits smaller values for b (approximately -0.1), while c is consistently higher, particularly for OPPS (approximately +0.3). Therefore, the sublimation residue has a slightly broader lobe, predominantly backscattered as for the compact samples, but with a higher contribution of forward scattering. In the sublimation experiment, we created a porous sample with a non-negligible roughness. Hence, $\bar{\theta}$ is a parameter of particular interest for the study of porosity and/or roughness. The $\bar{\theta}$ is described as the average slope angle or the macroscopic surface roughness \citep{Hapke_1993}. The sublimation residue of the two Phobos simulants showed a $\bar{\theta}$ larger than 20$\degree$, while the compact simulants had values of approximately 15$\degree$. However, the uncertainties associated with this parameter are relatively large, particularly for the sublimation residues, due to the fact that we were unable to acquire data with the same phase angle (120-130 $\degree$) as for the compact samples.\par
Although there is limited data to accurately constrain the opposition effect, we have computed the relative intensity of the opposition effect (ROIE) at 0.6 µm in a similar way to \cite{Beck_2012} and \cite{Wargnier_2024a}. The ROIE shows nearly no variation between the sublimation residue and the compact sample. Specifically, for the UTPS, the ROIE of the sublimation residue is 2.56, compared to 2.68 for the compact sample. For the OPPS, the ROIE of the sublimation residue is 2.44, whereas it was found to be 2.45 for the compact sample. Hence, from the available data, it appears that the opposition effect is not modified, or not significantly modified, when porosity is high. Looking at the Hapke opposition effect parameters (B$_{sh,0}$ and h$_{sh}$) were found to be slightly different between the compact and porous samples (Table \ref{tab:Hapke_parameters}). However, due to the high level of uncertainty in the parameters, it is difficult to obtain reliable values and make definitive conclusions. The porosity factor K was evaluated through the use of the relation with h$_{sh}$, given in \cite{Helfenstein_2011}. The opposition surge is typically associated with surface texture \citep{Naranen_2004} and theoretical works (e.g., \citealt{Hapke_2021}) predict a direct relation between opposition effect width and porosity. However, the fact that we found almost no relation between porosity and opposition effect parameters was already noticed by the experimental work of \cite{Shepard_2007}. Laboratory experiments conducted by \cite{Hapke_2021} found also no clear difference of the opposition effect width between a sample with various porosity. The last two experimental studies \citep{Shepard_2007, Hapke_2021} were unable to explain the unexpected results. All laboratory studies used different samples, with varying grain size distributions, porosity values, particles shapes, and albedos. These physical properties significantly impact the observed relationship between the opposition effect and porosity.

\begin{figure*}
\resizebox{\hsize}{!}{\includegraphics{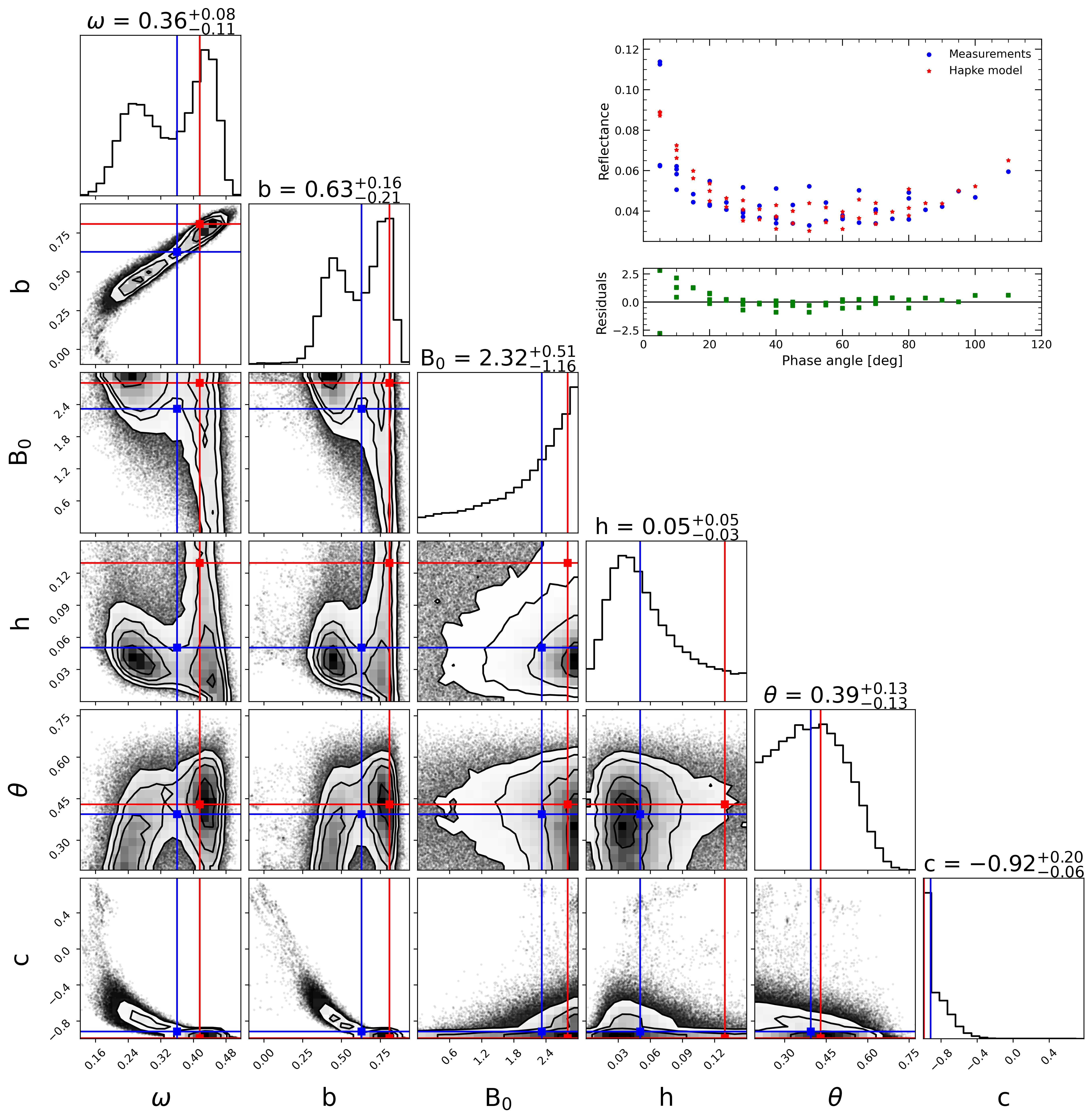}}
\caption{Posterior probability density function (PDF) of the Hapke parameters from the sublimation residue of the UTPS phase curve inversion. Each 1D histogram represents the 45000 accepted solutions. The 2D histograms shows the correlation between the parameters. The blue line represents the median for a given parameters and the red line is the Maximum Likelihood Estimation (MLE). For each parameter, we chose to use the median as the best-fit parameter. The values and the associated uncertainties are given as title of the 1D histogram. It should be noted that a bimodal distribution was obtained for two parameters, $\omega$ and b, resulting in significant uncertainties for these parameters. This highlights the challenges of fitting the Hapke model, where parameters are degenerated and correlated. Top right: Hapke modelisation using the best-fit parameters found from the MCMC inversion, compared to the experimental UTPS sublimation residue phase curve. The residuals are also plotted.}
\label{fig:mcmc_utps}
\end{figure*}

\begin{figure*}
\resizebox{\hsize}{!}{\includegraphics{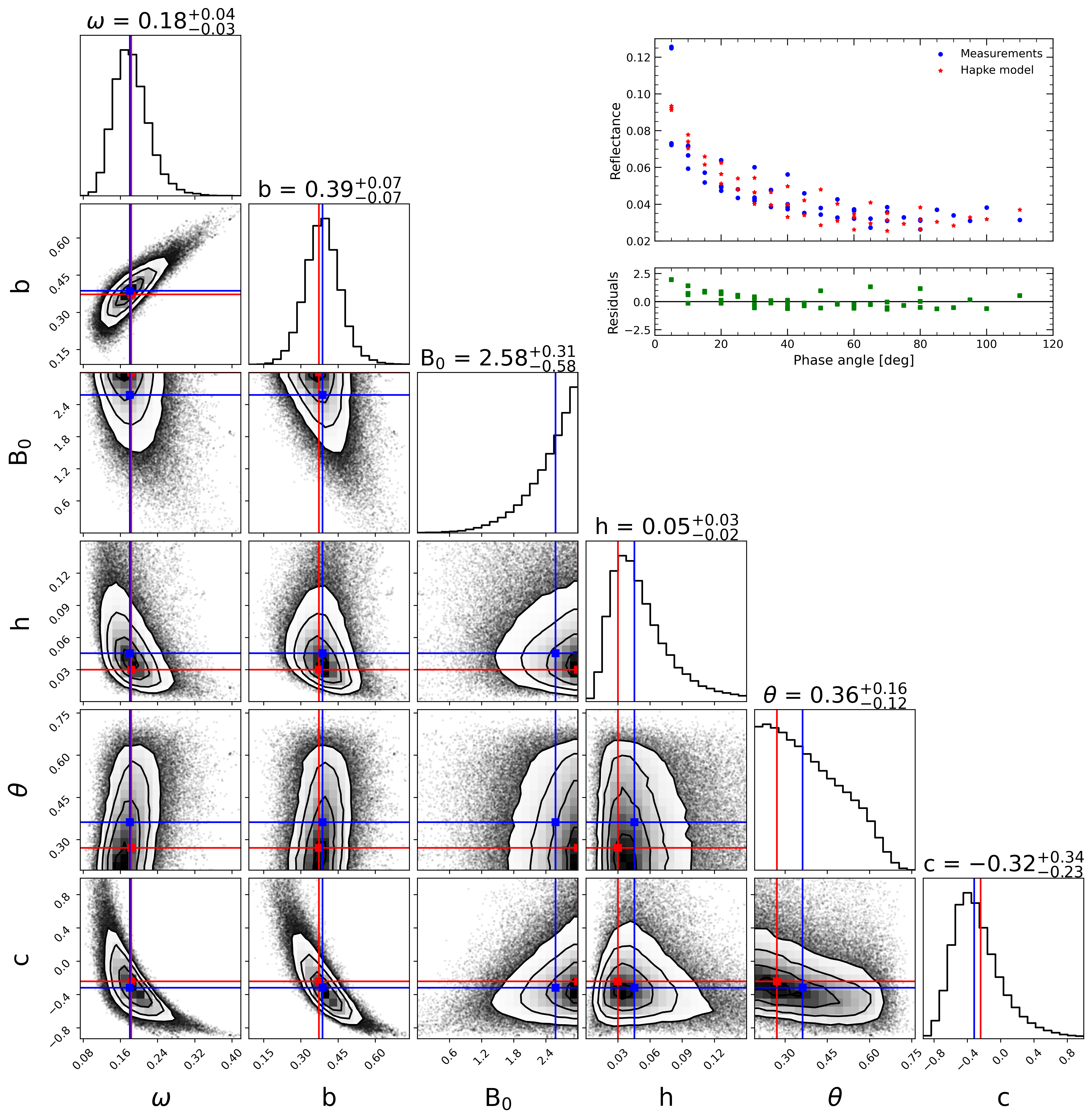}}
\caption{Posterior probability density function (PDF) of the Hapke parameters from the sublimation residue of the OPPS phase curve inversion. Each 1D histogram represents the 45000 accepted solutions. The 2D histograms shows the correlation between the parameters. The blue line represents the median for a given parameters and the red line is the Maximum Likelihood Estimation (MLE). For each parameter, we chose to use the median as the best-fit parameter. The values and the associated uncertainties are given as title of the 1D histogram. Top right: Hapke modelisation using the best-fit parameters found from the MCMC inversion, compared to the experimental OPPS sublimation residue phase curve. The residuals are also plotted.}
\label{fig:mcmc_mix98}
\end{figure*}

\begin{table*}
	\centering
	\caption{Hapke parameters of the two Phobos simulants derived from experimental phase curve at 0.6 µm measured with SHADOWS.}
	\label{tab:Hapke_parameters}
    \resizebox{\textwidth}{!}{
	\begin{tabular}{cccccccccc} 
		\hline
		\\[-1em]
		\textbf{Sample type} & \textbf{Sample} & \textbf{w} & \textbf{b} & \textbf{c} & \textbf{B$_{sh,0}$} & \textbf{h$_{sh}$} & \textbf{$\bar{\theta}$} & \textbf{Porosity} & \textbf{References}\\
		\hline
		\\[-1em]
		Powder & UTPS & 0.44$^{+0.03}_{-0.06}$ & 0.74$^{+0.07}_{-0.12}$ & -0.97$^{+0.04}_{-0.02}$ & 2.00$^{+0.72}_{-1.06}$ & 0.06$^{+0.05}_{-0.04}$ & 15.46$^{+2.29}_{-1.15}$ & 87\% & \cite{Wargnier_2024a}\\
		& & & & & & & &\\
		   & OPPS & 0.23$^{+0.06}_{-0.04}$ & 0.49$^{+0.11}_{-0.09}$ & -0.71$^{+0.17}_{-0.13}$ & 2.06$^{+0.67}_{-1.01}$ & 0.06$^{+0.05}_{-0.03}$ & 13.18$^{+2.86}_{-1.72}$ & 87\% & \cite{Wargnier_2024a}\\
		\\[-1em]
		\hline
		\\[-1em]
		  Porous & UTPS & 0.36$^{+0.08}_{-0.11}$ & 0.63$^{+0.16}_{-0.21}$ & -0.92$^{+0.20}_{-0.06}$ & 2.32$^{+0.51}_{-1.16}$ & 0.05$^{+0.05}_{-0.03}$ & 22.34$^{+9.17}_{-6.88}$ & 89\% & This work\\
		  (sub. res.)  & & & & & & & &\\
		  & OPPS & 0.18$^{+0.04}_{-0.03}$ & 0.39$^{+0.07}_{-0.07}$ & -0.32$^{+0.34}_{-0.23}$ & 2.58$^{+0.31}_{-0.58}$ & 0.05$^{+0.03}_{-0.02}$ & 20.62$^{+7.45}_{-7.45}$ & 89\% & This work\\
		\\[-1em]
        \hline
	\end{tabular}
	}
\end{table*}

\begin{figure*}
     \centering
     \begin{subfigure}[b]{0.49\textwidth}
         \centering
         \resizebox{\hsize}{!}{\includegraphics[width=\textwidth]{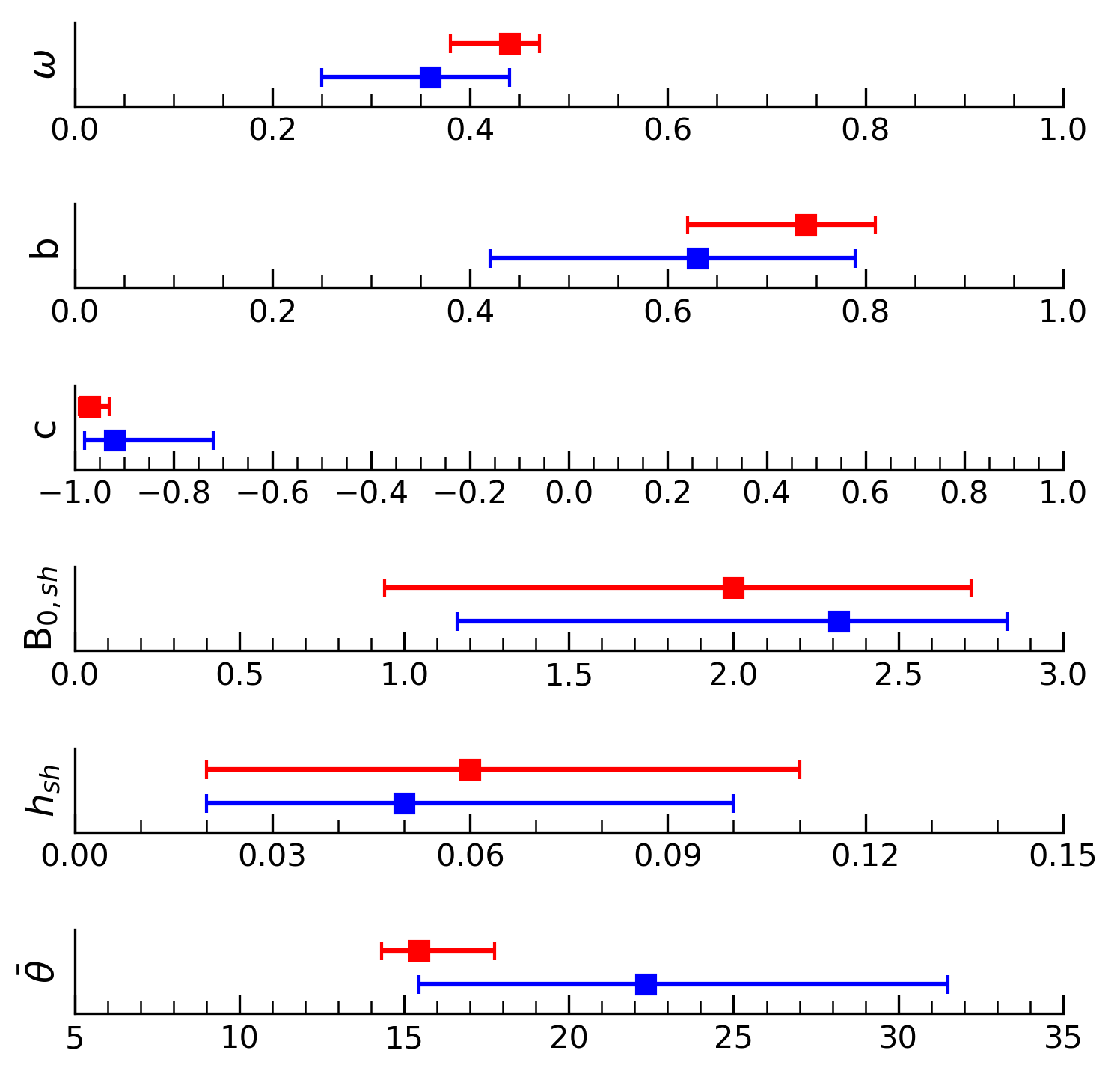}}
         \caption{UTPS}
         \label{fig:UTPSres_geometry}
     \end{subfigure}
     \hfill
     \begin{subfigure}[b]{0.49\textwidth}
         \centering
         \resizebox{\hsize}{!}{\includegraphics[width=\textwidth]{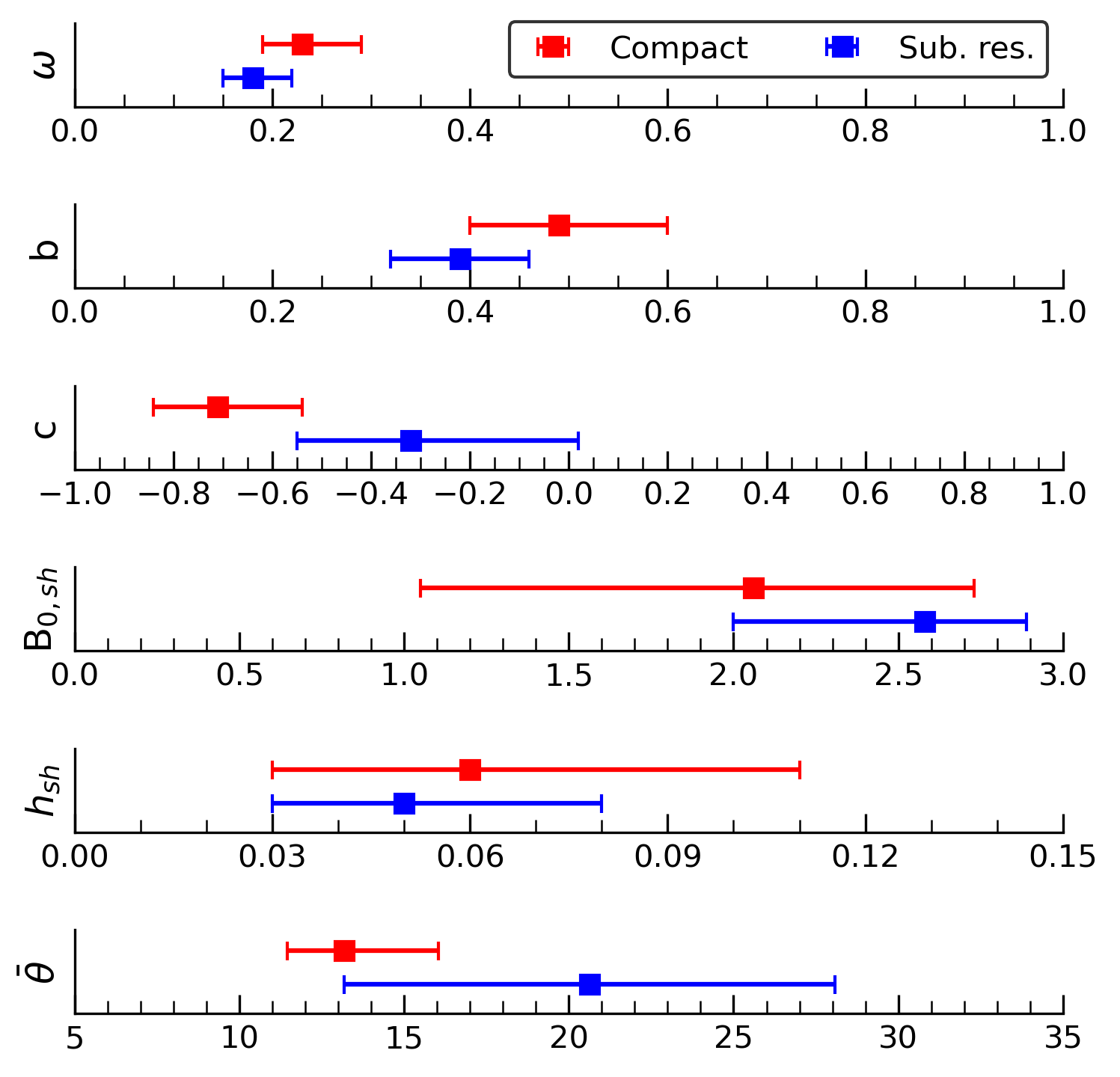}}
         \caption{OPPS}
         \label{fig:MIX98res_geometry}
     \end{subfigure}
     \caption{Hapke parameters evolution between compact (red) and porous sample (blue) after sublimation for the (a) UTPS (b) OPPS.}
     \label{fig:hapke_params_evolution}
\end{figure*}

\section{Discussion}
\subsection{Comparison of Phobos simulants' porosity with small bodies and meteorites}
The macro- and micro-porosity of Phobos' surface has not yet been well constrained due to the small number of observations. A recent study -- based on photometry analysis -- estimated an important roughness and a top-layer surface porosity larger than 80\% \citep{Fornasier_2023}. Comparison with known small bodies such as Ryugu, Bennu or 67P, which do not face the problem of atmospheric entry altering physical properties, is interesting. Remote-sensing observations reveal that the nucleus of comet 67P/Churyumov-Gerasimenko exhibits a high porosity up to 85\% (e.g., \citealt{Fornasier_2015, Hasselmann_2017, Kossacki_2018}). Additionally, Trojan asteroids exhibit extremely porous surfaces, with a regolith porosity of about 90\% \citep{Martin_2023b}. The two rubble-pile asteroids visited respectively by the Hayabusa2 spacecraft and by the OSIRIS-REx mission exhibit also a relatively important microporosity, which can reach up to 50-60\% \citep{Grott_2020, Biele_2020}. These values are similar to the obtained porous sublimation residue of the Phobos simulants. Note also that the slight variation of porosity obtained between the simulants (63\% for UTPS and 80\% for OPPS) is mainly due to their respective compositions, in particular the presence of expansible phyllosilicates (saponite) in OPPS, whereas the phyllosilicate component in the UTPS is represented by serpentine, which is a non-swelling phyllosilicate. Such high surface porosity of asteroids could be explained either by sublimation of water ice-dust mixtures as previously explained, or by grains that may fall back to the surface, therefore creating "fairy-castle" porous structures \citep{Hapke_1963, Emery_2006, Martin_2023b}. These structures are only possible in space environment, with extremely low gravity. Another possibility to explain the high porosity is the impact gardening. In particular, this process has been used to explain the high regolith porosity (80\%) of the Moon \citep{Hapke_2016, Pieters_2016}. \par
It is noteworthy that extraterrestrial samples were also measured on Earth including meteorites and asteroid samples. In general, Phobos exhibits spectral similarities with the Tagish Lake meteorite \citep{Pajola_2013}, which is typically associated with a D-type parent body \citep{Hiroi_2001} and displays the presence of hydrated minerals \citep{Zolensky_2002}. It exhibits porosity in the form of macroporosity ($\sim$40\%) as well as microporosity up to 15\% \citep{Beech_2010}. However, as a meteorite, atmospheric entry is likely to alter the porosity properties of the material. Moreover, the porosity of a meteorite is not identical to that of a regolith, which is estimated from remote-sensing observations. The environment of the surface also plays an important role in the physical properties of the regolith. When a surface originating from meteorites or returned samples falls back to Earth, the initial porosity is inevitably altered by the atmospheric entry and subsequently by the Earth's gravity. Returned samples of the primitive C-type asteroid (162173) Ryugu confirmed \textit{in situ} observations and demonstrated similar porosity \citep{Yada_2022}. The initial study of the extraterrestrial grains suggests the presence of a high microporosity, comparable to the microporosity we obtained for the two Phobos simulants. \par
Phobos would be expected to exhibit similar microporosity, particularly if it is primitive captured asteroid. Previous observations of asteroids have suggested significant surface bulk porosity and microporosity in surface grains (e.g., \citealt{Tatsumi_2018}), and the Hayabusa mission successfully returned a sample from the (25143) Itokawa S-type asteroid. However, the Itokawa sample only showed moderate microporosity \citep{Tsuchiyama_2014}. Thus, the Hayabusa2 sample of the Ryugu is the first asteroid sample to exhibit such high micro-porosity. The detailed analysis of the OSIRIS-REx sample of Bennu will provide further information about the micro-porosity of asteroids. Preliminary analysis of the B-type asteroid samples indicates at least a relatively high macroporosity, but lower than expected through remote sensing measurements \citep{Ryan_2024}. Such high porosity ($>$ 50-60\%) has also been observed in micrometeorites and comet dust from 67P and 81P (e.g., \citealt{Ishii_2008, Guttler_2019, Mannel_2019, Noguchi_2022, Engrand_2023}). The lithology of the sample studied in \cite{Noguchi_2022} indicates the presence of some phyllosilicates, including saponite. The aggregates we formed exhibit also similar morphological behavior, shape, and texture to some Apollo 11 soil samples \citep{McKay_1991}. It should be noted that the porosity of the extraterrestrial samples is not identical to that observed through remote-sensing. Nevertheless, the porosity of the grains is a crucial factor in understanding the nature of the extraterrestrial materials, and will play a role in the porosity estimated through remote-sensing in addition to porosity induced by grains organization at the surface (i.e., macro-porosity and macro-roughness).

\subsection{Evolution of band depth for porous materials}
The evolution of band depth in the VNIR spectral range of porous material has not yet been extensively studied. Previous studies \citep{Poch_2016, Sultana_2021} and this work indicate that porosity can influence band depth, with variations being influenced by various factors, including the physical and compositional properties of the samples. For example, in this study, porosity appears to have either no significant effect or a slight increase on the band depth for the two Phobos simulants. This finding is at odds with the systematic decrease in band depth for sublimation residue of olivine, pyroxene and smectite found in \cite{Sultana_2021}. This discrepancy could be attributed to the differences between the samples used in \cite{Sultana_2021} and those in this study. In particular, \cite{Sultana_2021} used pure mineral powders of hyperfine grains ($<$1 µm) with higher reflectance ($>$20\%), whereas this study used mixtures of silicates with some hyperfine grains of carbon and anthracite -- characterized by an extremely low reflectance ($<$5\%). These differences in absorption and scattering properties likely play a crucial role as more absorbent materials limit the number of multiple scatterings, making them more sensitive to the spatial organization of grains. \newline
It is noteworthy that \cite{Sultana_2021} observed more significant changes in the bands at 1 µm and 2 µm after sublimation, whereas the evolution of the 2.7 µm band was much less pronounced. Similarly, the spectra of porous materials in \cite{Poch_2016} show comparable variations in band depth behavior for different absorption bands. Additionally, the modification of porosity in Murchison (CM2) meteorite powder performed by \cite{Matsuoka_2023} shows no significant changes in the 2.7 µm band. Only a slight increase of approximately 7\% in the absorption band is observed when porosity varies from 20\% to 65\%, with no substantial modifications as a function of porosity. \newline
It is also crucial to note that, as previously stated, our measurements were performed under ambient pressure and room temperature. Therefore, the slight variations observed in the 2.7 µm band depth may only be attributed to the presence of absorbed atmospheric water. Conversely, if porosity had a slight effect of porosity on the 2.7 µm band depth, it would likely also affect the 3.4 µm organic bands. However, no variations were observed in the organic bands. Given the significance of porosity in small bodies, it is crucial to conduct further studies on its potential influence on band depth. 

\subsection{Is the Hapke model an efficient tool for studying porous planetary surfaces?}
As previously discussed, the Hapke model is an important tool for exploring planetary surface properties from photometric observations. The model is a powerful tool to fit, reproduce and compare spectro-photometric data. However, it is well known that the Hapke model parameters are highly degenerated and that several local minima can be found when fitting photometric data (e.g., \citealt{Gunderson_2006, Shepard_2011, Souchon_2011, Shkuratov_2012, Jost_2013, Fernando_2015, Feller_2016}). With the MCMC method implemented in \cite{Wargnier_2024a}, we explore the parameters space to try to obtain the best estimation of the Hapke parameters and their associated uncertainties. We found that the phase function parameters (b and c) show different behavior between a compact and a porous surface. \cite{Hapke_2008} introduced a porosity factor to account for the porous nature of many surfaces. The parameter was found to be related to the half-width of the opposition effect h$_{sh}$ \citep{Helfenstein_2011}. Nevertheless, it is not feasible to derive the opposition parameters unambiguously from our data due to the restricted angular sampling and the significant dispersion of reflectance values at small phase angles for varying incidence and emission angle combinations \citep{Hapke_2008}. Therefore, the porosity correction can only be given with large uncertainties. Interestingly, the estimated porosity of the surface can also be derived from h$_{sh}$, but like the porosity factor, it faces the same issue of large uncertainties in determining the opposition effect parameters. For the OPPS sample, we determined an estimated porosity, from the Hapke model, of 89\% for the sublimation residue and 87\% for the compact surface (Table \ref{tab:Hapke_parameters}). \par
However, interpreting the estimated porosity of the Hapke model in this case may be difficult despite the slightly higher porosity measured for sublimation residue ($\sim$80\% for OPPS and $\sim$63\% for UTPS sublimation residues and $\sim$49\% for OPPS and $\sim$44\% for UTPS compact samples). In the other hand, \cite{Helfenstein_2011} were able to retrieve relatively accurately the measured porosity from Hapke modeling. One of the reason that could explain this difference with our study is that \cite{Hapke_2008} and \cite{Helfenstein_2011} model take into account only the macroporosity through the use of the filling factor, while we created macroporosity associated to an important microporosity in this study. Microporosity is likely to have effects on the spectro-photometric behavior \citep{Sultana_2021} and the porosity at this scale, to the best of our knowledge, is not included in any radiative transfer model. \newline 
The porous samples studied here are composed of large aggregates that created a rough surface (Fig. \ref{fig:optical_mic_img_ressub}). Interestingly, this increase of the macroscopic surface roughness is also visible with the obtained $\bar{\theta}$ Hapke parameter. Microscopic roughness is also increased for the porous simulants but this small scale phenomenon is not included in the Hapke theory. Despite quantitative analysis cannot be made and measurements at larger phase angle are missing in our dataset, the macroscopic surface roughness $\bar{\theta}$ seems to be a relevant quantity for planetary surface studies.
\par
Hence, the Hapke model can give an indication of the bulk porosity and macroscopic roughness of a planetary surface, especially by relative comparison with other planetary surfaces. However, it is important to treat the results of Hapke modelling with caution, as the derived porosity is subject to large uncertainties and as the model is not able to give absolute measurements of the surface properties from the derived Hapke parameters. Further development to implement micro-porosity in radiative transfer model is needed.

\subsection{A high porosity to explain the blue unit of Phobos?}
The blue unit on Phobos indicates a surface with higher reflectance and a less steep red slope than the red unit \citep{Fraeman_2012}. This unit is mostly seen around the Stickney crater \citep{Murchie_1991, Murchie_1996}, particularly on its northeastern rim. The blue unit is also present around other craters (for the albedo map of Phobos, see \citealt{Fornasier_2023}), particularly to the south and east of Stickney, including the Grildrig, Todd, Sharpless, and Wendell craters. The exact cause of the difference between blue and red unit remains unknown today. The blue unit on Phobos may be the result of compositional and/or physical differences on the surface. Considering that the blue material is closely associated with crater ejecta, it seems likely that the Phobos blue unit resulted from impact processes. However, not all impact ejecta are "blue" and some red "patches" have been found inside the blue unit, such as at the floor of the Stickney crater \citep{Thomas_2011, Fornasier_2023}. Therefore, it has been suggested that Phobos might be a heterogeneous object consisting of internal blocks of red and blue units \citep{Basilevsky_2014}.\par
Bluer area were also observed on comet 67P (which has a spectral behavior similar to Phobos in the 0.4 - 2.5 µm range) and associated with terrains enriched in water ice \citep{DeSanctis_2015, Fornasier_2016}. Moreover, blue-sloped materials were observed predominantly inside and around young impact craters on the dwarf planet Ceres \citep{Stephan_2017}. In this instance, the blue units are supported by either ultrafine grains \citep{Stephan_2017} and/or highly porous phyllosilicates \citep{Schroder_2021}. In the latter case, porous filament structures of phyllosilicates would have been created by impacts that froze the initial phyllosilicates and led to a slow sublimation of the water ice. This theory was proposed by \cite{Schroder_2021} based on a water ice sublimation experiment on a Ceres analog. This explanation could also work in the case of Phobos as the surface temperature of Ceres ($\sim$ 170 K, \citealt{Li_2020}) is included in the surface temperature range on Phobos (130-350 K, \citealt{Giuranna_2011}); and if Phobos was H$_{2}$O-rich, which could be the case if it is a captured comet as suggested by \cite{Fornasier_2023}. Also, the presence of phyllosilicates on Phobos has been proposed several times: (1) considering MIR observations with PFS and TES, \cite{Giuranna_2011} note that the Stickney rim spectra are particularly consistent with the presence of phyllosilicates, (2) from the CRISM instrument, visible and near-infrared spectra of the Phobos blue and red units show a possible 2.7 µm band related to the presence of phyllosilicates. But the impact melt mechanism of water ice and phyllosilicates to explain the possible presence of porous materials on the top surface layer of Phobos is only possible in the case of a water ice interior of Phobos, as proposed by \cite{Fanale_1989}, which would melt to liquid water upon impact. Also, in this theory, how to explain the blue unit patches in the red unit, in the floor of the Stickney crater for example? Can they also be explained by the sublimation of water ice? None of the spectroscopic and photometric observations are sufficiently spatially resolved to observe if these patches correspond to rougher and more porous area in comparison with the red Stickney interior. \par
Space weathering may also have contributed to the formation of the two distinct units. According to \cite{Pajola_2018}, the blue units seem to correspond to fresher surface regions. The porous texture created by the water ice sublimation of phyllosilicate mixture could be destroyed by space weathering processes, resulting in a reddening of the spectral slope and a darkening of the initial "blue" spectrum (e.g., \citealt{Vernazza_2013, Lantz_2017, Lantz_2018}), therefore creating the Phobos red unit. The observations of a weak 0.65 µm absorption band in the Phobos red unit spectrum but not in the blue unit spectrum can also be explained by space weathering \citep{Fraeman_2014, Mason_2023}. Given that the red unit corresponds to a more altered surface, it suggests that the 0.65 µm absorption could be attributed to ion implantation at the surface of Phobos from the solar wind. This ion implantation in Fe-bearing materials is expected to result in the formation of iron nanophase particles, which could be the origin of the 0.65 µm absorption band observed in the Phobos red unit. Although this formation of a porous surface layer could be possible from sublimation, other processes could generate porosity on Phobos. This sublimation experiment was solely a means of obtaining microporosity.\par
Similar spectral slope behavior was observed in the porosity experiments on the Murchison meteorite conducted by \cite{Matsuoka_2023}, where the meteorite powder exhibits a spectral bluing with increasing porosity. However, the results of this experiment also demonstrates a distinct trend compared to our experiment, with decreasing reflectance when increasing porosity. A similar decreasing reflectance is observed on porosity experiment on Murchison meteorite performed by \cite{Cloutis_2018} and of various geological materials by \cite{Kar_2016}. Therefore, if we expect that porosity may be at the origin of the blue unit on Phobos, the porosity should increase the reflectance of Phobos material, as evidenced by our experiment. The parameter that controls the influence of the physical structure of the sample on the reflectance is the density of material interfaces in the first layers: the evolution of the reflectance level will depend on the scale of the porous structure, especially in comparison with the wavelength. The studies carried out by \cite{Matsuoka_2023} and \cite{Cloutis_2018} mainly modified the packing density, resulted in the creation of only macro-porosity in their Murchison samples. Other porosity studies that conducted the sublimation experiment \citep{Sultana_2021, Bockelee_2024} created both macro- and micro- porosity as in our work. \cite{Bockelee_2024} is of particular interest because extremely high porosity ($>$80\%) was introduced into the CM2 Aguas Zarcas carbonaceous chondrite. The findings indicate that the porous powder of Aguas Zarcas exhibits a significantly higher reflectance (approximately three times higher) compared to the compact sample. This results aligns with our observations on the Phobos simulants. It demonstrates that the reflectance is highly dependent on the spatial scale of the porosity and, more precisely, on the size of the facets of the sample in relation to the incident wavelength, even when using a different sample (and one relatively close to the Murchison meteorite used in \cite{Cloutis_2018} and \cite{Matsuoka_2023}). However, it should be also noted that the simple mineral sublimation residue sample (e.g., olivine and saponite) measured in \cite{Sultana_2021} can be brighter or darker depending on the sample and the wavelength of interest. The relationship between reflectance and porosity is complex and likely influenced by multiple factors as observed also in \cite{Hapke_1998} and \cite{Kar_2016}, including the surface composition and the surface roughness. In particular, opaque materials may play a significant role in determining the spectroscopic properties behavior and the evolution when modifying porosity. Furthermore, the creation of large porous aggregates resulting from the water ice sublimation experiment led to a significant increase in surface roughness which is likely an important factor influencing the spectrophotometric properties. (Fig. \ref{fig:optical_mic_img_ressub}).\par
In addition to porosity, a blueing of the spectral slope can be explained in several other ways. It has been suggested that space weathering on certain types of material could lead to a blueing of the slope \citep{Vernazza_2013, Lantz_2018}. Additionally, the physical properties of the surface could also lead to a similar result. The presence of hyperfine grains, porosity \citep{Sultana_2023} or hyperfine asperities on compact rocks \citep{Clark_2008, Beck_2021} were shown to affect the light scattering regime. If the scatterers are smaller than the wavelength and optically isolated, they will scatter according to the Rayleigh regime \citep{Beck_2021, Schroder_2021, Sultana_2023}. This modification of the regime will result in a brightening and bluing of the spectra \citep{Brown_2014}. When looking at the SEM images (Fig. \ref{fig:comparison_subres_UTPS}), it appears that the UTPS sublimation residue exhibits particles that are not sufficiently small, and the scattering centers are not separated by enough distance, i.e. larger than the wavelength, and are therefore not independent \citep{Clark_2008}. The modifications induced by the sublimation experiment are hence insufficient to scatter in the Rayleigh regime for the UTPS simulant. However, for the OPPS, the SEM images (Fig. \ref{fig:comparison_subres_UTPS}) reveal that the texture is different with saponite filaments created during the sublimation process. Despite the limited resolution of our SEM images, it is likely that the majority of these filaments are smaller than the wavelength. The obtained high porosity of the sample indicates that the scatterers are optically separated, leading to scattering in the Rayleigh regime and a bluing of the slope. \par
Some areas of Phobos' surface such as craters rims and grooves associated to the blue unit \citep{Thomas_1979, Avanesov_1991, Simonelli_1998} appear to exhibit different phase function, and in particular different opposition effect, than the surrounding surface. This behavior could be due to a different texture of this blue units compared to the red unit with variation of surface porosity and roughness \citep{Thomas_1979}. Recent photometric results \citep{Fornasier_2023}, based on the Hapke's theory, suggest an initial layer porosity of roughly 80\% and fractal aggregates at the surface, congruent with the porosity value and the SEM images of the sublimation residue obtained in this study. However, despite the small increase for porous samples, derivation of the opposition effect intensity from the Hapke model (B$_{sh,0}$) does not seem to show significant modifications that could explain differences between a compact and a porous surface. \par
\begin{figure}
\centering
\resizebox{0.5\hsize}{!}{\includegraphics{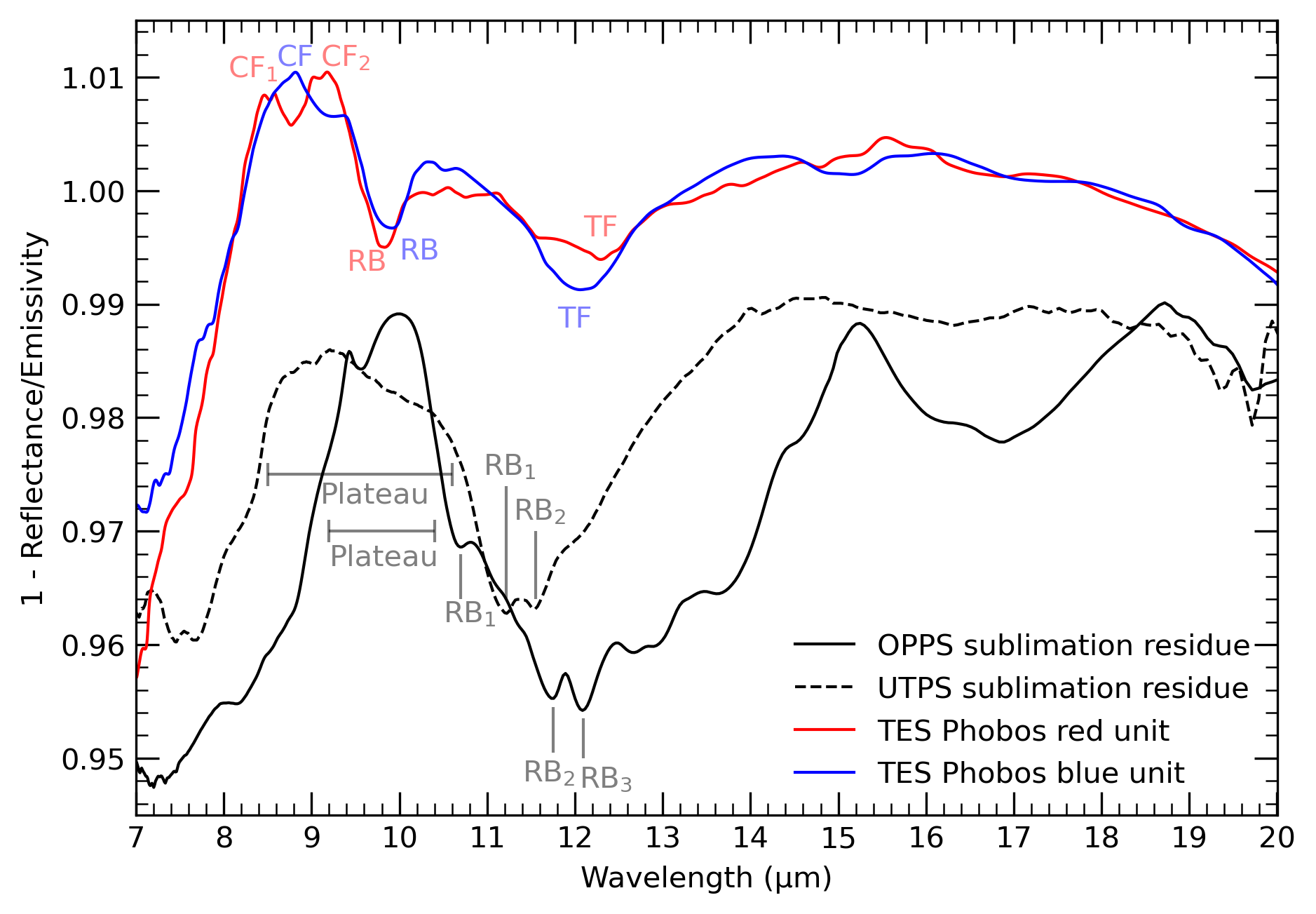}}
\caption{Mid-infrared spectra of the two porous sublimation residue of Phobos simulants compared to TES Phobos spectra of the red and blue unit \citep{Glotch_2018}. Sublimation residue spectra are given in "1- Reflectance" and the Phobos spectra in emissivity.}
\label{fig:comparison_MIR}
\end{figure}
\begin{figure}
\centering
\resizebox{0.5\hsize}{!}{\includegraphics{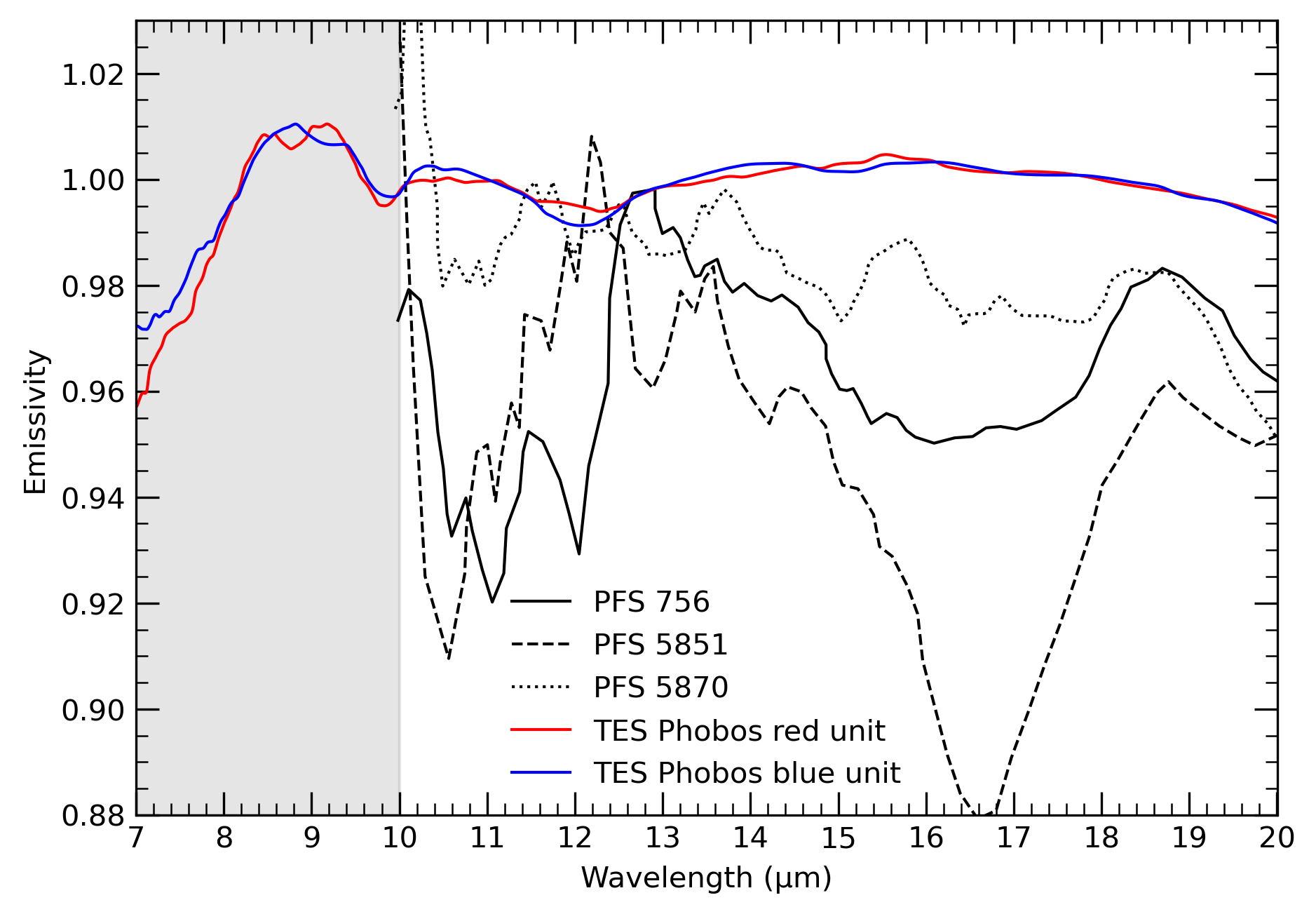}}
\caption{Mid-infrared spectra of Phobos from TES of the red and blue unit \citep{Glotch_2018} and from PFS \citep{Giuranna_2011}. The gray area corresponds to an inaccessible wavelength range of the PFS instrument.}
\label{fig:MIR_phobos}
\end{figure}

We also compared the mid-infrared spectra of the sublimation residues with the red and blue unit spectra of Phobos (Fig. \ref{fig:comparison_MIR}). We observed that OPPS and UTPS show different positions of the main features for example with a maximum emissivity at $\sim$9 µm for the UTPS and at $\sim$10 µm for the OPPS, but also with a local minimum emissivity at $\sim$11 µm for the UTPS and $\sim$12 µm for the OPPS. This position of the minimum of emissivity of OPPS is in agreement with the position of the transparency feature observed on the Phobos spectra. It is also important to note that the 10-µm plateau observed in the mid-infrared range of the porous Phobos simulants we created is not present at the same wavelength than in MIR Phobos spectra \citep{Glotch_2018}. The spectral contrast of the 10-µm plateau of Phobos is comparable between the two units with a value of about 1.8\%. This can be compared with the spectral contrast of 3.4\% observed for OPPS sublimation residue and 2.3\% for UTPS sublimation residue. This raises questions about whether the signature on Phobos is solely related to its composition or if it is also affected by its porosity. Can the 10-µm plateau be shifted in wavelength? Is the peak after the reststrahlen band at 9.82 µm in the Phobos spectra linked to porosity? Mid-infrared observations of primitive asteroids (e.g., \citealt{Licandro_2011, Marchis_2012, Vernazza_2013, Lowry_2022, Martin_2023b, Humes_2024}) shows different shapes of the 10-µm plateau from rounded to trapezoidal, and mainly ranging from 8 to 12 µm, with sometimes high asymmetry. The center of the plateau is likely located between 9.5 to 11 µm, and the spectral contrast of asteroids mainly ranges from 4 to 12\% \citep{Martin_2023b, Humes_2024}. The parameters are generally consistent with the obtained spectra of porous Phobos simulants (Fig. \ref{fig:comparison_MIR}), though there is a systematic reduction in spectral contrast compared to that observed in porous asteroids exhibiting the 10 µm plateau. Answering the above questions is complicated for Phobos because mid-infrared behavior is not well understood and available data is not always of good quality. Indeed, spectra from \cite{Giuranna_2011} and \cite{Glotch_2018} appear to be very different (Fig. \ref{fig:MIR_phobos}), making interpretation of composition and texture difficult. Moreover, all these differences may be due to several cumulative effects: (1) the grain size may differ between the simulants and Phobos; (2) the relative grain size between silicates and opaque materials could also affect the spectral contrast \citep{Sultana_2023, Poggiali_2024}; (3) the Phobos porosity could be different from the porosity of the simulants; (4) the experimental reflectance spectra are compared with Phobos emissivity spectra using the Kirchhoff law, which has been found to modify both the absolute value, spectral contrast, and can slightly shifted the features \citep{Salisbury_1991, Hapke_2012a}. \newline
The behavior of the Phobos simulants is also complicated to understand because these are complex mixtures of four to six endmembers. Some experiments have been performed on single components to provide new hints about the effects of porosity in the mid-infrared \citep{Martin_2022, Martin_2023}. Olivine shows a 10-µm plateau which cover the wavelength from 10 to 12 µm \citep{Martin_2022}. The addition of porosity in pyroxene powders investigated in \cite{Martin_2023} shows interesting behavior because it is observed that the plateau may potentially take on various shapes and widths. Some pyroxene plateau starts at 9 µm and goes to 11 µm. Most of the pyroxene samples studied in \cite{Martin_2023} seems to exhibit a 10-µm plateau that is shifted toward smaller wavelength compared to olivine \citep{Martin_2022}. The position of the plateau appears to be mostly dependent on the composition of the silicates. But, how this plateau is modified by the composition? What happened for example for phyllosilicate? What is the behavior when mixing several types of silicates? These questions are important to interpret mid-infrared observations. The spectral contrast of the 10-µm plateau appears to increase with increasing porosity, in agreement with our observations. A reduction in spectral contrast was noted for both olivine and pyroxene as the grain size increased \citep{Martin_2022, Martin_2023}. This is also interesting to note that no effects of grain size on shift position were observed. \par
The origin of the blue and red units on Phobos should finally be deciphered by the observations conducted by the MMX spacecraft's suite of instruments, such as the MEGANE gamma rays and neutrons spectrometer \citep{Lawrence_2019}, TENGOO/OROCHI cameras \citep{Kameda_2021} and the MMX InfraRed Spectrometer (MIRS, \citealt{Barucci_2021}), as well as the returned sample from one or both units \citep{Usui_2020}.

\subsection{Implications for MMX observations of Phobos and Deimos}
\subsubsection{MIRS}
Porosity is an important parameter to be taken into account when studying planetary surfaces, affecting their spectroscopic and photometric properties. High porosity is typically found only in the first few millimetres/centimetres of the initial regolith layer \citep{Kiuchi_2014}. As discussed above, the spectral differences between the blue and red units on Phobos could be related, at least partially, to porosity. However, determining the difference in porosity of a surface region from remote sensing spectroscopic observations can be challenging because the changes in slope, reflectance and band depth are weak and cannot be linked to a single process. In the MIR range, the formation of the 10 µm-plateau provides valuable information about surface porosity. The presence or absence of a 10 µm emissivity peak will provide essential compositional, as well as textural, information for the future MIRS observations of Phobos and Deimos. Additionally, as noticed by \cite{Martin_2022} and \cite{Sultana_2023}, the formation of a plateau is possible for non-extreme porosity, but the spectral contrast of the plateau could give information about the porosity value of the surface. As porosity increases, the spectral contrast also increases, resulting in a higher plateau. This effect is also present in our spectra (Fig. \ref{fig:comparison_MIR}) with a higher plateau for OPPS compared to the UTPS simulant. Hence, observations in the MIR could help to interpret the results of VNIR spectroscopy. The limited mid-infrared observations of Phobos \citep{Giuranna_2011, Glotch_2018} did not result in any definitive composition conclusions because of the insufficient signal-to-noise ratio. However, there are no dedicated instruments on board the MMX spacecraft for this purpose, and the wavelength range of MIRS (0.9-3.6 µm) will not cover this potential feature. Some additional informations may be retrieved by the infrared radiometer (miniRAD) as well as from the NavCams, both onboard the MMX rover, IDEFIX \citep{Michel_2022, Ulamec_2023}. The JWST could also offer a unique opportunity to cover this unknown wavelength range on Phobos.

\subsubsection{OROCHI}
Following the methodology of \cite{Wargnier_2024a}, we resampled and converted our VNIR spectra to match the band center and bandwidth of the filters used by the OROCHI camera \citep{Kameda_2021}. No significant absorption bands are present in the OROCHI wavelength range of the sublimation residue of the Phobos simulants. Only one weak feature at 0.73 µm is visible particularly in UTPS sublimation residue spectrum, and due to Fe$^{2+}$ - Fe$^{3+}$ charge transfer linked to the presence of serpentinite. Similarly, no absorption bands were found in the compact OROCHI-like spectra of the UTPS and OPPS obtained in \cite{Wargnier_2024a}. Although there is no major absorption band present in the spectra, we can still examine the visible spectral slope (0.48-0.86 µm). The spectral slope was found to be -2.66 $\pm$ 0.05 \%/100nm for the OPPS and 2.10 $\pm$ 0.01 \%/100nm for the UTPS. \par
The slope value for the sublimation residue of the UTPS is slightly smaller but very similar to the value found for the compact UTPS (2.22 $\pm$ 0.06 \%/100nm). Despite the almost identical slope value, the spectra exhibit different behaviour. In particular, the sublimation residue shows a more linear increase in reflectance in the OROCHI wavelength range, whereas the compact UTPS sample exhibits a rapid increase in reflectance between 0.48 and 0.65, followed by a smaller and more linear increase.\par
For the OPPS, there is a significant difference in slope between porous and compact samples. Whereas the compact OPPS spectrum is red-sloped (3.37 $\pm$ 0.05 \%/100nm), the sublimation residue spectrum is blue with a negative slope value. The slope difference can be attributed, as seen previously, to a decrease in the size of the scatterer for the highly porous OPPS sample, which results in a modification of the scattering regime as it moves from Mie to Rayleigh theory \citep{Sultana_2021}. Such a blue slope is not expected on Phobos and Deimos because all observations have shown a visible red slope (e.g. \citealt{Fraeman_2012, Mason_2023}). However, it is important to note that the porosity obtained for OPPS is extremely high. It is questionable whether such high porosity can be expected for the upper layer of regolith on Phobos. The porosity obtained for UTPS is perhaps more realistic, and so looking at the slope could give an indication of the porosity. However, it will also be extremely difficult to correlate the variation in slope to a single process or property, such as porosity, because many other processes and/or properties can modify the slope, such as composition, grain size, and space weathering. Finally, it should be kept in mind that porosity can have a small effect on the slope, as seen in the UTPS, or a larger effect with an inversion of the slope, as seen in the OPPS.
\par
The photometric properties derived from both OROCHI and MIRS will probably provide the most valuable information about the surface. In particular, the phase curve will allow us to derive Hapke parameters and identify differences between porous and compact surfaces in terms of the scattering description parameters (b and c) and the roughness parameter ($\bar{\theta}$). Although phase reddening does not exhibit any clear differences between a compact and porous surface, it could be an interesting additional observable for MIRS and OROCHI to determine if the difference between blue and red units is attributable to porosity alone.

\begin{figure}[h!]
\centering
\resizebox{0.5\hsize}{!}{\includegraphics{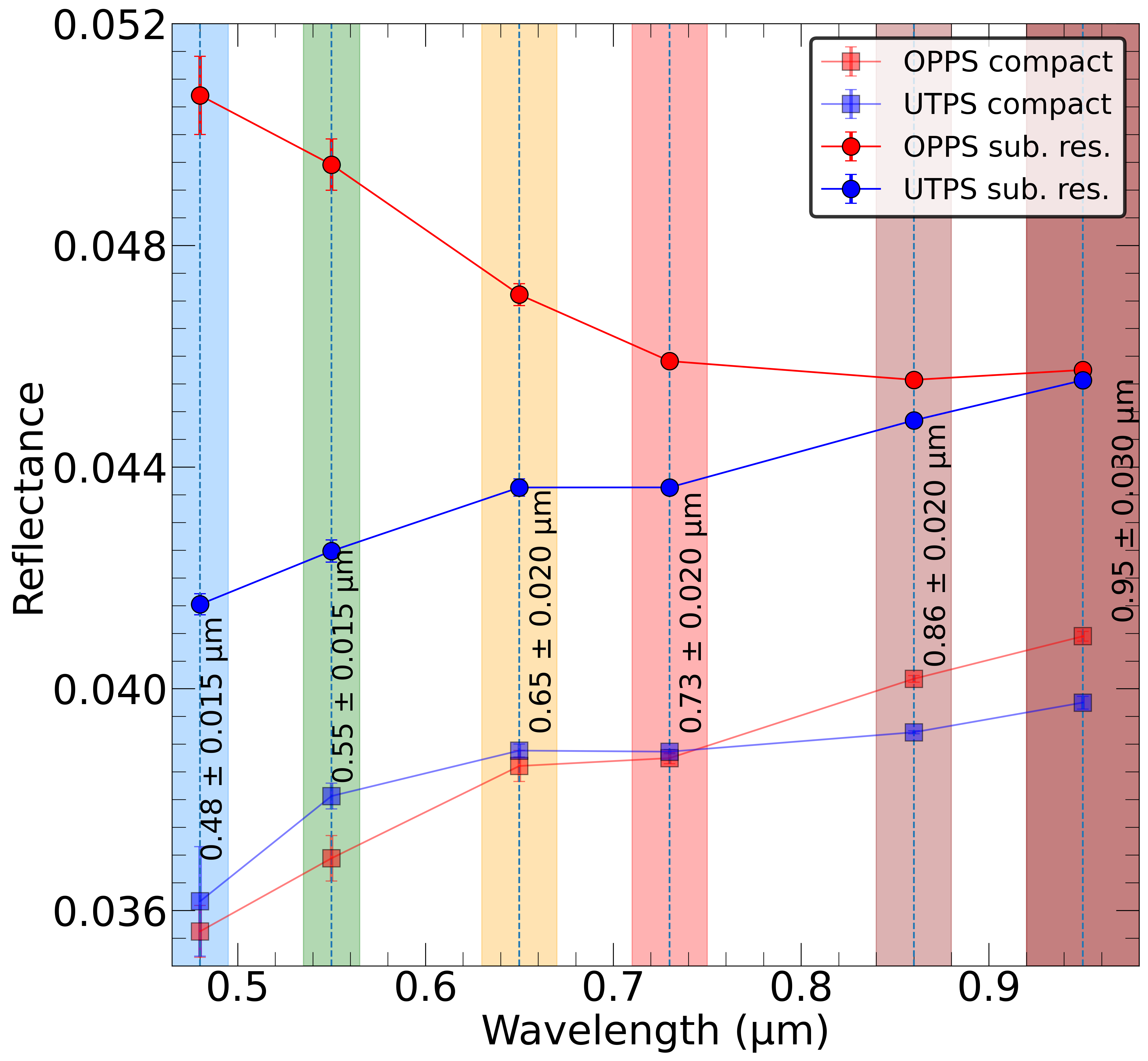}}
\caption{Comparison of the converted OROCHI spectra obtained from SHADOWS and SHINE measurements of the sublimation residues and the compact samples. Data for the compact samples were obtained from \cite{Wargnier_2024a}. The OROCHI spectra were obtained assuming rectangular filters.}
\label{fig:orochi}
\end{figure}

\section{Conclusion}
The porosity of a planetary surface is a crucial parameter to consider when interpreting remote-sensing observations as it can significantly affect the data on a large wavelength range. However, this effect has not been extensively studied in the literature, and some uncertainties remain regarding its spectrophotometric effects. In this work, we aim to provide new insights for the future interpretation of observations from the instruments onboard the Martian Moon eXploration mission. In general this study will be useful for the study of any small and/or airless body observations. The main results of this laboratory porosity study were as follows:
\begin{itemize}
    \item We investigated the effect of porosity for the UTPS-TB and the OPPS Phobos spectroscopic simulants. We create intra-mixtures of ice particles and simulant grains and create porosity by sublimation of the water ice (sublimation residue). These sublimation residues appear to be composed of large porous aggregates, therefore showing porosity and rugosity at both microscopic and macroscopic scales. We found that the visible and near-infrared wavelength ranges are only marginally affected by porosity. The reflectance appears brighter ($\sim$20\%) for the sublimation residues and, in the case of the OPPS spectrum, the VNIR spectral slope is slightly bluer. This bluing is likely due to the very high porosity we obtained for the OPPS (> 80\%) compared to the one obtained for the UTPS (> 63\%) induced respectively by the presence and absence of expandable clay mineral. Therefore, high porosity appears as one possibility among others to explain the origin of the Phobos blue unit. In the mid-infrared range, the Christiansen feature is modified. The peak is larger for the porous samples leading to the creation of a 10 µm-plateau. However, the position of the feature nor the position of the reststrahlen band and the transparency features are modified.
    \item Although it is challenging to attribute spectrophotometric modifications to a single property, the porous sample shows variations of the roughness parameter ($\bar{\theta}$) and of the phase function parameters (b and c). We also observed interesting variations of the phase reddening. The phase reddening could provide additional information about the surface of the texture and will be a pivotal observable for the on-board instruments.
\end{itemize}
Our results shows the huge interest of having a broad wavelength range from the visible to mid-infrared for the interpretation of remote-sensing observations of small bodies surface.

\section*{Acknowledgements}
This work was carried out in support for the MIRS instrument onboard the future MMX mission, with the financial support of the Centre National d’Etudes Spatiales (CNES). The authors acknowledge Bernard Schmitt and Olivier Brissaud for developping, funding and maintaining the SHINE and CARBONIR setups at IPAG, as well as Daniel Genin for the 3D-printing of the sample holders used for the sublimation experiments.

\section*{Data availability}
The data will be made available on the Zenodo repository (doi: 10.5281/zenodo.11241566).

\printcredits

\bibliographystyle{cas-model2-names}

\bibliography{cas-refs}

\begin{thebibliography}{132}
\expandafter\ifx\csname natexlab\endcsname\relax\def\natexlab#1{#1}\fi
\providecommand{\url}[1]{\texttt{#1}}
\providecommand{\href}[2]{#2}
\providecommand{\path}[1]{#1}
\providecommand{\DOIprefix}{doi:}
\providecommand{\ArXivprefix}{arXiv:}
\providecommand{\URLprefix}{URL: }
\providecommand{\Pubmedprefix}{pmid:}
\providecommand{\doi}[1]{\href{http://dx.doi.org/#1}{\path{#1}}}
\providecommand{\Pubmed}[1]{\href{pmid:#1}{\path{#1}}}
\providecommand{\bibinfo}[2]{#2}
\ifx\xfnm\relax \def\xfnm[#1]{\unskip,\space#1}\fi
\bibitem[{{A'Hearn} et~al.(2005){A'Hearn}, {Belton}, {Delamere}, {Kissel},
  {Klaasen}, {McFadden}, {Meech}, {Melosh}, {Schultz}, {Sunshine}, {Thomas},
  {Veverka}, {Yeomans}, {Baca}, {Busko}, {Crockett}, {Collins}, {Desnoyer},
  {Eberhardy}, {Ernst}, {Farnham}, {Feaga}, {Groussin}, {Hampton}, {Ipatov},
  {Li}, {Lindler}, {Lisse}, {Mastrodemos}, {Owen}, {Richardson}, {Wellnitz} and
  {White}}]{AHearn_2005}
\bibinfo{author}{{A'Hearn}, M.F.}, \bibinfo{author}{{Belton}, M.J.S.},
  \bibinfo{author}{{Delamere}, W.A.}, \bibinfo{author}{{Kissel}, J.},
  \bibinfo{author}{{Klaasen}, K.P.}, \bibinfo{author}{{McFadden}, L.A.},
  \bibinfo{author}{{Meech}, K.J.}, \bibinfo{author}{{Melosh}, H.J.},
  \bibinfo{author}{{Schultz}, P.H.}, \bibinfo{author}{{Sunshine}, J.M.},
  \bibinfo{author}{{Thomas}, P.C.}, \bibinfo{author}{{Veverka}, J.},
  \bibinfo{author}{{Yeomans}, D.K.}, \bibinfo{author}{{Baca}, M.W.},
  \bibinfo{author}{{Busko}, I.}, \bibinfo{author}{{Crockett}, C.J.},
  \bibinfo{author}{{Collins}, S.M.}, \bibinfo{author}{{Desnoyer}, M.},
  \bibinfo{author}{{Eberhardy}, C.A.}, \bibinfo{author}{{Ernst}, C.M.},
  \bibinfo{author}{{Farnham}, T.L.}, \bibinfo{author}{{Feaga}, L.},
  \bibinfo{author}{{Groussin}, O.}, \bibinfo{author}{{Hampton}, D.},
  \bibinfo{author}{{Ipatov}, S.I.}, \bibinfo{author}{{Li}, J.Y.},
  \bibinfo{author}{{Lindler}, D.}, \bibinfo{author}{{Lisse}, C.M.},
  \bibinfo{author}{{Mastrodemos}, N.}, \bibinfo{author}{{Owen}, W.M.},
  \bibinfo{author}{{Richardson}, J.E.}, \bibinfo{author}{{Wellnitz}, D.D.},
  \bibinfo{author}{{White}, R.L.}, \bibinfo{year}{2005}.
\newblock \bibinfo{title}{{Deep Impact: Excavating Comet Tempel 1}}.
\newblock \bibinfo{journal}{Science} \bibinfo{volume}{310},
  \bibinfo{pages}{258--264}.
\newblock \DOIprefix\doi{10.1126/science.1118923}.
\bibitem[{{Allen} et~al.(1998){Allen}, {Morris}, {Jager}, {Golden},
  {Lindstrom}, {Lindstrom} and {Lockwood}}]{Allen_1998}
\bibinfo{author}{{Allen}, C.C.}, \bibinfo{author}{{Morris}, R.V.},
  \bibinfo{author}{{Jager}, K.M.}, \bibinfo{author}{{Golden}, D.C.},
  \bibinfo{author}{{Lindstrom}, D.J.}, \bibinfo{author}{{Lindstrom}, M.M.},
  \bibinfo{author}{{Lockwood}, J.P.}, \bibinfo{year}{1998}.
\newblock \bibinfo{title}{{Martian Regolith Simulant JSC Mars-1}}, in:
  \bibinfo{booktitle}{Lunar and Planetary Science Conference}, p.
  \bibinfo{pages}{1690}.
\bibitem[{{Avanesov} et~al.(1991){Avanesov}, {Zhukov}, {Ziman}, {Kostenko},
  {Kuzmin}, {Muravev}, {Fedotov}, {Bonev}, {Mishev}, {Petkov}, {Krumov},
  {Simeonov}, {Boycheva}, {Uzunov}, {Weide}, {Halmann}, {Possel}, {Head},
  {Murchie}, {Schkuratov}, {Berghanel}, {Danz}, {Mangoldt}, {Pihan},
  {Weidlich}, {Lumme}, {Muinonen}, {Peltoniemi}, {Duxbury}, {Murray},
  {Herkenhoff}, {Fanale}, {Irvine} and {Smith}}]{Avanesov_1991}
\bibinfo{author}{{Avanesov}, G.}, \bibinfo{author}{{Zhukov}, B.},
  \bibinfo{author}{{Ziman}, Y.}, \bibinfo{author}{{Kostenko}, V.},
  \bibinfo{author}{{Kuzmin}, A.}, \bibinfo{author}{{Muravev}, V.},
  \bibinfo{author}{{Fedotov}, V.}, \bibinfo{author}{{Bonev}, B.},
  \bibinfo{author}{{Mishev}, D.}, \bibinfo{author}{{Petkov}, D.},
  \bibinfo{author}{{Krumov}, A.}, \bibinfo{author}{{Simeonov}, S.},
  \bibinfo{author}{{Boycheva}, V.}, \bibinfo{author}{{Uzunov}, Y.},
  \bibinfo{author}{{Weide}, G.G.}, \bibinfo{author}{{Halmann}, D.},
  \bibinfo{author}{{Possel}, W.}, \bibinfo{author}{{Head}, J.},
  \bibinfo{author}{{Murchie}, S.}, \bibinfo{author}{{Schkuratov}, Y.G.},
  \bibinfo{author}{{Berghanel}, R.}, \bibinfo{author}{{Danz}, M.},
  \bibinfo{author}{{Mangoldt}, T.}, \bibinfo{author}{{Pihan}, U.},
  \bibinfo{author}{{Weidlich}, U.}, \bibinfo{author}{{Lumme}, K.},
  \bibinfo{author}{{Muinonen}, K.}, \bibinfo{author}{{Peltoniemi}, J.},
  \bibinfo{author}{{Duxbury}, T.}, \bibinfo{author}{{Murray}, B.},
  \bibinfo{author}{{Herkenhoff}, K.}, \bibinfo{author}{{Fanale}, F.},
  \bibinfo{author}{{Irvine}, W.}, \bibinfo{author}{{Smith}, B.},
  \bibinfo{year}{1991}.
\newblock \bibinfo{title}{{Results of TV imaging of phobos (experiment
  VSK-FREGAT)}}.
\newblock \bibinfo{journal}{Planetary and Space Science} \bibinfo{volume}{39},
  \bibinfo{pages}{281--295}.
\newblock \DOIprefix\doi{10.1016/0032-0633(91)90150-9}.
\bibitem[{{Badyukov}(2020)}]{Badyukov_2020}
\bibinfo{author}{{Badyukov}, D.D.}, \bibinfo{year}{2020}.
\newblock \bibinfo{title}{{Micrometeoroids: the Flux on the Moon and a Source
  of Volatiles}}.
\newblock \bibinfo{journal}{Solar System Research} \bibinfo{volume}{54},
  \bibinfo{pages}{263--274}.
\newblock \DOIprefix\doi{10.1134/S0038094620040024}.
\bibitem[{{Barucci} et~al.(2021){Barucci}, {Reess}, {Bernardi},
  {Doressoundiram}, {Fornasier}, {Le Du}, {Iwata}, {Nakagawa}, {Nakamura},
  {Andr{\'e}}, {Aoki}, {Arai}, {Baldit}, {Beck}, {Buey}, {Canalias},
  {Castelnau}, {Charnoz}, {Chaussidon}, {Chapron}, {Ciarletti}, {Delbo},
  {Dubois}, {Gauffre}, {Gautier}, {Genda}, {Hassen-Khodja}, {Hervet}, {Hyodo},
  {Imbert}, {Imamura}, {Jorda}, {Kameda}, {Kouach}, {Kouyama}, {Kuroda},
  {Kurokawa}, {Lapaw}, {Lasue}, {Le Deit}, {Ledot}, {Leyrat}, {Le Ruyet},
  {Matsuoka}, {Merlin}, {Miyamoto}, {Moynier}, {Nguyen Tuong}, {Ogohara},
  {Osawa}, {Parisot}, {Pistre}, {Quertier}, {Raymond}, {Rocard}, {Sakanoi},
  {Sato}, {Sawyer}, {Tache}, {Tr{\'e}moli{\`e}res}, {Tsuchiya}, {Vernazza} and
  {Zeganadin}}]{Barucci_2021}
\bibinfo{author}{{Barucci}, M.A.}, \bibinfo{author}{{Reess}, J.M.},
  \bibinfo{author}{{Bernardi}, P.}, \bibinfo{author}{{Doressoundiram}, A.},
  \bibinfo{author}{{Fornasier}, S.}, \bibinfo{author}{{Le Du}, M.},
  \bibinfo{author}{{Iwata}, T.}, \bibinfo{author}{{Nakagawa}, H.},
  \bibinfo{author}{{Nakamura}, T.}, \bibinfo{author}{{Andr{\'e}}, Y.},
  \bibinfo{author}{{Aoki}, S.}, \bibinfo{author}{{Arai}, T.},
  \bibinfo{author}{{Baldit}, E.}, \bibinfo{author}{{Beck}, P.},
  \bibinfo{author}{{Buey}, J.T.}, \bibinfo{author}{{Canalias}, E.},
  \bibinfo{author}{{Castelnau}, M.}, \bibinfo{author}{{Charnoz}, S.},
  \bibinfo{author}{{Chaussidon}, M.}, \bibinfo{author}{{Chapron}, F.},
  \bibinfo{author}{{Ciarletti}, V.}, \bibinfo{author}{{Delbo}, M.},
  \bibinfo{author}{{Dubois}, B.}, \bibinfo{author}{{Gauffre}, S.},
  \bibinfo{author}{{Gautier}, T.}, \bibinfo{author}{{Genda}, H.},
  \bibinfo{author}{{Hassen-Khodja}, R.}, \bibinfo{author}{{Hervet}, G.},
  \bibinfo{author}{{Hyodo}, R.}, \bibinfo{author}{{Imbert}, C.},
  \bibinfo{author}{{Imamura}, T.}, \bibinfo{author}{{Jorda}, L.},
  \bibinfo{author}{{Kameda}, S.}, \bibinfo{author}{{Kouach}, D.},
  \bibinfo{author}{{Kouyama}, T.}, \bibinfo{author}{{Kuroda}, T.},
  \bibinfo{author}{{Kurokawa}, H.}, \bibinfo{author}{{Lapaw}, L.},
  \bibinfo{author}{{Lasue}, J.}, \bibinfo{author}{{Le Deit}, L.},
  \bibinfo{author}{{Ledot}, A.}, \bibinfo{author}{{Leyrat}, C.},
  \bibinfo{author}{{Le Ruyet}, B.}, \bibinfo{author}{{Matsuoka}, M.},
  \bibinfo{author}{{Merlin}, F.}, \bibinfo{author}{{Miyamoto}, H.},
  \bibinfo{author}{{Moynier}, F.}, \bibinfo{author}{{Nguyen Tuong}, N.},
  \bibinfo{author}{{Ogohara}, K.}, \bibinfo{author}{{Osawa}, T.},
  \bibinfo{author}{{Parisot}, J.}, \bibinfo{author}{{Pistre}, L.},
  \bibinfo{author}{{Quertier}, B.}, \bibinfo{author}{{Raymond}, S.N.},
  \bibinfo{author}{{Rocard}, F.}, \bibinfo{author}{{Sakanoi}, T.},
  \bibinfo{author}{{Sato}, T.M.}, \bibinfo{author}{{Sawyer}, E.},
  \bibinfo{author}{{Tache}, F.}, \bibinfo{author}{{Tr{\'e}moli{\`e}res}, S.},
  \bibinfo{author}{{Tsuchiya}, F.}, \bibinfo{author}{{Vernazza}, P.},
  \bibinfo{author}{{Zeganadin}, D.}, \bibinfo{year}{2021}.
\newblock \bibinfo{title}{{MIRS: an imaging spectrometer for the MMX mission}}.
\newblock \bibinfo{journal}{Earth, Planets and Space} \bibinfo{volume}{73},
  \bibinfo{pages}{211}.
\newblock \DOIprefix\doi{10.1186/s40623-021-01423-2}.
\bibitem[{{Basilevsky} and {Keller}(2006)}]{Basilevsky_2006}
\bibinfo{author}{{Basilevsky}, A.T.}, \bibinfo{author}{{Keller}, H.U.},
  \bibinfo{year}{2006}.
\newblock \bibinfo{title}{{Comet nuclei: Morphology and implied processes of
  surface modification}}.
\newblock \bibinfo{journal}{Planetary and Space Science} \bibinfo{volume}{54},
  \bibinfo{pages}{808--829}.
\newblock \DOIprefix\doi{10.1016/j.pss.2006.05.001}.
\bibitem[{Basilevsky et~al.(2014)Basilevsky, Lorenz, Shingareva, Head, Ramsley
  and Zubarev}]{Basilevsky_2014}
\bibinfo{author}{Basilevsky, A.T.}, \bibinfo{author}{Lorenz, C.A.},
  \bibinfo{author}{Shingareva, T.V.}, \bibinfo{author}{Head, J.W.},
  \bibinfo{author}{Ramsley, K.R.}, \bibinfo{author}{Zubarev, A.E.},
  \bibinfo{year}{2014}.
\newblock \bibinfo{title}{The surface geology and geomorphology of {P}hobos}.
\newblock \bibinfo{journal}{Planetary and Space Science} \bibinfo{volume}{102},
  \bibinfo{pages}{95--118}.
\bibitem[{{Beck} and {Poch}(2021)}]{Beck_2021}
\bibinfo{author}{{Beck}, P.}, \bibinfo{author}{{Poch}, O.},
  \bibinfo{year}{2021}.
\newblock \bibinfo{title}{{Origins of colors variability among C-cluster
  main-belt asteroids}}.
\newblock \bibinfo{journal}{Icarus} \bibinfo{volume}{365},
  \bibinfo{pages}{114494}.
\newblock \DOIprefix\doi{10.1016/j.icarus.2021.114494},
  \href{http://arxiv.org/abs/2105.01912}{\tt arXiv:2105.01912}.
\bibitem[{{Beck} et~al.(2012){Beck}, {Pommerol}, {Thomas}, {Schmitt}, {Moynier}
  and {Barrat}}]{Beck_2012}
\bibinfo{author}{{Beck}, P.}, \bibinfo{author}{{Pommerol}, A.},
  \bibinfo{author}{{Thomas}, N.}, \bibinfo{author}{{Schmitt}, B.},
  \bibinfo{author}{{Moynier}, F.}, \bibinfo{author}{{Barrat}, J.A.},
  \bibinfo{year}{2012}.
\newblock \bibinfo{title}{{Photometry of meteorites}}.
\newblock \bibinfo{journal}{Icarus} \bibinfo{volume}{218},
  \bibinfo{pages}{364--377}.
\newblock \DOIprefix\doi{10.1016/j.icarus.2011.12.005}.
\bibitem[{Beech and Coulson(2010)}]{Beech_2010}
\bibinfo{author}{Beech, M.}, \bibinfo{author}{Coulson, I.},
  \bibinfo{year}{2010}.
\newblock \bibinfo{title}{The tagish lake meteorite: microstructural porosity
  variations and implications for parent body identification}.
\bibitem[{{Biele} et~al.(2020){Biele}, {Burke}, {Grott}, {Ryan},
  {DellaGiustina}, {Rozitis}, {Michel}, {Schroeder} and {Neumann}}]{Biele_2020}
\bibinfo{author}{{Biele}, J.}, \bibinfo{author}{{Burke}, K.N.},
  \bibinfo{author}{{Grott}, M.}, \bibinfo{author}{{Ryan}, A.J.},
  \bibinfo{author}{{DellaGiustina}, D.}, \bibinfo{author}{{Rozitis}, B.},
  \bibinfo{author}{{Michel}, P.}, \bibinfo{author}{{Schroeder}, S.},
  \bibinfo{author}{{Neumann}, W.}, \bibinfo{year}{2020}.
\newblock \bibinfo{title}{{Macroporosity and Grain Density of Rubble Pile
  Asteroid (101955) Bennu}}, in: \bibinfo{booktitle}{AGU Fall Meeting
  Abstracts}, pp. \bibinfo{pages}{P037--04}.
\bibitem[{{Bockel{\'e}e-Morvan} et~al.(2024){Bockel{\'e}e-Morvan}, {Lellouch},
  {Poch}, {Quirico}, {Cazaux}, {de Pater}, {Fouchet}, {Fry},
  {Rodriguez-Ovalle}, {Tosi}, {Wong}, {Boshuizen}, {de Kleer}, {Fletcher},
  {Meunier}, {Mura}, {Roth}, {Saur}, {Schmitt}, {Trumbo}, {Brown},
  {O'Donoghue}, {Orton} and {Showalter}}]{Bockelee_2024}
\bibinfo{author}{{Bockel{\'e}e-Morvan}, D.}, \bibinfo{author}{{Lellouch}, E.},
  \bibinfo{author}{{Poch}, O.}, \bibinfo{author}{{Quirico}, E.},
  \bibinfo{author}{{Cazaux}, S.}, \bibinfo{author}{{de Pater}, I.},
  \bibinfo{author}{{Fouchet}, T.}, \bibinfo{author}{{Fry}, P.M.},
  \bibinfo{author}{{Rodriguez-Ovalle}, P.}, \bibinfo{author}{{Tosi}, F.},
  \bibinfo{author}{{Wong}, M.H.}, \bibinfo{author}{{Boshuizen}, I.},
  \bibinfo{author}{{de Kleer}, K.}, \bibinfo{author}{{Fletcher}, L.N.},
  \bibinfo{author}{{Meunier}, L.}, \bibinfo{author}{{Mura}, A.},
  \bibinfo{author}{{Roth}, L.}, \bibinfo{author}{{Saur}, J.},
  \bibinfo{author}{{Schmitt}, B.}, \bibinfo{author}{{Trumbo}, S.K.},
  \bibinfo{author}{{Brown}, M.E.}, \bibinfo{author}{{O'Donoghue}, J.},
  \bibinfo{author}{{Orton}, G.S.}, \bibinfo{author}{{Showalter}, M.R.},
  \bibinfo{year}{2024}.
\newblock \bibinfo{title}{{Composition and thermal properties of Ganymede's
  surface from JWST/NIRSpec and MIRI observations}}.
\newblock \bibinfo{journal}{Astronomy and Astrophysics} \bibinfo{volume}{681},
  \bibinfo{pages}{A27}.
\newblock \DOIprefix\doi{10.1051/0004-6361/202347326},
  \href{http://arxiv.org/abs/2310.13982}{\tt arXiv:2310.13982}.
\bibitem[{{Britt} et~al.(2004){Britt}, {Boice}, {Buratti}, {Campins}, {Nelson},
  {Oberst}, {Sandel}, {Stern}, {Soderblom} and {Thomas}}]{Britt_2004}
\bibinfo{author}{{Britt}, D.T.}, \bibinfo{author}{{Boice}, D.C.},
  \bibinfo{author}{{Buratti}, B.J.}, \bibinfo{author}{{Campins}, H.},
  \bibinfo{author}{{Nelson}, R.M.}, \bibinfo{author}{{Oberst}, J.},
  \bibinfo{author}{{Sandel}, B.R.}, \bibinfo{author}{{Stern}, S.A.},
  \bibinfo{author}{{Soderblom}, L.A.}, \bibinfo{author}{{Thomas}, N.},
  \bibinfo{year}{2004}.
\newblock \bibinfo{title}{{The morphology and surface processes of Comet 19/P
  Borrelly}}.
\newblock \bibinfo{journal}{Icarus} \bibinfo{volume}{167},
  \bibinfo{pages}{45--53}.
\newblock \DOIprefix\doi{10.1016/j.icarus.2003.09.004}.
\bibitem[{{Britt} et~al.(2002){Britt}, {Yeomans}, {Housen} and
  {Consolmagno}}]{Britt_2002}
\bibinfo{author}{{Britt}, D.T.}, \bibinfo{author}{{Yeomans}, D.},
  \bibinfo{author}{{Housen}, K.}, \bibinfo{author}{{Consolmagno}, G.},
  \bibinfo{year}{2002}.
\newblock \bibinfo{title}{{Asteroid Density, Porosity, and Structure}}, in:
  \bibinfo{booktitle}{Asteroids III}, pp. \bibinfo{pages}{485--500}.
\bibitem[{{Brown}(2014)}]{Brown_2014}
\bibinfo{author}{{Brown}, A.J.}, \bibinfo{year}{2014}.
\newblock \bibinfo{title}{{Spectral bluing induced by small particles under the
  Mie and Rayleigh regimes}}.
\newblock \bibinfo{journal}{Icarus} \bibinfo{volume}{239},
  \bibinfo{pages}{85--95}.
\newblock \DOIprefix\doi{10.1016/j.icarus.2014.05.042}.
\bibitem[{{Cambioni} et~al.(2021){Cambioni}, {Delbo}, {Poggiali}, {Avdellidou},
  {Ryan}, {Deshapriya}, {Asphaug}, {Ballouz}, {Barucci}, {Bennett}, {Bottke},
  {Brucato}, {Burke}, {Cloutis}, {DellaGiustina}, {Emery}, {Rozitis}, {Walsh}
  and {Lauretta}}]{Cambioni_2021}
\bibinfo{author}{{Cambioni}, S.}, \bibinfo{author}{{Delbo}, M.},
  \bibinfo{author}{{Poggiali}, G.}, \bibinfo{author}{{Avdellidou}, C.},
  \bibinfo{author}{{Ryan}, A.J.}, \bibinfo{author}{{Deshapriya}, J.D.P.},
  \bibinfo{author}{{Asphaug}, E.}, \bibinfo{author}{{Ballouz}, R.L.},
  \bibinfo{author}{{Barucci}, M.A.}, \bibinfo{author}{{Bennett}, C.A.},
  \bibinfo{author}{{Bottke}, W.F.}, \bibinfo{author}{{Brucato}, J.R.},
  \bibinfo{author}{{Burke}, K.N.}, \bibinfo{author}{{Cloutis}, E.},
  \bibinfo{author}{{DellaGiustina}, D.N.}, \bibinfo{author}{{Emery}, J.P.},
  \bibinfo{author}{{Rozitis}, B.}, \bibinfo{author}{{Walsh}, K.J.},
  \bibinfo{author}{{Lauretta}, D.S.}, \bibinfo{year}{2021}.
\newblock \bibinfo{title}{{Fine-regolith production on asteroids controlled by
  rock porosity}}.
\newblock \bibinfo{journal}{Nature} \bibinfo{volume}{598},
  \bibinfo{pages}{49--52}.
\newblock \DOIprefix\doi{10.1038/s41586-021-03816-5}.
\bibitem[{{Clark} et~al.(1977){Clark}, {Baird}, {Rose}, {Toulmin}, {Christian},
  {Kelliher}, {Castro}, {Rowe}, {Keil} and {Huss}}]{Clark_1977}
\bibinfo{author}{{Clark}, B.C.}, \bibinfo{author}{{Baird}, A.K.},
  \bibinfo{author}{{Rose}, H.J.}, \bibinfo{author}{{Toulmin}, P.},
  \bibinfo{author}{{Christian}, R.P.}, \bibinfo{author}{{Kelliher}, W.C.},
  \bibinfo{author}{{Castro}, A.J.}, \bibinfo{author}{{Rowe}, C.D.},
  \bibinfo{author}{{Keil}, K.}, \bibinfo{author}{{Huss}, G.R.},
  \bibinfo{year}{1977}.
\newblock \bibinfo{title}{{The Viking X ray fluorescence experiment: Analytical
  methods and early results}}.
\newblock \bibinfo{journal}{Journal of Geophysics Research}
  \bibinfo{volume}{82}, \bibinfo{pages}{4577--4594}.
\newblock \DOIprefix\doi{10.1029/JS082i028p04577}.
\bibitem[{Clark et~al.(2008)Clark, Curchin, Jaumann, Cruikshank, Brown, Hoefen,
  Stephan, Moore, Buratti, Baines, Nicholson and Nelson}]{Clark_2008}
\bibinfo{author}{Clark, R.N.}, \bibinfo{author}{Curchin, J.M.},
  \bibinfo{author}{Jaumann, R.}, \bibinfo{author}{Cruikshank, D.P.},
  \bibinfo{author}{Brown, R.H.}, \bibinfo{author}{Hoefen, T.M.},
  \bibinfo{author}{Stephan, K.}, \bibinfo{author}{Moore, J.M.},
  \bibinfo{author}{Buratti, B.J.}, \bibinfo{author}{Baines, K.H.},
  \bibinfo{author}{Nicholson, P.D.}, \bibinfo{author}{Nelson, R.M.},
  \bibinfo{year}{2008}.
\newblock \bibinfo{title}{Compositional mapping of saturn's satellite dione
  with cassini vims and implications of dark material in the saturn system}.
\newblock \bibinfo{journal}{Icarus} \bibinfo{volume}{193},
  \bibinfo{pages}{372--386}.
\newblock \URLprefix
  \url{https://www.sciencedirect.com/science/article/pii/S0019103507003740},
  \DOIprefix\doi{https://doi.org/10.1016/j.icarus.2007.08.035}.
  \bibinfo{note}{saturn's Icy Satellites from Cassini}.
\bibitem[{{Cloutis} et~al.(2018){Cloutis}, {Pietrasz}, {Kiddell}, {Izawa},
  {Vernazza}, {Burbine}, {DeMeo}, {Tait}, {Bell}, {Mann}, {Applin} and
  {Reddy}}]{Cloutis_2018}
\bibinfo{author}{{Cloutis}, E.A.}, \bibinfo{author}{{Pietrasz}, V.B.},
  \bibinfo{author}{{Kiddell}, C.}, \bibinfo{author}{{Izawa}, M.R.M.},
  \bibinfo{author}{{Vernazza}, P.}, \bibinfo{author}{{Burbine}, T.H.},
  \bibinfo{author}{{DeMeo}, F.}, \bibinfo{author}{{Tait}, K.T.},
  \bibinfo{author}{{Bell}, J.F.}, \bibinfo{author}{{Mann}, P.},
  \bibinfo{author}{{Applin}, D.M.}, \bibinfo{author}{{Reddy}, V.},
  \bibinfo{year}{2018}.
\newblock \bibinfo{title}{{Spectral reflectance ``deconstruction'' of the
  Murchison CM2 carbonaceous chondrite and implications for spectroscopic
  investigations of dark asteroids}}.
\newblock \bibinfo{journal}{Icarus} \bibinfo{volume}{305},
  \bibinfo{pages}{203--224}.
\newblock \DOIprefix\doi{10.1016/j.icarus.2018.01.015}.
\bibitem[{{Consolmagno} et~al.(2008){Consolmagno}, {Britt} and
  {Macke}}]{Consolmagno_2008}
\bibinfo{author}{{Consolmagno}, G.}, \bibinfo{author}{{Britt}, D.},
  \bibinfo{author}{{Macke}, R.}, \bibinfo{year}{2008}.
\newblock \bibinfo{title}{{The significance of meteorite density and
  porosity}}.
\newblock \bibinfo{journal}{Chemie der Erde / Geochemistry}
  \bibinfo{volume}{68}, \bibinfo{pages}{1--29}.
\newblock \DOIprefix\doi{10.1016/j.chemer.2008.01.003}.
\bibitem[{{De Sanctis} et~al.(2015){De Sanctis}, {Capaccioni}, {Ciarniello},
  {Filacchione}, {Formisano}, {Mottola}, {Raponi}, {Tosi},
  {Bockel{\'e}e-Morvan}, {Erard}, {Leyrat}, {Schmitt}, {Ammannito}, {Arnold},
  {Barucci}, {Combi}, {Capria}, {Cerroni}, {Ip}, {Kuehrt}, {McCord}, {Palomba},
  {Beck}, {Quirico}, {VIRTIS Team}, {Piccioni}, {Bellucci}, {Fulchignoni},
  {Jaumann}, {Stephan}, {Longobardo}, {Mennella}, {Migliorini}, {Benkhoff},
  {Bibring}, {Blanco}, {Blecka}, {Carlson}, {Carsenty}, {Colangeli}, {Combes},
  {Crovisier}, {Drossart}, {Encrenaz}, {Federico}, {Fink}, {Fonti}, {Irwin},
  {Langevin}, {Magni}, {Moroz}, {Orofino}, {Schade}, {Taylor}, {Tiphene},
  {Tozzi}, {Biver}, {Bonal}, {Combe}, {Despan}, {Flamini}, {Fornasier},
  {Frigeri}, {Grassi}, {Gudipati}, {Mancarella}, {Markus}, {Merlin}, {Orosei},
  {Rinaldi}, {Cartacci}, {Cicchetti}, {Giuppi}, {Hello}, {Henry}, {Jacquinod},
  {Rees}, {Noschese}, {Politi} and {Peter}}]{DeSanctis_2015}
\bibinfo{author}{{De Sanctis}, M.C.}, \bibinfo{author}{{Capaccioni}, F.},
  \bibinfo{author}{{Ciarniello}, M.}, \bibinfo{author}{{Filacchione}, G.},
  \bibinfo{author}{{Formisano}, M.}, \bibinfo{author}{{Mottola}, S.},
  \bibinfo{author}{{Raponi}, A.}, \bibinfo{author}{{Tosi}, F.},
  \bibinfo{author}{{Bockel{\'e}e-Morvan}, D.}, \bibinfo{author}{{Erard}, S.},
  \bibinfo{author}{{Leyrat}, C.}, \bibinfo{author}{{Schmitt}, B.},
  \bibinfo{author}{{Ammannito}, E.}, \bibinfo{author}{{Arnold}, G.},
  \bibinfo{author}{{Barucci}, M.A.}, \bibinfo{author}{{Combi}, M.},
  \bibinfo{author}{{Capria}, M.T.}, \bibinfo{author}{{Cerroni}, P.},
  \bibinfo{author}{{Ip}, W.H.}, \bibinfo{author}{{Kuehrt}, E.},
  \bibinfo{author}{{McCord}, T.B.}, \bibinfo{author}{{Palomba}, E.},
  \bibinfo{author}{{Beck}, P.}, \bibinfo{author}{{Quirico}, E.},
  \bibinfo{author}{{VIRTIS Team}}, \bibinfo{author}{{Piccioni}, G.},
  \bibinfo{author}{{Bellucci}, G.}, \bibinfo{author}{{Fulchignoni}, M.},
  \bibinfo{author}{{Jaumann}, R.}, \bibinfo{author}{{Stephan}, K.},
  \bibinfo{author}{{Longobardo}, A.}, \bibinfo{author}{{Mennella}, V.},
  \bibinfo{author}{{Migliorini}, A.}, \bibinfo{author}{{Benkhoff}, J.},
  \bibinfo{author}{{Bibring}, J.P.}, \bibinfo{author}{{Blanco}, A.},
  \bibinfo{author}{{Blecka}, M.}, \bibinfo{author}{{Carlson}, R.},
  \bibinfo{author}{{Carsenty}, U.}, \bibinfo{author}{{Colangeli}, L.},
  \bibinfo{author}{{Combes}, M.}, \bibinfo{author}{{Crovisier}, J.},
  \bibinfo{author}{{Drossart}, P.}, \bibinfo{author}{{Encrenaz}, T.},
  \bibinfo{author}{{Federico}, C.}, \bibinfo{author}{{Fink}, U.},
  \bibinfo{author}{{Fonti}, S.}, \bibinfo{author}{{Irwin}, P.},
  \bibinfo{author}{{Langevin}, Y.}, \bibinfo{author}{{Magni}, G.},
  \bibinfo{author}{{Moroz}, L.}, \bibinfo{author}{{Orofino}, V.},
  \bibinfo{author}{{Schade}, U.}, \bibinfo{author}{{Taylor}, F.},
  \bibinfo{author}{{Tiphene}, D.}, \bibinfo{author}{{Tozzi}, G.P.},
  \bibinfo{author}{{Biver}, N.}, \bibinfo{author}{{Bonal}, L.},
  \bibinfo{author}{{Combe}, J.P.}, \bibinfo{author}{{Despan}, D.},
  \bibinfo{author}{{Flamini}, E.}, \bibinfo{author}{{Fornasier}, S.},
  \bibinfo{author}{{Frigeri}, A.}, \bibinfo{author}{{Grassi}, D.},
  \bibinfo{author}{{Gudipati}, M.S.}, \bibinfo{author}{{Mancarella}, F.},
  \bibinfo{author}{{Markus}, K.}, \bibinfo{author}{{Merlin}, F.},
  \bibinfo{author}{{Orosei}, R.}, \bibinfo{author}{{Rinaldi}, G.},
  \bibinfo{author}{{Cartacci}, M.}, \bibinfo{author}{{Cicchetti}, A.},
  \bibinfo{author}{{Giuppi}, S.}, \bibinfo{author}{{Hello}, Y.},
  \bibinfo{author}{{Henry}, F.}, \bibinfo{author}{{Jacquinod}, S.},
  \bibinfo{author}{{Rees}, J.M.}, \bibinfo{author}{{Noschese}, R.},
  \bibinfo{author}{{Politi}, R.}, \bibinfo{author}{{Peter}, G.},
  \bibinfo{year}{2015}.
\newblock \bibinfo{title}{{The diurnal cycle of water ice on comet
  67P/Churyumov-Gerasimenko}}.
\newblock \bibinfo{journal}{Nature} \bibinfo{volume}{525},
  \bibinfo{pages}{500--503}.
\newblock \DOIprefix\doi{10.1038/nature14869}.
\bibitem[{{Emery} et~al.(2006){Emery}, {Cruikshank} and {Van
  Cleve}}]{Emery_2006}
\bibinfo{author}{{Emery}, J.P.}, \bibinfo{author}{{Cruikshank}, D.P.},
  \bibinfo{author}{{Van Cleve}, J.}, \bibinfo{year}{2006}.
\newblock \bibinfo{title}{{Thermal emission spectroscopy (5.2 38
  {\ensuremath{\mu}}m) of three Trojan asteroids with the Spitzer Space
  Telescope: Detection of fine-grained silicates}}.
\newblock \bibinfo{journal}{Icarus} \bibinfo{volume}{182},
  \bibinfo{pages}{496--512}.
\newblock \DOIprefix\doi{10.1016/j.icarus.2006.01.011}.
\bibitem[{{Engrand} et~al.(2023){Engrand}, {Lasue}, {Wooden} and
  {Zolensky}}]{Engrand_2023}
\bibinfo{author}{{Engrand}, C.}, \bibinfo{author}{{Lasue}, J.},
  \bibinfo{author}{{Wooden}, D.H.}, \bibinfo{author}{{Zolensky}, M.E.},
  \bibinfo{year}{2023}.
\newblock \bibinfo{title}{{Chemical and physical properties of cometary dust}}.
\newblock \bibinfo{journal}{arXiv e-prints} ,
  \bibinfo{pages}{arXiv:2305.03417}\DOIprefix\doi{10.48550/arXiv.2305.03417},
  \href{http://arxiv.org/abs/2305.03417}{\tt arXiv:2305.03417}.
\bibitem[{Fanale and Salvail(1989)}]{Fanale_1989}
\bibinfo{author}{Fanale, F.P.}, \bibinfo{author}{Salvail, J.R.},
  \bibinfo{year}{1989}.
\newblock \bibinfo{title}{Loss of water from phobos}.
\newblock \bibinfo{journal}{Geophysical Research Letters} \bibinfo{volume}{16},
  \bibinfo{pages}{287--290}.
\newblock \URLprefix
  \url{https://agupubs.onlinelibrary.wiley.com/doi/abs/10.1029/GL016i004p00287},
  \DOIprefix\doi{https://doi.org/10.1029/GL016i004p00287},
  \href{http://arxiv.org/abs/https://agupubs.onlinelibrary.wiley.com/doi/pdf/10.1029/GL016i004p00287}{\tt
  arXiv:https://agupubs.onlinelibrary.wiley.com/doi/pdf/10.1029/GL016i004p00287}.
\bibitem[{{Feller} et~al.(2016){Feller}, {Fornasier}, {Hasselmann}, {Barucci},
  {Preusker}, {Scholten}, {Jorda}, {Pommerol}, {Jost}, {Poch}, {ElMaary},
  {Thomas}, {Belskaya}, {Pajola}, {Sierks}, {Barbieri}, {Lamy}, {Koschny},
  {Rickman}, {Rodrigo}, {Agarwal}, {A'Hearn}, {Bertaux}, {Bertini},
  {Boudreault}, {Cremonese}, {Da Deppo}, {Davidsson}, {Debei}, {De Cecco},
  {Deller}, {Fulle}, {Giquel}, {Groussin}, {Gutierrez}, {G{\"u}ttler},
  {Hofmann}, {Hviid}, {Keller}, {Ip}, {Knollenberg}, {Kovacs}, {Kramm},
  {K{\"u}hrt}, {K{\"u}ppers}, {Lara}, {Lazzarin}, {Leyrat}, {Lopez Moreno},
  {Marzari}, {Masoumzadeh}, {Mottola}, {Naletto}, {Perna}, {Oklay}, {Shi},
  {Tubiana} and {Vincent}}]{Feller_2016}
\bibinfo{author}{{Feller}, C.}, \bibinfo{author}{{Fornasier}, S.},
  \bibinfo{author}{{Hasselmann}, P.H.}, \bibinfo{author}{{Barucci}, A.},
  \bibinfo{author}{{Preusker}, F.}, \bibinfo{author}{{Scholten}, F.},
  \bibinfo{author}{{Jorda}, L.}, \bibinfo{author}{{Pommerol}, A.},
  \bibinfo{author}{{Jost}, B.}, \bibinfo{author}{{Poch}, O.},
  \bibinfo{author}{{ElMaary}, M.R.}, \bibinfo{author}{{Thomas}, N.},
  \bibinfo{author}{{Belskaya}, I.}, \bibinfo{author}{{Pajola}, M.},
  \bibinfo{author}{{Sierks}, H.}, \bibinfo{author}{{Barbieri}, C.},
  \bibinfo{author}{{Lamy}, P.L.}, \bibinfo{author}{{Koschny}, D.},
  \bibinfo{author}{{Rickman}, H.}, \bibinfo{author}{{Rodrigo}, R.},
  \bibinfo{author}{{Agarwal}, J.}, \bibinfo{author}{{A'Hearn}, M.},
  \bibinfo{author}{{Bertaux}, J.L.}, \bibinfo{author}{{Bertini}, I.},
  \bibinfo{author}{{Boudreault}, S.}, \bibinfo{author}{{Cremonese}, G.},
  \bibinfo{author}{{Da Deppo}, V.}, \bibinfo{author}{{Davidsson}, B.J.R.},
  \bibinfo{author}{{Debei}, S.}, \bibinfo{author}{{De Cecco}, M.},
  \bibinfo{author}{{Deller}, J.}, \bibinfo{author}{{Fulle}, M.},
  \bibinfo{author}{{Giquel}, A.}, \bibinfo{author}{{Groussin}, O.},
  \bibinfo{author}{{Gutierrez}, P.J.}, \bibinfo{author}{{G{\"u}ttler}, C.},
  \bibinfo{author}{{Hofmann}, M.}, \bibinfo{author}{{Hviid}, S.F.},
  \bibinfo{author}{{Keller}, H.}, \bibinfo{author}{{Ip}, W.H.},
  \bibinfo{author}{{Knollenberg}, J.}, \bibinfo{author}{{Kovacs}, G.},
  \bibinfo{author}{{Kramm}, J.R.}, \bibinfo{author}{{K{\"u}hrt}, E.},
  \bibinfo{author}{{K{\"u}ppers}, M.}, \bibinfo{author}{{Lara}, M.L.},
  \bibinfo{author}{{Lazzarin}, M.}, \bibinfo{author}{{Leyrat}, C.},
  \bibinfo{author}{{Lopez Moreno}, J.J.}, \bibinfo{author}{{Marzari}, F.},
  \bibinfo{author}{{Masoumzadeh}, N.}, \bibinfo{author}{{Mottola}, S.},
  \bibinfo{author}{{Naletto}, G.}, \bibinfo{author}{{Perna}, D.},
  \bibinfo{author}{{Oklay}, N.}, \bibinfo{author}{{Shi}, X.},
  \bibinfo{author}{{Tubiana}, C.}, \bibinfo{author}{{Vincent}, J.B.},
  \bibinfo{year}{2016}.
\newblock \bibinfo{title}{{Decimetre-scaled spectrophotometric properties of
  the nucleus of comet 67P/Churyumov-Gerasimenko from OSIRIS observations}}.
\newblock \bibinfo{journal}{Monthly Notices of the Royal Astronomical Society}
  \bibinfo{volume}{462}, \bibinfo{pages}{S287--S303}.
\newblock \DOIprefix\doi{10.1093/mnras/stw2511},
  \href{http://arxiv.org/abs/1611.00012}{\tt arXiv:1611.00012}.
\bibitem[{{Fernando} et~al.(2015){Fernando}, {Schmidt}, {Pilorget}, {Pinet},
  {Ceamanos}, {Dout{\'e}}, {Daydou} and {Costard}}]{Fernando_2015}
\bibinfo{author}{{Fernando}, J.}, \bibinfo{author}{{Schmidt}, F.},
  \bibinfo{author}{{Pilorget}, C.}, \bibinfo{author}{{Pinet}, P.},
  \bibinfo{author}{{Ceamanos}, X.}, \bibinfo{author}{{Dout{\'e}}, S.},
  \bibinfo{author}{{Daydou}, Y.}, \bibinfo{author}{{Costard}, F.},
  \bibinfo{year}{2015}.
\newblock \bibinfo{title}{{Characterization and mapping of surface physical
  properties of Mars from CRISM multi-angular data: Application to Gusev Crater
  and Meridiani Planum}}.
\newblock \bibinfo{journal}{Icarus} \bibinfo{volume}{253},
  \bibinfo{pages}{271--295}.
\newblock \DOIprefix\doi{10.1016/j.icarus.2015.03.012},
  \href{http://arxiv.org/abs/1408.5301}{\tt arXiv:1408.5301}.
\bibitem[{{Fornasier} et~al.(2015){Fornasier}, {Hasselmann}, {Barucci},
  {Feller}, {Besse}, {Leyrat}, {Lara}, {Gutierrez}, {Oklay}, {Tubiana},
  {Scholten}, {Sierks}, {Barbieri}, {Lamy}, {Rodrigo}, {Koschny}, {Rickman},
  {Keller}, {Agarwal}, {A'Hearn}, {Bertaux}, {Bertini}, {Cremonese}, {Da
  Deppo}, {Davidsson}, {Debei}, {De Cecco}, {Fulle}, {Groussin}, {G{\"u}ttler},
  {Hviid}, {Ip}, {Jorda}, {Knollenberg}, {Kovacs}, {Kramm}, {K{\"u}hrt},
  {K{\"u}ppers}, {La Forgia}, {Lazzarin}, {Lopez Moreno}, {Marzari}, {Matz},
  {Michalik}, {Moreno}, {Mottola}, {Naletto}, {Pajola}, {Pommerol}, {Preusker},
  {Shi}, {Snodgrass}, {Thomas} and {Vincent}}]{Fornasier_2015}
\bibinfo{author}{{Fornasier}, S.}, \bibinfo{author}{{Hasselmann}, P.H.},
  \bibinfo{author}{{Barucci}, M.A.}, \bibinfo{author}{{Feller}, C.},
  \bibinfo{author}{{Besse}, S.}, \bibinfo{author}{{Leyrat}, C.},
  \bibinfo{author}{{Lara}, L.}, \bibinfo{author}{{Gutierrez}, P.J.},
  \bibinfo{author}{{Oklay}, N.}, \bibinfo{author}{{Tubiana}, C.},
  \bibinfo{author}{{Scholten}, F.}, \bibinfo{author}{{Sierks}, H.},
  \bibinfo{author}{{Barbieri}, C.}, \bibinfo{author}{{Lamy}, P.L.},
  \bibinfo{author}{{Rodrigo}, R.}, \bibinfo{author}{{Koschny}, D.},
  \bibinfo{author}{{Rickman}, H.}, \bibinfo{author}{{Keller}, H.U.},
  \bibinfo{author}{{Agarwal}, J.}, \bibinfo{author}{{A'Hearn}, M.F.},
  \bibinfo{author}{{Bertaux}, J.L.}, \bibinfo{author}{{Bertini}, I.},
  \bibinfo{author}{{Cremonese}, G.}, \bibinfo{author}{{Da Deppo}, V.},
  \bibinfo{author}{{Davidsson}, B.}, \bibinfo{author}{{Debei}, S.},
  \bibinfo{author}{{De Cecco}, M.}, \bibinfo{author}{{Fulle}, M.},
  \bibinfo{author}{{Groussin}, O.}, \bibinfo{author}{{G{\"u}ttler}, C.},
  \bibinfo{author}{{Hviid}, S.F.}, \bibinfo{author}{{Ip}, W.},
  \bibinfo{author}{{Jorda}, L.}, \bibinfo{author}{{Knollenberg}, J.},
  \bibinfo{author}{{Kovacs}, G.}, \bibinfo{author}{{Kramm}, R.},
  \bibinfo{author}{{K{\"u}hrt}, E.}, \bibinfo{author}{{K{\"u}ppers}, M.},
  \bibinfo{author}{{La Forgia}, F.}, \bibinfo{author}{{Lazzarin}, M.},
  \bibinfo{author}{{Lopez Moreno}, J.J.}, \bibinfo{author}{{Marzari}, F.},
  \bibinfo{author}{{Matz}, K.D.}, \bibinfo{author}{{Michalik}, H.},
  \bibinfo{author}{{Moreno}, F.}, \bibinfo{author}{{Mottola}, S.},
  \bibinfo{author}{{Naletto}, G.}, \bibinfo{author}{{Pajola}, M.},
  \bibinfo{author}{{Pommerol}, A.}, \bibinfo{author}{{Preusker}, F.},
  \bibinfo{author}{{Shi}, X.}, \bibinfo{author}{{Snodgrass}, C.},
  \bibinfo{author}{{Thomas}, N.}, \bibinfo{author}{{Vincent}, J.B.},
  \bibinfo{year}{2015}.
\newblock \bibinfo{title}{Spectrophotometric properties of the nucleus of comet
  67{P}/{C}huryumov-{G}erasimenko from the {OSIRIS} instrument onboard the
  {ROSETTA} spacecraft}.
\newblock \bibinfo{journal}{Astronomy and Astrophysics} \bibinfo{volume}{583}.
\bibitem[{{Fornasier} et~al.(2020){Fornasier}, {Hasselmann}, {Deshapriya},
  {Barucci}, {Clark}, {Praet}, {Hamilton}, {Simon}, {Li}, {Cloutis}, {Merlin},
  {Zou} and {Lauretta}}]{Fornasier_2020}
\bibinfo{author}{{Fornasier}, S.}, \bibinfo{author}{{Hasselmann}, P.H.},
  \bibinfo{author}{{Deshapriya}, J.D.P.}, \bibinfo{author}{{Barucci}, M.A.},
  \bibinfo{author}{{Clark}, B.E.}, \bibinfo{author}{{Praet}, A.},
  \bibinfo{author}{{Hamilton}, V.E.}, \bibinfo{author}{{Simon}, A.},
  \bibinfo{author}{{Li}, J.Y.}, \bibinfo{author}{{Cloutis}, E.A.},
  \bibinfo{author}{{Merlin}, F.}, \bibinfo{author}{{Zou}, X.D.},
  \bibinfo{author}{{Lauretta}, D.S.}, \bibinfo{year}{2020}.
\newblock \bibinfo{title}{{Phase reddening on asteroid Bennu from visible and
  near-infrared spectroscopy}}.
\newblock \bibinfo{journal}{Astronomy and Astrophysics} \bibinfo{volume}{644},
  \bibinfo{pages}{A142}.
\newblock \DOIprefix\doi{10.1051/0004-6361/202039552},
  \href{http://arxiv.org/abs/2011.09339}{\tt arXiv:2011.09339}.
\bibitem[{{Fornasier} et~al.(2016){Fornasier}, {Mottola}, {Keller}, {Barucci},
  {Davidsson}, {Feller}, {Deshapriya}, {Sierks}, {Barbieri}, {Lamy}, {Rodrigo},
  {Koschny}, {Rickman}, {A'Hearn}, {Agarwal}, {Bertaux}, {Bertini}, {Besse},
  {Cremonese}, {Da Deppo}, {Debei}, {De Cecco}, {Deller}, {El-Maarry}, {Fulle},
  {Groussin}, {Gutierrez}, {G{\"u}ttler}, {Hofmann}, {Hviid}, {Ip}, {Jorda},
  {Knollenberg}, {Kovacs}, {Kramm}, {K{\"u}hrt}, {K{\"u}ppers}, {Lara},
  {Lazzarin}, {Moreno}, {Marzari}, {Massironi}, {Naletto}, {Oklay}, {Pajola},
  {Pommerol}, {Preusker}, {Scholten}, {Shi}, {Thomas}, {Toth}, {Tubiana} and
  {Vincent}}]{Fornasier_2016}
\bibinfo{author}{{Fornasier}, S.}, \bibinfo{author}{{Mottola}, S.},
  \bibinfo{author}{{Keller}, H.U.}, \bibinfo{author}{{Barucci}, M.A.},
  \bibinfo{author}{{Davidsson}, B.}, \bibinfo{author}{{Feller}, C.},
  \bibinfo{author}{{Deshapriya}, J.D.P.}, \bibinfo{author}{{Sierks}, H.},
  \bibinfo{author}{{Barbieri}, C.}, \bibinfo{author}{{Lamy}, P.L.},
  \bibinfo{author}{{Rodrigo}, R.}, \bibinfo{author}{{Koschny}, D.},
  \bibinfo{author}{{Rickman}, H.}, \bibinfo{author}{{A'Hearn}, M.},
  \bibinfo{author}{{Agarwal}, J.}, \bibinfo{author}{{Bertaux}, J.L.},
  \bibinfo{author}{{Bertini}, I.}, \bibinfo{author}{{Besse}, S.},
  \bibinfo{author}{{Cremonese}, G.}, \bibinfo{author}{{Da Deppo}, V.},
  \bibinfo{author}{{Debei}, S.}, \bibinfo{author}{{De Cecco}, M.},
  \bibinfo{author}{{Deller}, J.}, \bibinfo{author}{{El-Maarry}, M.R.},
  \bibinfo{author}{{Fulle}, M.}, \bibinfo{author}{{Groussin}, O.},
  \bibinfo{author}{{Gutierrez}, P.J.}, \bibinfo{author}{{G{\"u}ttler}, C.},
  \bibinfo{author}{{Hofmann}, M.}, \bibinfo{author}{{Hviid}, S.F.},
  \bibinfo{author}{{Ip}, W.H.}, \bibinfo{author}{{Jorda}, L.},
  \bibinfo{author}{{Knollenberg}, J.}, \bibinfo{author}{{Kovacs}, G.},
  \bibinfo{author}{{Kramm}, R.}, \bibinfo{author}{{K{\"u}hrt}, E.},
  \bibinfo{author}{{K{\"u}ppers}, M.}, \bibinfo{author}{{Lara}, M.L.},
  \bibinfo{author}{{Lazzarin}, M.}, \bibinfo{author}{{Moreno}, J.J.L.},
  \bibinfo{author}{{Marzari}, F.}, \bibinfo{author}{{Massironi}, M.},
  \bibinfo{author}{{Naletto}, G.}, \bibinfo{author}{{Oklay}, N.},
  \bibinfo{author}{{Pajola}, M.}, \bibinfo{author}{{Pommerol}, A.},
  \bibinfo{author}{{Preusker}, F.}, \bibinfo{author}{{Scholten}, F.},
  \bibinfo{author}{{Shi}, X.}, \bibinfo{author}{{Thomas}, N.},
  \bibinfo{author}{{Toth}, I.}, \bibinfo{author}{{Tubiana}, C.},
  \bibinfo{author}{{Vincent}, J.B.}, \bibinfo{year}{2016}.
\newblock \bibinfo{title}{{Rosetta{\textquoteright}s comet
  67P/Churyumov-Gerasimenko sheds its dusty mantle to reveal its icy nature}}.
\newblock \bibinfo{journal}{Science} \bibinfo{volume}{354},
  \bibinfo{pages}{1566--1570}.
\newblock \DOIprefix\doi{10.1126/science.aag2671}.
\bibitem[{{Fornasier} et~al.(2024){Fornasier}, {Wargnier}, {Hasselmann},
  {Tirsch}, {Matz}, {Doressoundiram}, {Gautier} and {Barucci}}]{Fornasier_2023}
\bibinfo{author}{{Fornasier}, S.}, \bibinfo{author}{{Wargnier}, A.},
  \bibinfo{author}{{Hasselmann}, P.H.}, \bibinfo{author}{{Tirsch}, D.},
  \bibinfo{author}{{Matz}, K.D.}, \bibinfo{author}{{Doressoundiram}, A.},
  \bibinfo{author}{{Gautier}, T.}, \bibinfo{author}{{Barucci}, M.A.},
  \bibinfo{year}{2024}.
\newblock \bibinfo{title}{{Phobos photometric properties from Mars Express HRSC
  observations}}.
\newblock \bibinfo{journal}{Astronomy and Astrophysics} \bibinfo{volume}{686},
  \bibinfo{pages}{A203}.
\newblock \DOIprefix\doi{10.1051/0004-6361/202449220},
  \href{http://arxiv.org/abs/2403.12156}{\tt arXiv:2403.12156}.
\bibitem[{{Fraeman} et~al.(2012){Fraeman}, {Arvidson}, {Murchie}, {Rivkin},
  {Bibring}, {Choo}, {Gondet}, {Humm}, {Kuzmin}, {Manaud} and
  {Zabalueva}}]{Fraeman_2012}
\bibinfo{author}{{Fraeman}, A.A.}, \bibinfo{author}{{Arvidson}, R.E.},
  \bibinfo{author}{{Murchie}, S.L.}, \bibinfo{author}{{Rivkin}, A.},
  \bibinfo{author}{{Bibring}, J.P.}, \bibinfo{author}{{Choo}, T.H.},
  \bibinfo{author}{{Gondet}, B.}, \bibinfo{author}{{Humm}, D.},
  \bibinfo{author}{{Kuzmin}, R.O.}, \bibinfo{author}{{Manaud}, N.},
  \bibinfo{author}{{Zabalueva}, E.V.}, \bibinfo{year}{2012}.
\newblock \bibinfo{title}{{Analysis of disk-resolved OMEGA and CRISM spectral
  observations of Phobos and Deimos}}.
\newblock \bibinfo{journal}{Journal of Geophysical Research (Planets)}
  \bibinfo{volume}{117}, \bibinfo{pages}{E00J15}.
\newblock \DOIprefix\doi{10.1029/2012JE004137}.
\bibitem[{{Fraeman} et~al.(2014){Fraeman}, {Murchie}, {Arvidson}, {Clark},
  {Morris}, {Rivkin} and {Vilas}}]{Fraeman_2014}
\bibinfo{author}{{Fraeman}, A.A.}, \bibinfo{author}{{Murchie}, S.L.},
  \bibinfo{author}{{Arvidson}, R.E.}, \bibinfo{author}{{Clark}, R.N.},
  \bibinfo{author}{{Morris}, R.V.}, \bibinfo{author}{{Rivkin}, A.S.},
  \bibinfo{author}{{Vilas}, F.}, \bibinfo{year}{2014}.
\newblock \bibinfo{title}{{Spectral absorptions on Phobos and Deimos in the
  visible/near infrared wavelengths and their compositional constraints}}.
\newblock \bibinfo{journal}{Icarus} \bibinfo{volume}{229},
  \bibinfo{pages}{196--205}.
\newblock \DOIprefix\doi{10.1016/j.icarus.2013.11.021}.
\bibitem[{{Giuranna} et~al.(2011){Giuranna}, {Roush}, {Duxbury}, {Hogan},
  {Carli}, {Geminale} and {Formisano}}]{Giuranna_2011}
\bibinfo{author}{{Giuranna}, M.}, \bibinfo{author}{{Roush}, T.L.},
  \bibinfo{author}{{Duxbury}, T.}, \bibinfo{author}{{Hogan}, R.C.},
  \bibinfo{author}{{Carli}, C.}, \bibinfo{author}{{Geminale}, A.},
  \bibinfo{author}{{Formisano}, V.}, \bibinfo{year}{2011}.
\newblock \bibinfo{title}{{Compositional interpretation of PFS/MEx and TES/MGS
  thermal infrared spectra of Phobos}}.
\newblock \bibinfo{journal}{Planetary and Space Science} \bibinfo{volume}{59},
  \bibinfo{pages}{1308--1325}.
\newblock \DOIprefix\doi{10.1016/j.pss.2011.01.019}.
\bibitem[{{Glotch} et~al.(2018){Glotch}, {Edwards}, {Yesiltas}, {Shirley},
  {McDougall}, {Kling}, {Bandfield} and {Herd}}]{Glotch_2018}
\bibinfo{author}{{Glotch}, T.D.}, \bibinfo{author}{{Edwards}, C.S.},
  \bibinfo{author}{{Yesiltas}, M.}, \bibinfo{author}{{Shirley}, K.A.},
  \bibinfo{author}{{McDougall}, D.S.}, \bibinfo{author}{{Kling}, A.M.},
  \bibinfo{author}{{Bandfield}, J.L.}, \bibinfo{author}{{Herd}, C.D.K.},
  \bibinfo{year}{2018}.
\newblock \bibinfo{title}{{MGS-TES Spectra Suggest a Basaltic Component in the
  Regolith of Phobos}}.
\newblock \bibinfo{journal}{Journal of Geophysical Research (Planets)}
  \bibinfo{volume}{123}, \bibinfo{pages}{2467--2484}.
\newblock \DOIprefix\doi{10.1029/2018JE005647}.
\bibitem[{Grisolle(2013)}]{Grisolle_2013}
\bibinfo{author}{Grisolle, F.}, \bibinfo{year}{2013}.
\newblock \bibinfo{title}{{Les condensats saisonniers de Mars : {\'e}tude
  exp{\'e}rimentale de la formation et du m{\'e}tamorphisme de glaces de CO2}}.
\newblock \bibinfo{type}{Theses}. {Universit{\'e} de Grenoble}.
\newblock \URLprefix \url{https://theses.hal.science/tel-01167247}.
\bibitem[{{Grott} et~al.(2020){Grott}, {Biele}, {Michel}, {Sugita},
  {Schr{\"o}der}, {Sakatani}, {Neumann}, {Kameda}, {Michikami} and
  {Honda}}]{Grott_2020}
\bibinfo{author}{{Grott}, M.}, \bibinfo{author}{{Biele}, J.},
  \bibinfo{author}{{Michel}, P.}, \bibinfo{author}{{Sugita}, S.},
  \bibinfo{author}{{Schr{\"o}der}, S.}, \bibinfo{author}{{Sakatani}, N.},
  \bibinfo{author}{{Neumann}, W.}, \bibinfo{author}{{Kameda}, S.},
  \bibinfo{author}{{Michikami}, T.}, \bibinfo{author}{{Honda}, C.},
  \bibinfo{year}{2020}.
\newblock \bibinfo{title}{{Macroporosity and Grain Density of Rubble Pile
  Asteroid (162173) Ryugu}}.
\newblock \bibinfo{journal}{Journal of Geophysical Research (Planets)}
  \bibinfo{volume}{125}, \bibinfo{pages}{e06519}.
\newblock \DOIprefix\doi{10.1029/2020JE00651910.1002/essoar.10503201.2}.
\bibitem[{{Groussin} et~al.(2019){Groussin}, {Attree}, {Brouet}, {Ciarletti},
  {Davidsson}, {Filacchione}, {Fischer}, {Gundlach}, {Knapmeyer},
  {Knollenberg}, {Kokotanekova}, {K{\"u}hrt}, {Leyrat}, {Marshall}, {Pelivan},
  {Skorov}, {Snodgrass}, {Spohn} and {Tosi}}]{Groussin_2019}
\bibinfo{author}{{Groussin}, O.}, \bibinfo{author}{{Attree}, N.},
  \bibinfo{author}{{Brouet}, Y.}, \bibinfo{author}{{Ciarletti}, V.},
  \bibinfo{author}{{Davidsson}, B.}, \bibinfo{author}{{Filacchione}, G.},
  \bibinfo{author}{{Fischer}, H.H.}, \bibinfo{author}{{Gundlach}, B.},
  \bibinfo{author}{{Knapmeyer}, M.}, \bibinfo{author}{{Knollenberg}, J.},
  \bibinfo{author}{{Kokotanekova}, R.}, \bibinfo{author}{{K{\"u}hrt}, E.},
  \bibinfo{author}{{Leyrat}, C.}, \bibinfo{author}{{Marshall}, D.},
  \bibinfo{author}{{Pelivan}, I.}, \bibinfo{author}{{Skorov}, Y.},
  \bibinfo{author}{{Snodgrass}, C.}, \bibinfo{author}{{Spohn}, T.},
  \bibinfo{author}{{Tosi}, F.}, \bibinfo{year}{2019}.
\newblock \bibinfo{title}{{The Thermal, Mechanical, Structural, and Dielectric
  Properties of Cometary Nuclei After Rosetta}}.
\newblock \bibinfo{journal}{Space Science Reviews} \bibinfo{volume}{215},
  \bibinfo{pages}{29}.
\newblock \DOIprefix\doi{10.1007/s11214-019-0594-x},
  \href{http://arxiv.org/abs/1905.01156}{\tt arXiv:1905.01156}.
\bibitem[{{Gunderson} et~al.(2006){Gunderson}, {Thomas} and
  {Whitby}}]{Gunderson_2006}
\bibinfo{author}{{Gunderson}, K.}, \bibinfo{author}{{Thomas}, N.},
  \bibinfo{author}{{Whitby}, J.A.}, \bibinfo{year}{2006}.
\newblock \bibinfo{title}{{First measurements with the Physikalisches Institut
  Radiometric Experiment (PHIRE)}}.
\newblock \bibinfo{journal}{Planetary and Space Science} \bibinfo{volume}{54},
  \bibinfo{pages}{1046--1056}.
\newblock \DOIprefix\doi{10.1016/j.pss.2005.12.020}.
\bibitem[{{G{\"u}ttler} et~al.(2019){G{\"u}ttler}, {Mannel}, {Rotundi},
  {Merouane}, {Fulle}, {Bockel{\'e}e-Morvan}, {Lasue}, {Levasseur-Regourd},
  {Blum}, {Naletto}, {Sierks}, {Hilchenbach}, {Tubiana}, {Capaccioni},
  {Paquette}, {Flandes}, {Moreno}, {Agarwal}, {Bodewits}, {Bertini}, {Tozzi},
  {Hornung}, {Langevin}, {Kr{\"u}ger}, {Longobardo}, {Della Corte}, {T{\'o}th},
  {Filacchione}, {Ivanovski}, {Mottola} and {Rinaldi}}]{Guttler_2019}
\bibinfo{author}{{G{\"u}ttler}, C.}, \bibinfo{author}{{Mannel}, T.},
  \bibinfo{author}{{Rotundi}, A.}, \bibinfo{author}{{Merouane}, S.},
  \bibinfo{author}{{Fulle}, M.}, \bibinfo{author}{{Bockel{\'e}e-Morvan}, D.},
  \bibinfo{author}{{Lasue}, J.}, \bibinfo{author}{{Levasseur-Regourd}, A.C.},
  \bibinfo{author}{{Blum}, J.}, \bibinfo{author}{{Naletto}, G.},
  \bibinfo{author}{{Sierks}, H.}, \bibinfo{author}{{Hilchenbach}, M.},
  \bibinfo{author}{{Tubiana}, C.}, \bibinfo{author}{{Capaccioni}, F.},
  \bibinfo{author}{{Paquette}, J.A.}, \bibinfo{author}{{Flandes}, A.},
  \bibinfo{author}{{Moreno}, F.}, \bibinfo{author}{{Agarwal}, J.},
  \bibinfo{author}{{Bodewits}, D.}, \bibinfo{author}{{Bertini}, I.},
  \bibinfo{author}{{Tozzi}, G.P.}, \bibinfo{author}{{Hornung}, K.},
  \bibinfo{author}{{Langevin}, Y.}, \bibinfo{author}{{Kr{\"u}ger}, H.},
  \bibinfo{author}{{Longobardo}, A.}, \bibinfo{author}{{Della Corte}, V.},
  \bibinfo{author}{{T{\'o}th}, I.}, \bibinfo{author}{{Filacchione}, G.},
  \bibinfo{author}{{Ivanovski}, S.L.}, \bibinfo{author}{{Mottola}, S.},
  \bibinfo{author}{{Rinaldi}, G.}, \bibinfo{year}{2019}.
\newblock \bibinfo{title}{{Synthesis of the morphological description of
  cometary dust at comet 67P/Churyumov-Gerasimenko}}.
\newblock \bibinfo{journal}{Astronomy and Astrophysics} \bibinfo{volume}{630},
  \bibinfo{pages}{A24}.
\newblock \DOIprefix\doi{10.1051/0004-6361/201834751},
  \href{http://arxiv.org/abs/1902.10634}{\tt arXiv:1902.10634}.
\bibitem[{{Hapke}(1981)}]{Hapke_1981}
\bibinfo{author}{{Hapke}, B.}, \bibinfo{year}{1981}.
\newblock \bibinfo{title}{{Bidirectional reflectance spectroscopy. I -
  Theory}}.
\newblock \bibinfo{journal}{Journal of Geophysics Research}
  \bibinfo{volume}{86}, \bibinfo{pages}{3039--3054}.
\newblock \DOIprefix\doi{10.1029/JB086iB04p03039}.
\bibitem[{Hapke(1993)}]{Hapke_1993}
\bibinfo{author}{Hapke, B.}, \bibinfo{year}{1993}.
\newblock \bibinfo{title}{Theory of reflectance and emittance spectroscopy}.
\bibitem[{{Hapke}(1999)}]{Hapke_1999}
\bibinfo{author}{{Hapke}, B.}, \bibinfo{year}{1999}.
\newblock \bibinfo{title}{{Scattering and diffraction of light by particles in
  planetary regoliths.}}
\newblock \bibinfo{journal}{Journal of Quantitiative Spectroscopy and Radiative
  Transfer} \bibinfo{volume}{61}, \bibinfo{pages}{565--581}.
\newblock \DOIprefix\doi{10.1016/S0022-4073(98)00042-9}.
\bibitem[{Hapke(2008)}]{Hapke_2008}
\bibinfo{author}{Hapke, B.}, \bibinfo{year}{2008}.
\newblock \bibinfo{title}{Bidirectional reflectance spectroscopy. 6. {E}ffects
  of porosity}.
\newblock \bibinfo{journal}{Icarus} \bibinfo{volume}{195},
  \bibinfo{pages}{918--926}.
\bibitem[{{Hapke}(2012)}]{Hapke_2012a}
\bibinfo{author}{{Hapke}, B.}, \bibinfo{year}{2012}.
\newblock \bibinfo{title}{{Theory of Reflectance and Emittance Spectroscopy}}.
\newblock \DOIprefix\doi{10.1017/CBO9781139025683}.
\bibitem[{{Hapke}(2021)}]{Hapke_2021}
\bibinfo{author}{{Hapke}, B.}, \bibinfo{year}{2021}.
\newblock \bibinfo{title}{{Bidirectional reflectance spectroscopy 8. The
  angular width of the opposition effect in regolith-like media}}.
\newblock \bibinfo{journal}{Icarus} \bibinfo{volume}{354},
  \bibinfo{pages}{114105}.
\newblock \DOIprefix\doi{10.1016/j.icarus.2020.114105}.
\bibitem[{{Hapke} et~al.(1998){Hapke}, {Nelson} and {Smythe}}]{Hapke_1998}
\bibinfo{author}{{Hapke}, B.}, \bibinfo{author}{{Nelson}, R.},
  \bibinfo{author}{{Smythe}, W.}, \bibinfo{year}{1998}.
\newblock \bibinfo{title}{{The Opposition Effect of the Moon: Coherent
  Backscatter and Shadow Hiding}}.
\newblock \bibinfo{journal}{Icarus} \bibinfo{volume}{133},
  \bibinfo{pages}{89--97}.
\newblock \DOIprefix\doi{10.1006/icar.1998.5907}.
\bibitem[{{Hapke} and {Sato}(2016)}]{Hapke_2016}
\bibinfo{author}{{Hapke}, B.}, \bibinfo{author}{{Sato}, H.},
  \bibinfo{year}{2016}.
\newblock \bibinfo{title}{{The porosity of the upper lunar regolith}}.
\newblock \bibinfo{journal}{Icarus} \bibinfo{volume}{273},
  \bibinfo{pages}{75--83}.
\newblock \DOIprefix\doi{10.1016/j.icarus.2015.10.031}.
\bibitem[{{Hapke} and {van Horn}(1963)}]{Hapke_1963}
\bibinfo{author}{{Hapke}, B.}, \bibinfo{author}{{van Horn}, H.},
  \bibinfo{year}{1963}.
\newblock \bibinfo{title}{{Photometric studies of complex surfaces, with
  applications to the Moon}}.
\newblock \bibinfo{journal}{Journal of Geophysics Research}
  \bibinfo{volume}{68}, \bibinfo{pages}{4545--4570}.
\newblock \DOIprefix\doi{10.1029/JZ068i015p04545}.
\bibitem[{{Hapke} and {Wells}(1981)}]{Hapke_Wells_1981}
\bibinfo{author}{{Hapke}, B.}, \bibinfo{author}{{Wells}, E.},
  \bibinfo{year}{1981}.
\newblock \bibinfo{title}{{Bidirectional reflectance spectroscopy. 2.
  Experiments and observations.}}
\newblock \bibinfo{journal}{Journal of Geophysics Research}
  \bibinfo{volume}{86}, \bibinfo{pages}{3055--3060}.
\newblock \DOIprefix\doi{10.1029/JB086iB04p03055}.
\bibitem[{{Hartmann} et~al.(2001){Hartmann}, {Anguita}, {de la Casa}, {Berman}
  and {Ryan}}]{Hartmann_2001}
\bibinfo{author}{{Hartmann}, W.K.}, \bibinfo{author}{{Anguita}, J.},
  \bibinfo{author}{{de la Casa}, M.A.}, \bibinfo{author}{{Berman}, D.C.},
  \bibinfo{author}{{Ryan}, E.V.}, \bibinfo{year}{2001}.
\newblock \bibinfo{title}{{Martian Cratering 7: The Role of Impact Gardening}}.
\newblock \bibinfo{journal}{Icarus} \bibinfo{volume}{149},
  \bibinfo{pages}{37--53}.
\newblock \DOIprefix\doi{10.1006/icar.2000.6532}.
\bibitem[{{Hasselmann} et~al.(2017){Hasselmann}, {Barucci}, {Fornasier},
  {Feller}, {Deshapriya}, {Fulchignoni}, {Jost}, {Sierks}, {Barbieri}, {Lamy},
  {Rodrigo}, {Koschny}, {Rickman}, {A'Hearn}, {Bertaux}, {Bertini},
  {Cremonese}, {Da Deppo}, {Davidsson}, {Debei}, {De Cecco}, {Deller}, {Fulle},
  {Gaskell}, {Groussin}, {Gutierrez}, {G{\"u}ttler}, {Hofmann}, {Hviid}, {Ip},
  {Jorda}, {Keller}, {Knollenberg}, {Kovacs}, {Kramm}, {K{\"u}hrt},
  {K{\"u}ppers}, {Lara}, {Lazzarin}, {Lopez-Moreno}, {Marzari}, {Mottola},
  {Naletto}, {Oklay}, {Pommerol}, {Thomas}, {Tubiana} and
  {Vincent}}]{Hasselmann_2017}
\bibinfo{author}{{Hasselmann}, P.H.}, \bibinfo{author}{{Barucci}, M.A.},
  \bibinfo{author}{{Fornasier}, S.}, \bibinfo{author}{{Feller}, C.},
  \bibinfo{author}{{Deshapriya}, J.D.P.}, \bibinfo{author}{{Fulchignoni}, M.},
  \bibinfo{author}{{Jost}, B.}, \bibinfo{author}{{Sierks}, H.},
  \bibinfo{author}{{Barbieri}, C.}, \bibinfo{author}{{Lamy}, P.L.},
  \bibinfo{author}{{Rodrigo}, R.}, \bibinfo{author}{{Koschny}, D.},
  \bibinfo{author}{{Rickman}, H.}, \bibinfo{author}{{A'Hearn}, M.},
  \bibinfo{author}{{Bertaux}, J.L.}, \bibinfo{author}{{Bertini}, I.},
  \bibinfo{author}{{Cremonese}, G.}, \bibinfo{author}{{Da Deppo}, V.},
  \bibinfo{author}{{Davidsson}, B.}, \bibinfo{author}{{Debei}, S.},
  \bibinfo{author}{{De Cecco}, M.}, \bibinfo{author}{{Deller}, J.},
  \bibinfo{author}{{Fulle}, M.}, \bibinfo{author}{{Gaskell}, R.W.},
  \bibinfo{author}{{Groussin}, O.}, \bibinfo{author}{{Gutierrez}, P.J.},
  \bibinfo{author}{{G{\"u}ttler}, C.}, \bibinfo{author}{{Hofmann}, M.},
  \bibinfo{author}{{Hviid}, S.F.}, \bibinfo{author}{{Ip}, W.H.},
  \bibinfo{author}{{Jorda}, L.}, \bibinfo{author}{{Keller}, H.U.},
  \bibinfo{author}{{Knollenberg}, J.}, \bibinfo{author}{{Kovacs}, G.},
  \bibinfo{author}{{Kramm}, R.}, \bibinfo{author}{{K{\"u}hrt}, E.},
  \bibinfo{author}{{K{\"u}ppers}, M.}, \bibinfo{author}{{Lara}, M.L.},
  \bibinfo{author}{{Lazzarin}, M.}, \bibinfo{author}{{Lopez-Moreno}, J.J.},
  \bibinfo{author}{{Marzari}, F.}, \bibinfo{author}{{Mottola}, S.},
  \bibinfo{author}{{Naletto}, G.}, \bibinfo{author}{{Oklay}, N.},
  \bibinfo{author}{{Pommerol}, A.}, \bibinfo{author}{{Thomas}, N.},
  \bibinfo{author}{{Tubiana}, C.}, \bibinfo{author}{{Vincent}, J.B.},
  \bibinfo{year}{2017}.
\newblock \bibinfo{title}{{The opposition effect of 67P/Churyumov-Gerasimenko
  on post-perihelion Rosetta images}}.
\newblock \bibinfo{journal}{Monthly Notices of the Royal Astronomical Society}
  \bibinfo{volume}{469}, \bibinfo{pages}{S550--S567}.
\newblock \DOIprefix\doi{10.1093/mnras/stx1834}.
\bibitem[{{Helfenstein} and {Shepard}(2011)}]{Helfenstein_2011}
\bibinfo{author}{{Helfenstein}, P.}, \bibinfo{author}{{Shepard}, M.K.},
  \bibinfo{year}{2011}.
\newblock \bibinfo{title}{{Testing the Hapke photometric model: Improved
  inversion and the porosity correction}}.
\newblock \bibinfo{journal}{Icarus} \bibinfo{volume}{215},
  \bibinfo{pages}{83--100}.
\newblock \DOIprefix\doi{10.1016/j.icarus.2011.07.002}.
\bibitem[{{Hiroi} et~al.(2001){Hiroi}, {Zolensky} and {Pieters}}]{Hiroi_2001}
\bibinfo{author}{{Hiroi}, T.}, \bibinfo{author}{{Zolensky}, M.E.},
  \bibinfo{author}{{Pieters}, C.M.}, \bibinfo{year}{2001}.
\newblock \bibinfo{title}{{The Tagish Lake Meteorite: A Possible Sample from a
  D-Type Asteroid}}.
\newblock \bibinfo{journal}{Science} \bibinfo{volume}{293},
  \bibinfo{pages}{2234--2236}.
\newblock \DOIprefix\doi{10.1126/science.1063734}.
\bibitem[{{Humes} et~al.(2024){Humes}, {Martin}, {Thomas} and
  {Emery}}]{Humes_2024}
\bibinfo{author}{{Humes}, O.A.}, \bibinfo{author}{{Martin}, A.C.},
  \bibinfo{author}{{Thomas}, C.A.}, \bibinfo{author}{{Emery}, J.P.},
  \bibinfo{year}{2024}.
\newblock \bibinfo{title}{{Comparative Mid-infrared Spectroscopy of Dark,
  Primitive Asteroids: Does Shared Taxonomic Class Indicate Shared Silicate
  Composition?}}
\newblock \bibinfo{journal}{Planetary Science Journal} \bibinfo{volume}{5},
  \bibinfo{pages}{108}.
\newblock \DOIprefix\doi{10.3847/PSJ/ad3a69},
  \href{http://arxiv.org/abs/2404.19388}{\tt arXiv:2404.19388}.
\bibitem[{{Ishii} et~al.(2008){Ishii}, {Bradley}, {Dai}, {Chi}, {Kearsley},
  {Burchell}, {Browning} and {Molster}}]{Ishii_2008}
\bibinfo{author}{{Ishii}, H.A.}, \bibinfo{author}{{Bradley}, J.P.},
  \bibinfo{author}{{Dai}, Z.R.}, \bibinfo{author}{{Chi}, M.},
  \bibinfo{author}{{Kearsley}, A.T.}, \bibinfo{author}{{Burchell}, M.J.},
  \bibinfo{author}{{Browning}, N.D.}, \bibinfo{author}{{Molster}, F.},
  \bibinfo{year}{2008}.
\newblock \bibinfo{title}{{Comparison of Comet 81P/Wild 2 Dust with
  Interplanetary Dust from Comets}}.
\newblock \bibinfo{journal}{Science} \bibinfo{volume}{319},
  \bibinfo{pages}{447}.
\newblock \DOIprefix\doi{10.1126/science.1150683}.
\bibitem[{{Jost} et~al.(2013){Jost}, {Gundlach}, {Pommerol}, {Oesert}, {Gorb},
  {Blum} and {Thomas}}]{Jost_2013}
\bibinfo{author}{{Jost}, B.}, \bibinfo{author}{{Gundlach}, B.},
  \bibinfo{author}{{Pommerol}, A.}, \bibinfo{author}{{Oesert}, J.},
  \bibinfo{author}{{Gorb}, S.N.}, \bibinfo{author}{{Blum}, J.},
  \bibinfo{author}{{Thomas}, N.}, \bibinfo{year}{2013}.
\newblock \bibinfo{title}{{Micrometer-sized ice particles for planetary-science
  experiments - II. Bidirectional reflectance}}.
\newblock \bibinfo{journal}{Icarus} \bibinfo{volume}{225},
  \bibinfo{pages}{352--366}.
\newblock \DOIprefix\doi{10.1016/j.icarus.2013.04.007}.
\bibitem[{{Kameda} et~al.(2021){Kameda}, {Ozaki}, {Enya}, {Fuse}, {Kouyama},
  {Sakatani}, {Suzuki}, {Osada}, {Kato}, {Miyamoto}, {Yamazaki}, {Nakamura},
  {Okamoto}, {Ishimaru}, {Hong}, {Ishibashi}, {Takashima}, {Ishigami}, {Kuo},
  {Abe}, {Goda}, {Murao}, {Fujishima}, {Aoyama}, {Hagiwara}, {Mizumoto},
  {Tanaka}, {Murakami}, {Matsumoto}, {Tanaka} and {Sakuta}}]{Kameda_2021}
\bibinfo{author}{{Kameda}, S.}, \bibinfo{author}{{Ozaki}, M.},
  \bibinfo{author}{{Enya}, K.}, \bibinfo{author}{{Fuse}, R.},
  \bibinfo{author}{{Kouyama}, T.}, \bibinfo{author}{{Sakatani}, N.},
  \bibinfo{author}{{Suzuki}, H.}, \bibinfo{author}{{Osada}, N.},
  \bibinfo{author}{{Kato}, H.}, \bibinfo{author}{{Miyamoto}, H.},
  \bibinfo{author}{{Yamazaki}, A.}, \bibinfo{author}{{Nakamura}, T.},
  \bibinfo{author}{{Okamoto}, T.}, \bibinfo{author}{{Ishimaru}, T.},
  \bibinfo{author}{{Hong}, P.}, \bibinfo{author}{{Ishibashi}, K.},
  \bibinfo{author}{{Takashima}, T.}, \bibinfo{author}{{Ishigami}, R.},
  \bibinfo{author}{{Kuo}, C.L.}, \bibinfo{author}{{Abe}, S.},
  \bibinfo{author}{{Goda}, Y.}, \bibinfo{author}{{Murao}, H.},
  \bibinfo{author}{{Fujishima}, S.}, \bibinfo{author}{{Aoyama}, T.},
  \bibinfo{author}{{Hagiwara}, K.}, \bibinfo{author}{{Mizumoto}, S.},
  \bibinfo{author}{{Tanaka}, N.}, \bibinfo{author}{{Murakami}, K.},
  \bibinfo{author}{{Matsumoto}, M.}, \bibinfo{author}{{Tanaka}, K.},
  \bibinfo{author}{{Sakuta}, H.}, \bibinfo{year}{2021}.
\newblock \bibinfo{title}{{Design of telescopic nadir imager for geomorphology
  (TENGOO) and observation of surface reflectance by optical chromatic imager
  (OROCHI) for the Martian Moons Exploration (MMX)}}.
\newblock \bibinfo{journal}{Earth, Planets and Space} \bibinfo{volume}{73},
  \bibinfo{pages}{218}.
\newblock \DOIprefix\doi{10.1186/s40623-021-01462-9}.
\bibitem[{{Kar} et~al.(2016){Kar}, {Sen} and {Gupta}}]{Kar_2016}
\bibinfo{author}{{Kar}, A.}, \bibinfo{author}{{Sen}, A.K.},
  \bibinfo{author}{{Gupta}, R.}, \bibinfo{year}{2016}.
\newblock \bibinfo{title}{{Laboratory photometry of regolith analogues: Effect
  of porosity}}.
\newblock \bibinfo{journal}{Icarus} \bibinfo{volume}{277},
  \bibinfo{pages}{300--310}.
\newblock \DOIprefix\doi{10.1016/j.icarus.2016.05.024}.
\bibitem[{{Keller} et~al.(2015){Keller}, {Mottola}, {Davidsson},
  {Schr{\"o}der}, {Skorov}, {K{\"u}hrt}, {Groussin}, {Pajola}, {Hviid},
  {Preusker}, {Scholten}, {A'Hearn}, {Sierks}, {Barbieri}, {Lamy}, {Rodrigo},
  {Koschny}, {Rickman}, {Barucci}, {Bertaux}, {Bertini}, {Cremonese}, {Da
  Deppo}, {Debei}, {De Cecco}, {Fornasier}, {Fulle}, {Guti{\'e}rrez}, {Ip},
  {Jorda}, {Knollenberg}, {Kramm}, {K{\"u}ppers}, {Lara}, {Lazzarin}, {Lopez
  Moreno}, {Marzari}, {Michalik}, {Naletto}, {Sabau}, {Thomas}, {Vincent},
  {Wenzel}, {Agarwal}, {G{\"u}ttler}, {Oklay} and {Tubiana}}]{Keller_2015}
\bibinfo{author}{{Keller}, H.U.}, \bibinfo{author}{{Mottola}, S.},
  \bibinfo{author}{{Davidsson}, B.}, \bibinfo{author}{{Schr{\"o}der}, S.E.},
  \bibinfo{author}{{Skorov}, Y.}, \bibinfo{author}{{K{\"u}hrt}, E.},
  \bibinfo{author}{{Groussin}, O.}, \bibinfo{author}{{Pajola}, M.},
  \bibinfo{author}{{Hviid}, S.F.}, \bibinfo{author}{{Preusker}, F.},
  \bibinfo{author}{{Scholten}, F.}, \bibinfo{author}{{A'Hearn}, M.F.},
  \bibinfo{author}{{Sierks}, H.}, \bibinfo{author}{{Barbieri}, C.},
  \bibinfo{author}{{Lamy}, P.}, \bibinfo{author}{{Rodrigo}, R.},
  \bibinfo{author}{{Koschny}, D.}, \bibinfo{author}{{Rickman}, H.},
  \bibinfo{author}{{Barucci}, M.A.}, \bibinfo{author}{{Bertaux}, J.L.},
  \bibinfo{author}{{Bertini}, I.}, \bibinfo{author}{{Cremonese}, G.},
  \bibinfo{author}{{Da Deppo}, V.}, \bibinfo{author}{{Debei}, S.},
  \bibinfo{author}{{De Cecco}, M.}, \bibinfo{author}{{Fornasier}, S.},
  \bibinfo{author}{{Fulle}, M.}, \bibinfo{author}{{Guti{\'e}rrez}, P.J.},
  \bibinfo{author}{{Ip}, W.H.}, \bibinfo{author}{{Jorda}, L.},
  \bibinfo{author}{{Knollenberg}, J.}, \bibinfo{author}{{Kramm}, J.R.},
  \bibinfo{author}{{K{\"u}ppers}, M.}, \bibinfo{author}{{Lara}, L.M.},
  \bibinfo{author}{{Lazzarin}, M.}, \bibinfo{author}{{Lopez Moreno}, J.J.},
  \bibinfo{author}{{Marzari}, F.}, \bibinfo{author}{{Michalik}, H.},
  \bibinfo{author}{{Naletto}, G.}, \bibinfo{author}{{Sabau}, L.},
  \bibinfo{author}{{Thomas}, N.}, \bibinfo{author}{{Vincent}, J.B.},
  \bibinfo{author}{{Wenzel}, K.P.}, \bibinfo{author}{{Agarwal}, J.},
  \bibinfo{author}{{G{\"u}ttler}, C.}, \bibinfo{author}{{Oklay}, N.},
  \bibinfo{author}{{Tubiana}, C.}, \bibinfo{year}{2015}.
\newblock \bibinfo{title}{{Insolation, erosion, and morphology of comet
  67P/Churyumov-Gerasimenko}}.
\newblock \bibinfo{journal}{Astronomy and Astrophysics} \bibinfo{volume}{583},
  \bibinfo{pages}{A34}.
\newblock \DOIprefix\doi{10.1051/0004-6361/201525964}.
\bibitem[{{Kiuchi} and {Nakamura}(2014)}]{Kiuchi_2014}
\bibinfo{author}{{Kiuchi}, M.}, \bibinfo{author}{{Nakamura}, A.M.},
  \bibinfo{year}{2014}.
\newblock \bibinfo{title}{{Relationship between regolith particle size and
  porosity on small bodies}}.
\newblock \bibinfo{journal}{Icarus} \bibinfo{volume}{239},
  \bibinfo{pages}{291--293}.
\newblock \DOIprefix\doi{10.1016/j.icarus.2014.05.029}.
\bibitem[{{Kossacki} and {Czechowski}(2018)}]{Kossacki_2018}
\bibinfo{author}{{Kossacki}, K.J.}, \bibinfo{author}{{Czechowski}, L.},
  \bibinfo{year}{2018}.
\newblock \bibinfo{title}{{Comet 67p/Churyumov-Gerasimenko, possible origin of
  the depression Hatmehit}}.
\newblock \bibinfo{journal}{Icarus} \bibinfo{volume}{305},
  \bibinfo{pages}{1--14}.
\newblock \DOIprefix\doi{10.1016/j.icarus.2017.12.027}.
\bibitem[{{Kuramoto} et~al.(2022){Kuramoto}, {Kawakatsu}, {Fujimoto}, {Araya},
  {Barucci}, {Genda}, {Hirata}, {Ikeda}, {Imamura}, {Helbert}, {Kameda},
  {Kobayashi}, {Kusano}, {Lawrence}, {Matsumoto}, {Michel}, {Miyamoto},
  {Morota}, {Nakagawa}, {Nakamura}, {Ogawa}, {Otake}, {Ozaki}, {Russell},
  {Sasaki}, {Sawada}, {Senshu}, {Tachibana}, {Terada}, {Ulamec}, {Usui},
  {Wada}, {Watanabe} and {Yokota}}]{Kuramoto_2022}
\bibinfo{author}{{Kuramoto}, K.}, \bibinfo{author}{{Kawakatsu}, Y.},
  \bibinfo{author}{{Fujimoto}, M.}, \bibinfo{author}{{Araya}, A.},
  \bibinfo{author}{{Barucci}, M.A.}, \bibinfo{author}{{Genda}, H.},
  \bibinfo{author}{{Hirata}, N.}, \bibinfo{author}{{Ikeda}, H.},
  \bibinfo{author}{{Imamura}, T.}, \bibinfo{author}{{Helbert}, J.},
  \bibinfo{author}{{Kameda}, S.}, \bibinfo{author}{{Kobayashi}, M.},
  \bibinfo{author}{{Kusano}, H.}, \bibinfo{author}{{Lawrence}, D.J.},
  \bibinfo{author}{{Matsumoto}, K.}, \bibinfo{author}{{Michel}, P.},
  \bibinfo{author}{{Miyamoto}, H.}, \bibinfo{author}{{Morota}, T.},
  \bibinfo{author}{{Nakagawa}, H.}, \bibinfo{author}{{Nakamura}, T.},
  \bibinfo{author}{{Ogawa}, K.}, \bibinfo{author}{{Otake}, H.},
  \bibinfo{author}{{Ozaki}, M.}, \bibinfo{author}{{Russell}, S.},
  \bibinfo{author}{{Sasaki}, S.}, \bibinfo{author}{{Sawada}, H.},
  \bibinfo{author}{{Senshu}, H.}, \bibinfo{author}{{Tachibana}, S.},
  \bibinfo{author}{{Terada}, N.}, \bibinfo{author}{{Ulamec}, S.},
  \bibinfo{author}{{Usui}, T.}, \bibinfo{author}{{Wada}, K.},
  \bibinfo{author}{{Watanabe}, S.i.}, \bibinfo{author}{{Yokota}, S.},
  \bibinfo{year}{2022}.
\newblock \bibinfo{title}{{Martian moons exploration MMX: sample return mission
  to Phobos elucidating formation processes of habitable planets}}.
\newblock \bibinfo{journal}{Earth, Planets and Space} \bibinfo{volume}{74},
  \bibinfo{pages}{12}.
\newblock \DOIprefix\doi{10.1186/s40623-021-01545-7}.
\bibitem[{{Lantz} et~al.(2018){Lantz}, {Binzel} and {DeMeo}}]{Lantz_2018}
\bibinfo{author}{{Lantz}, C.}, \bibinfo{author}{{Binzel}, R.P.},
  \bibinfo{author}{{DeMeo}, F.E.}, \bibinfo{year}{2018}.
\newblock \bibinfo{title}{{Space weathering trends on carbonaceous asteroids: A
  possible explanation for Bennu's blue slope?}}
\newblock \bibinfo{journal}{Icarus} \bibinfo{volume}{302},
  \bibinfo{pages}{10--17}.
\newblock \DOIprefix\doi{10.1016/j.icarus.2017.11.010}.
\bibitem[{{Lantz} et~al.(2017){Lantz}, {Brunetto}, {Barucci}, {Fornasier},
  {Baklouti}, {Bour{\c{c}}ois} and {Godard}}]{Lantz_2017}
\bibinfo{author}{{Lantz}, C.}, \bibinfo{author}{{Brunetto}, R.},
  \bibinfo{author}{{Barucci}, M.A.}, \bibinfo{author}{{Fornasier}, S.},
  \bibinfo{author}{{Baklouti}, D.}, \bibinfo{author}{{Bour{\c{c}}ois}, J.},
  \bibinfo{author}{{Godard}, M.}, \bibinfo{year}{2017}.
\newblock \bibinfo{title}{{Ion irradiation of carbonaceous chondrites: A new
  view of space weathering on primitive asteroids}}.
\newblock \bibinfo{journal}{Icarus} \bibinfo{volume}{285},
  \bibinfo{pages}{43--57}.
\newblock \DOIprefix\doi{10.1016/j.icarus.2016.12.019}.
\bibitem[{{Lawrence} et~al.(2019){Lawrence}, {Peplowski}, {Beck}, {Burks},
  {Chabot}, {Cully}, {Elphic}, {Ernst}, {Fix}, {Goldsten}, {Hoffer}, {Kusano},
  {Murchie}, {Schratz}, {Usui} and {Yokley}}]{Lawrence_2019}
\bibinfo{author}{{Lawrence}, D.J.}, \bibinfo{author}{{Peplowski}, P.N.},
  \bibinfo{author}{{Beck}, A.W.}, \bibinfo{author}{{Burks}, M.T.},
  \bibinfo{author}{{Chabot}, N.L.}, \bibinfo{author}{{Cully}, M.J.},
  \bibinfo{author}{{Elphic}, R.C.}, \bibinfo{author}{{Ernst}, C.M.},
  \bibinfo{author}{{Fix}, S.}, \bibinfo{author}{{Goldsten}, J.O.},
  \bibinfo{author}{{Hoffer}, E.M.}, \bibinfo{author}{{Kusano}, H.},
  \bibinfo{author}{{Murchie}, S.L.}, \bibinfo{author}{{Schratz}, B.C.},
  \bibinfo{author}{{Usui}, T.}, \bibinfo{author}{{Yokley}, Z.W.},
  \bibinfo{year}{2019}.
\newblock \bibinfo{title}{{Measuring the Elemental Composition of Phobos: The
  Mars-moon Exploration with GAmma rays and NEutrons (MEGANE) Investigation for
  the Martian Moons eXploration (MMX) Mission}}.
\newblock \bibinfo{journal}{Earth and Space Science} \bibinfo{volume}{6},
  \bibinfo{pages}{2605--2623}.
\newblock \DOIprefix\doi{10.1029/2019EA000811}.
\bibitem[{Li et~al.(2020)Li, Moullet, Titus, Hsieh and Sykes}]{Li_2020}
\bibinfo{author}{Li, J.Y.}, \bibinfo{author}{Moullet, A.},
  \bibinfo{author}{Titus, T.N.}, \bibinfo{author}{Hsieh, H.H.},
  \bibinfo{author}{Sykes, M.V.}, \bibinfo{year}{2020}.
\newblock \bibinfo{title}{Disk-integrated thermal properties of ceres measured
  at millimeter wavelengths}.
\newblock \bibinfo{journal}{The Astronomical Journal} \bibinfo{volume}{159},
  \bibinfo{pages}{215}.
\newblock \URLprefix \url{https://dx.doi.org/10.3847/1538-3881/ab8305},
  \DOIprefix\doi{10.3847/1538-3881/ab8305}.
\bibitem[{{Licandro} et~al.(2011){Licandro}, {Campins}, {Kelley}, {Hargrove},
  {Pinilla-Alonso}, {Cruikshank}, {Rivkin} and {Emery}}]{Licandro_2011}
\bibinfo{author}{{Licandro}, J.}, \bibinfo{author}{{Campins}, H.},
  \bibinfo{author}{{Kelley}, M.}, \bibinfo{author}{{Hargrove}, K.},
  \bibinfo{author}{{Pinilla-Alonso}, N.}, \bibinfo{author}{{Cruikshank}, D.},
  \bibinfo{author}{{Rivkin}, A.S.}, \bibinfo{author}{{Emery}, J.},
  \bibinfo{year}{2011}.
\newblock \bibinfo{title}{{(65) Cybele: detection of small silicate grains,
  water-ice, and organics}}.
\newblock \bibinfo{journal}{Astronomy and Astrophysics} \bibinfo{volume}{525},
  \bibinfo{pages}{A34}.
\newblock \DOIprefix\doi{10.1051/0004-6361/201015339}.
\bibitem[{{Lowry} et~al.(2022){Lowry}, {Donaldson Hanna}, {Ito}, {Kelley},
  {Campins} and {Lindsay}}]{Lowry_2022}
\bibinfo{author}{{Lowry}, V.C.}, \bibinfo{author}{{Donaldson Hanna}, K.L.},
  \bibinfo{author}{{Ito}, G.}, \bibinfo{author}{{Kelley}, M.S.P.},
  \bibinfo{author}{{Campins}, H.}, \bibinfo{author}{{Lindsay}, S.},
  \bibinfo{year}{2022}.
\newblock \bibinfo{title}{{T-matrix and Hapke Modeling of the Thermal Infrared
  Spectra of Trojan Asteroids and (944) Hidalgo: Implications for Their
  Regolith Particle Size and Porosity}}.
\newblock \bibinfo{journal}{Planetary Science Journal} \bibinfo{volume}{3},
  \bibinfo{pages}{181}.
\newblock \DOIprefix\doi{10.3847/PSJ/ac7a30}.
\bibitem[{{Lumme} and {Bowell}(1981)}]{Lumme_1981}
\bibinfo{author}{{Lumme}, K.}, \bibinfo{author}{{Bowell}, E.},
  \bibinfo{year}{1981}.
\newblock \bibinfo{title}{{Radiative transfer in the surfaces of atmosphereless
  bodies. I. Theory.}}
\newblock \bibinfo{journal}{Astronomical Journal} \bibinfo{volume}{86},
  \bibinfo{pages}{1694--1721}.
\newblock \DOIprefix\doi{10.1086/113054}.
\bibitem[{{Mannel} et~al.(2019){Mannel}, {Bentley}, {Boakes}, {Jeszenszky},
  {Ehrenfreund}, {Engrand}, {Koeberl}, {Levasseur-Regourd}, {Romstedt},
  {Schmied}, {Torkar} and {Weber}}]{Mannel_2019}
\bibinfo{author}{{Mannel}, T.}, \bibinfo{author}{{Bentley}, M.S.},
  \bibinfo{author}{{Boakes}, P.D.}, \bibinfo{author}{{Jeszenszky}, H.},
  \bibinfo{author}{{Ehrenfreund}, P.}, \bibinfo{author}{{Engrand}, C.},
  \bibinfo{author}{{Koeberl}, C.}, \bibinfo{author}{{Levasseur-Regourd}, A.C.},
  \bibinfo{author}{{Romstedt}, J.}, \bibinfo{author}{{Schmied}, R.},
  \bibinfo{author}{{Torkar}, K.}, \bibinfo{author}{{Weber}, I.},
  \bibinfo{year}{2019}.
\newblock \bibinfo{title}{{Dust of comet 67P/Churyumov-Gerasimenko collected by
  Rosetta/MIDAS: classification and extension to the nanometer scale}}.
\newblock \bibinfo{journal}{Astronomy and Astrophysics} \bibinfo{volume}{630},
  \bibinfo{pages}{A26}.
\newblock \DOIprefix\doi{10.1051/0004-6361/201834851}.
\bibitem[{{Marchis} et~al.(2012){Marchis}, {Enriquez}, {Emery}, {Mueller},
  {Baek}, {Pollock}, {Assafin}, {Vieira Martins}, {Berthier}, {Vachier},
  {Cruikshank}, {Lim}, {Reichart}, {Ivarsen}, {Haislip} and
  {LaCluyze}}]{Marchis_2012}
\bibinfo{author}{{Marchis}, F.}, \bibinfo{author}{{Enriquez}, J.E.},
  \bibinfo{author}{{Emery}, J.P.}, \bibinfo{author}{{Mueller}, M.},
  \bibinfo{author}{{Baek}, M.}, \bibinfo{author}{{Pollock}, J.},
  \bibinfo{author}{{Assafin}, M.}, \bibinfo{author}{{Vieira Martins}, R.},
  \bibinfo{author}{{Berthier}, J.}, \bibinfo{author}{{Vachier}, F.},
  \bibinfo{author}{{Cruikshank}, D.P.}, \bibinfo{author}{{Lim}, L.F.},
  \bibinfo{author}{{Reichart}, D.E.}, \bibinfo{author}{{Ivarsen}, K.M.},
  \bibinfo{author}{{Haislip}, J.B.}, \bibinfo{author}{{LaCluyze}, A.P.},
  \bibinfo{year}{2012}.
\newblock \bibinfo{title}{{Multiple asteroid systems: Dimensions and thermal
  properties from Spitzer Space Telescope and ground-based observations}}.
\newblock \bibinfo{journal}{Icarus} \bibinfo{volume}{221},
  \bibinfo{pages}{1130--1161}.
\newblock \DOIprefix\doi{10.1016/j.icarus.2012.09.013},
  \href{http://arxiv.org/abs/1604.05384}{\tt arXiv:1604.05384}.
\bibitem[{{Martin} and {Emery}(2023)}]{Martin_2023b}
\bibinfo{author}{{Martin}, A.C.}, \bibinfo{author}{{Emery}, J.P.},
  \bibinfo{year}{2023}.
\newblock \bibinfo{title}{{MIR Spectra and Analysis of Jovian Trojan
  Asteroids}}.
\newblock \bibinfo{journal}{Planetary Science Journal} \bibinfo{volume}{4},
  \bibinfo{pages}{153}.
\newblock \DOIprefix\doi{10.3847/PSJ/aced0c}.
\bibitem[{{Martin} et~al.(2022){Martin}, {Emery} and {Loeffler}}]{Martin_2022}
\bibinfo{author}{{Martin}, A.C.}, \bibinfo{author}{{Emery}, J.P.},
  \bibinfo{author}{{Loeffler}, M.J.}, \bibinfo{year}{2022}.
\newblock \bibinfo{title}{{Spectral effects of regolith porosity in the mid-IR
  - Forsteritic olivine}}.
\newblock \bibinfo{journal}{Icarus} \bibinfo{volume}{378},
  \bibinfo{pages}{114921}.
\newblock \DOIprefix\doi{10.1016/j.icarus.2022.114921}.
\bibitem[{{Martin} et~al.(2023){Martin}, {Emery} and {Loeffler}}]{Martin_2023}
\bibinfo{author}{{Martin}, A.C.}, \bibinfo{author}{{Emery}, J.P.},
  \bibinfo{author}{{Loeffler}, M.J.}, \bibinfo{year}{2023}.
\newblock \bibinfo{title}{{Spectral effects of regolith porosity in the Mid-IR
  - Pyroxene}}.
\newblock \bibinfo{journal}{Icarus} \bibinfo{volume}{397},
  \bibinfo{pages}{115507}.
\newblock \DOIprefix\doi{10.1016/j.icarus.2023.115507}.
\bibitem[{Mason et~al.(2023)Mason, Patel, Pajola, Cloutis, Alday, Olsen,
  Marriner, Holmes, Sellers, Thomas, Almeida, Read, Nakagawa, Thomas, Ristic,
  Willame, Depiesse, Daerden, Vandaele, Lopez-Moreno and Bellucci}]{Mason_2023}
\bibinfo{author}{Mason, J.P.}, \bibinfo{author}{Patel, M.R.},
  \bibinfo{author}{Pajola, M.}, \bibinfo{author}{Cloutis, E.D.},
  \bibinfo{author}{Alday, J.}, \bibinfo{author}{Olsen, K.S.},
  \bibinfo{author}{Marriner, C.}, \bibinfo{author}{Holmes, J.A.},
  \bibinfo{author}{Sellers, G.}, \bibinfo{author}{Thomas, N.},
  \bibinfo{author}{Almeida, M.}, \bibinfo{author}{Read, M.},
  \bibinfo{author}{Nakagawa, H.}, \bibinfo{author}{Thomas, I.R.},
  \bibinfo{author}{Ristic, B.}, \bibinfo{author}{Willame, Y.},
  \bibinfo{author}{Depiesse, C.}, \bibinfo{author}{Daerden, F.},
  \bibinfo{author}{Vandaele, A.C.}, \bibinfo{author}{Lopez-Moreno, J.J.},
  \bibinfo{author}{Bellucci, G.}, \bibinfo{year}{2023}.
\newblock \bibinfo{title}{Ultraviolet and visible reflectance spectra of phobos
  and deimos as measured by the exomars-tgo/nomad-uvis spectrometer}.
\newblock \bibinfo{journal}{Journal of Geophysical Research: Planets}
  \bibinfo{volume}{128}, \bibinfo{pages}{e2023JE008002}.
\newblock \URLprefix
  \url{https://agupubs.onlinelibrary.wiley.com/doi/abs/10.1029/2023JE008002},
  \DOIprefix\doi{https://doi.org/10.1029/2023JE008002},
  \href{http://arxiv.org/abs/https://agupubs.onlinelibrary.wiley.com/doi/pdf/10.1029/2023JE008002}{\tt
  arXiv:https://agupubs.onlinelibrary.wiley.com/doi/pdf/10.1029/2023JE008002}.
  \bibinfo{note}{e2023JE008002 2023JE008002}.
\bibitem[{{Matsuoka} et~al.(2023){Matsuoka}, {Kagawa}, {Amano}, {Nakamura},
  {Tatsumi}, {Osawa}, {Hiroi}, {Milliken}, {Domingue}, {Takir}, {Brunetto},
  {Barucci}, {Kitazato}, {Sugita}, {Fujioka}, {Sasaki}, {Kobayashi}, {Iwata},
  {Morota}, {Yokota}, {Kouyama}, {Honda}, {Kameda}, {Cho}, {Yoshioka},
  {Sawada}, {Hayakawa}, {Sakatani}, {Yamada}, {Suzuki}, {Honda}, {Ogawa},
  {Shirai}, {Lantz}, {Rubino}, {Yurimoto}, {Noguchi}, {Okazaki}, {Yabuta},
  {Naraoka}, {Sakamoto}, {Tachibana}, {Yada}, {Nishimura}, {Nakato},
  {Miyazaki}, {Yogata}, {Abe}, {Okada}, {Usui}, {Yoshikawa}, {Saiki}, {Tanaka},
  {Terui}, {Nakazawa}, {Watanabe} and {Tsuda}}]{Matsuoka_2023}
\bibinfo{author}{{Matsuoka}, M.}, \bibinfo{author}{{Kagawa}, E.i.},
  \bibinfo{author}{{Amano}, K.}, \bibinfo{author}{{Nakamura}, T.},
  \bibinfo{author}{{Tatsumi}, E.}, \bibinfo{author}{{Osawa}, T.},
  \bibinfo{author}{{Hiroi}, T.}, \bibinfo{author}{{Milliken}, R.},
  \bibinfo{author}{{Domingue}, D.}, \bibinfo{author}{{Takir}, D.},
  \bibinfo{author}{{Brunetto}, R.}, \bibinfo{author}{{Barucci}, A.},
  \bibinfo{author}{{Kitazato}, K.}, \bibinfo{author}{{Sugita}, S.},
  \bibinfo{author}{{Fujioka}, Y.}, \bibinfo{author}{{Sasaki}, O.},
  \bibinfo{author}{{Kobayashi}, S.}, \bibinfo{author}{{Iwata}, T.},
  \bibinfo{author}{{Morota}, T.}, \bibinfo{author}{{Yokota}, Y.},
  \bibinfo{author}{{Kouyama}, T.}, \bibinfo{author}{{Honda}, R.},
  \bibinfo{author}{{Kameda}, S.}, \bibinfo{author}{{Cho}, Y.},
  \bibinfo{author}{{Yoshioka}, K.}, \bibinfo{author}{{Sawada}, H.},
  \bibinfo{author}{{Hayakawa}, M.}, \bibinfo{author}{{Sakatani}, N.},
  \bibinfo{author}{{Yamada}, M.}, \bibinfo{author}{{Suzuki}, H.},
  \bibinfo{author}{{Honda}, C.}, \bibinfo{author}{{Ogawa}, K.},
  \bibinfo{author}{{Shirai}, K.}, \bibinfo{author}{{Lantz}, C.},
  \bibinfo{author}{{Rubino}, S.}, \bibinfo{author}{{Yurimoto}, H.},
  \bibinfo{author}{{Noguchi}, T.}, \bibinfo{author}{{Okazaki}, R.},
  \bibinfo{author}{{Yabuta}, H.}, \bibinfo{author}{{Naraoka}, H.},
  \bibinfo{author}{{Sakamoto}, K.}, \bibinfo{author}{{Tachibana}, S.},
  \bibinfo{author}{{Yada}, T.}, \bibinfo{author}{{Nishimura}, M.},
  \bibinfo{author}{{Nakato}, A.}, \bibinfo{author}{{Miyazaki}, A.},
  \bibinfo{author}{{Yogata}, K.}, \bibinfo{author}{{Abe}, M.},
  \bibinfo{author}{{Okada}, T.}, \bibinfo{author}{{Usui}, T.},
  \bibinfo{author}{{Yoshikawa}, M.}, \bibinfo{author}{{Saiki}, T.},
  \bibinfo{author}{{Tanaka}, S.}, \bibinfo{author}{{Terui}, F.},
  \bibinfo{author}{{Nakazawa}, S.}, \bibinfo{author}{{Watanabe}, S.i.},
  \bibinfo{author}{{Tsuda}, Y.}, \bibinfo{year}{2023}.
\newblock \bibinfo{title}{{Space weathering acts strongly on the uppermost
  surface of Ryugu}}.
\newblock \bibinfo{journal}{Communications Earth and Environment}
  \bibinfo{volume}{4}, \bibinfo{pages}{335}.
\newblock \DOIprefix\doi{10.1038/s43247-023-00991-3}.
\bibitem[{{McKay} et~al.(1991){McKay}, {Heiken}, {Basu}, {Blanford}, {Simon},
  {Reedy}, {French} and {Papike}}]{McKay_1991}
\bibinfo{author}{{McKay}, D.S.}, \bibinfo{author}{{Heiken}, G.},
  \bibinfo{author}{{Basu}, A.}, \bibinfo{author}{{Blanford}, G.},
  \bibinfo{author}{{Simon}, S.}, \bibinfo{author}{{Reedy}, R.},
  \bibinfo{author}{{French}, B.M.}, \bibinfo{author}{{Papike}, J.},
  \bibinfo{year}{1991}.
\newblock \bibinfo{title}{{The Lunar Regolith}}, in: \bibinfo{editor}{{Heiken},
  G.H.}, \bibinfo{editor}{{Vaniman}, D.T.}, \bibinfo{editor}{{French}, B.M.}
  (Eds.), \bibinfo{booktitle}{Lunar Sourcebook, A User's Guide to the Moon},
  pp. \bibinfo{pages}{285--356}.
\bibitem[{{Michel} et~al.(2022){Michel}, {Ulamec}, {B{\"o}ttger}, {Grott},
  {Murdoch}, {Vernazza}, {Sunday}, {Zhang}, {Valette}, {Castellani}, {Biele},
  {Tardivel}, {Groussin}, {Jorda}, {Knollenberg}, {Grundmann}, {Arrat}, {Pont},
  {Mary}, {Grebenstein}, {Miyamoto}, {Nakamura}, {Wada}, {Yoshikawa} and
  {Kuramoto}}]{Michel_2022}
\bibinfo{author}{{Michel}, P.}, \bibinfo{author}{{Ulamec}, S.},
  \bibinfo{author}{{B{\"o}ttger}, U.}, \bibinfo{author}{{Grott}, M.},
  \bibinfo{author}{{Murdoch}, N.}, \bibinfo{author}{{Vernazza}, P.},
  \bibinfo{author}{{Sunday}, C.}, \bibinfo{author}{{Zhang}, Y.},
  \bibinfo{author}{{Valette}, R.}, \bibinfo{author}{{Castellani}, R.},
  \bibinfo{author}{{Biele}, J.}, \bibinfo{author}{{Tardivel}, S.},
  \bibinfo{author}{{Groussin}, O.}, \bibinfo{author}{{Jorda}, L.},
  \bibinfo{author}{{Knollenberg}, J.}, \bibinfo{author}{{Grundmann}, J.T.},
  \bibinfo{author}{{Arrat}, D.}, \bibinfo{author}{{Pont}, G.},
  \bibinfo{author}{{Mary}, S.}, \bibinfo{author}{{Grebenstein}, M.},
  \bibinfo{author}{{Miyamoto}, H.}, \bibinfo{author}{{Nakamura}, T.},
  \bibinfo{author}{{Wada}, K.}, \bibinfo{author}{{Yoshikawa}, K.},
  \bibinfo{author}{{Kuramoto}, K.}, \bibinfo{year}{2022}.
\newblock \bibinfo{title}{{The MMX rover: performing in situ surface
  investigations on Phobos}}.
\newblock \bibinfo{journal}{Earth, Planets and Space} \bibinfo{volume}{74},
  \bibinfo{pages}{2}.
\newblock \DOIprefix\doi{10.1186/s40623-021-01464-7}.
\bibitem[{{Mishchenko}(1992)}]{Mishchenko_1992}
\bibinfo{author}{{Mishchenko}, M.I.}, \bibinfo{year}{1992}.
\newblock \bibinfo{title}{{The Angular Width of the Coherent Backscatter
  Opposition Effect - an Application to Icy Outer Planet Satellites}}.
\newblock \bibinfo{journal}{Astrophysics and Space Science}
  \bibinfo{volume}{194}, \bibinfo{pages}{327--333}.
\newblock \DOIprefix\doi{10.1007/BF00644001}.
\bibitem[{{Mishchenko} et~al.(1999){Mishchenko}, {Dlugach}, {Yanovitskij} and
  {Zakharova}}]{Mishchenko_1999}
\bibinfo{author}{{Mishchenko}, M.I.}, \bibinfo{author}{{Dlugach}, Z.M.},
  \bibinfo{author}{{Yanovitskij}, E.G.}, \bibinfo{author}{{Zakharova}, N.T.},
  \bibinfo{year}{1999}.
\newblock \bibinfo{title}{{Bidirectional reflectance of flat, optically thick
  particulate layers: an efficient radiative transfer solution and applications
  to snow and soil surfaces.}}
\newblock \bibinfo{journal}{Journal of Quantitiative Spectroscopy and Radiative
  Transfer} \bibinfo{volume}{63}, \bibinfo{pages}{409--432}.
\newblock \DOIprefix\doi{10.1016/S0022-4073(99)00028-X}.
\bibitem[{{Miyamoto} et~al.(2021){Miyamoto}, {Niihara}, {Wada}, {Ogawa},
  {Senshu}, {Michel}, {Kikuchi}, {Hemmi}, {Nakamura}, {Nakamura}, {Hirata},
  {Sasaki}, {Asphaug}, {Britt}, {Abell}, {Ballouz}, {Banouin}, {Baresi},
  {Barucci}, {Biele}, {Grott}, {Hino}, {Hong}, {Imada}, {Kameda}, {Kobayashi},
  {Libourel}, {Mogi}, {Murdoch}, {Nishio}, {Okamoto}, {Ota}, {Otsuki}, {Otto},
  {Sakatani}, {Shimizu}, {Takemura}, {Terada}, {Tsukamoto}, {Usui} and
  {Willner}}]{Miyamoto_2021}
\bibinfo{author}{{Miyamoto}, H.}, \bibinfo{author}{{Niihara}, T.},
  \bibinfo{author}{{Wada}, K.}, \bibinfo{author}{{Ogawa}, K.},
  \bibinfo{author}{{Senshu}, H.}, \bibinfo{author}{{Michel}, P.},
  \bibinfo{author}{{Kikuchi}, H.}, \bibinfo{author}{{Hemmi}, R.},
  \bibinfo{author}{{Nakamura}, T.}, \bibinfo{author}{{Nakamura}, A.M.},
  \bibinfo{author}{{Hirata}, N.}, \bibinfo{author}{{Sasaki}, S.},
  \bibinfo{author}{{Asphaug}, E.}, \bibinfo{author}{{Britt}, D.T.},
  \bibinfo{author}{{Abell}, P.A.}, \bibinfo{author}{{Ballouz}, R.L.},
  \bibinfo{author}{{Banouin}, O.S.}, \bibinfo{author}{{Baresi}, N.},
  \bibinfo{author}{{Barucci}, M.A.}, \bibinfo{author}{{Biele}, J.},
  \bibinfo{author}{{Grott}, M.}, \bibinfo{author}{{Hino}, H.},
  \bibinfo{author}{{Hong}, P.K.}, \bibinfo{author}{{Imada}, T.},
  \bibinfo{author}{{Kameda}, S.}, \bibinfo{author}{{Kobayashi}, M.},
  \bibinfo{author}{{Libourel}, G.}, \bibinfo{author}{{Mogi}, K.},
  \bibinfo{author}{{Murdoch}, N.}, \bibinfo{author}{{Nishio}, Y.},
  \bibinfo{author}{{Okamoto}, S.}, \bibinfo{author}{{Ota}, Y.},
  \bibinfo{author}{{Otsuki}, M.}, \bibinfo{author}{{Otto}, K.A.},
  \bibinfo{author}{{Sakatani}, N.}, \bibinfo{author}{{Shimizu}, Y.},
  \bibinfo{author}{{Takemura}, T.}, \bibinfo{author}{{Terada}, N.},
  \bibinfo{author}{{Tsukamoto}, M.}, \bibinfo{author}{{Usui}, T.},
  \bibinfo{author}{{Willner}, K.}, \bibinfo{year}{2021}.
\newblock \bibinfo{title}{{Surface environment of Phobos and Phobos simulant
  UTPS}}.
\newblock \bibinfo{journal}{Earth, Planets and Space} \bibinfo{volume}{73},
  \bibinfo{pages}{214}.
\newblock \DOIprefix\doi{10.1186/s40623-021-01406-3}.
\bibitem[{{M{\"o}hlmann}(2002)}]{Mohlmann_2002}
\bibinfo{author}{{M{\"o}hlmann}, D.}, \bibinfo{year}{2002}.
\newblock \bibinfo{title}{{Influence of porosity on the thermal behaviour of
  comet surfaces}}.
\newblock \bibinfo{journal}{Advances in Space Research} \bibinfo{volume}{29},
  \bibinfo{pages}{691--704}.
\newblock \DOIprefix\doi{10.1016/S0273-1177(02)00003-0}.
\bibitem[{{Molaro} et~al.(2020){Molaro}, {Hergenrother}, {Chesley}, {Walsh},
  {Hanna}, {Haberle}, {Schwartz}, {Ballouz}, {Bottke}, {Campins} and
  {Lauretta}}]{Molaro_2020}
\bibinfo{author}{{Molaro}, J.L.}, \bibinfo{author}{{Hergenrother}, C.W.},
  \bibinfo{author}{{Chesley}, S.R.}, \bibinfo{author}{{Walsh}, K.J.},
  \bibinfo{author}{{Hanna}, R.D.}, \bibinfo{author}{{Haberle}, C.W.},
  \bibinfo{author}{{Schwartz}, S.R.}, \bibinfo{author}{{Ballouz}, R.L.},
  \bibinfo{author}{{Bottke}, W.F.}, \bibinfo{author}{{Campins}, H.J.},
  \bibinfo{author}{{Lauretta}, D.S.}, \bibinfo{year}{2020}.
\newblock \bibinfo{title}{{Thermal Fatigue as a Driving Mechanism for Activity
  on Asteroid Bennu}}.
\newblock \bibinfo{journal}{Journal of Geophysical Research (Planets)}
  \bibinfo{volume}{125}, \bibinfo{pages}{e06325}.
\newblock \DOIprefix\doi{10.1029/2019JE00632510.1002/essoar.10501385.2}.
\bibitem[{Murchie and Erard(1996)}]{Murchie_1996}
\bibinfo{author}{Murchie, S.}, \bibinfo{author}{Erard, S.},
  \bibinfo{year}{1996}.
\newblock \bibinfo{title}{Spectral properties and heterogeneity of {P}hobos
  from measurements by {P}hobos 2}.
\newblock \bibinfo{journal}{Icarus} \bibinfo{volume}{123},
  \bibinfo{pages}{63--86}.
\bibitem[{Murchie et~al.(1991)Murchie, Britt, Head, Pratt, Fisher, Zhukov,
  Kuzmin, Ksanfomality, Zharkov, Nikitin, Fanale, Blaney, Bell and
  Robinson}]{Murchie_1991}
\bibinfo{author}{Murchie, S.L.}, \bibinfo{author}{Britt, D.T.},
  \bibinfo{author}{Head, J.W.}, \bibinfo{author}{Pratt, S.F.},
  \bibinfo{author}{Fisher, P.C.}, \bibinfo{author}{Zhukov, B.S.},
  \bibinfo{author}{Kuzmin, A.A.}, \bibinfo{author}{Ksanfomality, L.V.},
  \bibinfo{author}{Zharkov, A.V.}, \bibinfo{author}{Nikitin, G.E.},
  \bibinfo{author}{Fanale, F.P.}, \bibinfo{author}{Blaney, D.L.},
  \bibinfo{author}{Bell, J.F.}, \bibinfo{author}{Robinson, M.S.},
  \bibinfo{year}{1991}.
\newblock \bibinfo{title}{Color heterogeneity of the surface of {P}hobos:
  Relationships to geological features and comparison to meteorites analogs}.
\newblock \bibinfo{journal}{Journal of Geophysical Research}
  \bibinfo{volume}{96}, \bibinfo{pages}{5295--5945}.
\bibitem[{{N{\"a}r{\"a}nen} et~al.(2004){N{\"a}r{\"a}nen}, {Kaasalainen},
  {Peltoniemi}, {Heikkil{\"a}}, {Granvik} and {Saarinen}}]{Naranen_2004}
\bibinfo{author}{{N{\"a}r{\"a}nen}, J.}, \bibinfo{author}{{Kaasalainen}, S.},
  \bibinfo{author}{{Peltoniemi}, J.}, \bibinfo{author}{{Heikkil{\"a}}, S.},
  \bibinfo{author}{{Granvik}, M.}, \bibinfo{author}{{Saarinen}, V.},
  \bibinfo{year}{2004}.
\newblock \bibinfo{title}{{Laboratory photometry of planetary regolith analogs.
  II. Surface roughness and extremes of packing density}}.
\newblock \bibinfo{journal}{Astronomy and Astrophysics} \bibinfo{volume}{426},
  \bibinfo{pages}{1103--1109}.
\newblock \DOIprefix\doi{10.1051/0004-6361:20040556}.
\bibitem[{{Noguchi} et~al.(2022){Noguchi}, {Matsumoto}, {Yabuta}, {Kobayashi},
  {Miyake}, {Naraoka}, {Okazaki}, {Imae}, {Yamaguchi}, {Kilcoyne}, {Takeichi}
  and {Takahashi}}]{Noguchi_2022}
\bibinfo{author}{{Noguchi}, T.}, \bibinfo{author}{{Matsumoto}, R.},
  \bibinfo{author}{{Yabuta}, H.}, \bibinfo{author}{{Kobayashi}, H.},
  \bibinfo{author}{{Miyake}, A.}, \bibinfo{author}{{Naraoka}, H.},
  \bibinfo{author}{{Okazaki}, R.}, \bibinfo{author}{{Imae}, N.},
  \bibinfo{author}{{Yamaguchi}, A.}, \bibinfo{author}{{Kilcoyne}, A.L.D.},
  \bibinfo{author}{{Takeichi}, Y.}, \bibinfo{author}{{Takahashi}, Y.},
  \bibinfo{year}{2022}.
\newblock \bibinfo{title}{{Antarctic micrometeorite composed of CP and CS
  IDP-like material: A micro-breccia originated from a partially ice-melted
  comet-like small body}}.
\newblock \bibinfo{journal}{Meteoritics \& Planetary Science}
  \bibinfo{volume}{57}, \bibinfo{pages}{2042--2062}.
\newblock \DOIprefix\doi{10.1111/maps.13919}.
\bibitem[{{Ohtake} et~al.(2010){Ohtake}, {Matsunaga}, {Yokota}, {Yamamoto},
  {Ogawa}, {Morota}, {Honda}, {Haruyama}, {Kitazato}, {Takeda}, {Iwasaki},
  {Nakamura}, {Hiroi}, {Kodama} and {Otake}}]{Ohtake_2010}
\bibinfo{author}{{Ohtake}, M.}, \bibinfo{author}{{Matsunaga}, T.},
  \bibinfo{author}{{Yokota}, Y.}, \bibinfo{author}{{Yamamoto}, S.},
  \bibinfo{author}{{Ogawa}, Y.}, \bibinfo{author}{{Morota}, T.},
  \bibinfo{author}{{Honda}, C.}, \bibinfo{author}{{Haruyama}, J.},
  \bibinfo{author}{{Kitazato}, K.}, \bibinfo{author}{{Takeda}, H.},
  \bibinfo{author}{{Iwasaki}, A.}, \bibinfo{author}{{Nakamura}, R.},
  \bibinfo{author}{{Hiroi}, T.}, \bibinfo{author}{{Kodama}, S.},
  \bibinfo{author}{{Otake}, H.}, \bibinfo{year}{2010}.
\newblock \bibinfo{title}{{Deriving the Absolute Reflectance of Lunar Surface
  Using SELENE (Kaguya) Multiband Imager Data}}.
\newblock \bibinfo{journal}{Space Science Reviews} \bibinfo{volume}{154},
  \bibinfo{pages}{57--77}.
\newblock \DOIprefix\doi{10.1007/s11214-010-9689-0}.
\bibitem[{{Pajola} et~al.(2013){Pajola}, {Lazzarin}, {Dalle Ore}, {Cruikshank},
  {Roush}, {Magrin}, {Bertini}, {La Forgia} and {Barbieri}}]{Pajola_2013}
\bibinfo{author}{{Pajola}, M.}, \bibinfo{author}{{Lazzarin}, M.},
  \bibinfo{author}{{Dalle Ore}, C.M.}, \bibinfo{author}{{Cruikshank}, D.P.},
  \bibinfo{author}{{Roush}, T.L.}, \bibinfo{author}{{Magrin}, S.},
  \bibinfo{author}{{Bertini}, I.}, \bibinfo{author}{{La Forgia}, F.},
  \bibinfo{author}{{Barbieri}, C.}, \bibinfo{year}{2013}.
\newblock \bibinfo{title}{{Phobos as a D-type Captured Asteroid, Spectral
  Modeling from 0.25 to 4.0 {\ensuremath{\mu}}m}}.
\newblock \bibinfo{journal}{Astrophysical Journal} \bibinfo{volume}{777},
  \bibinfo{pages}{127}.
\newblock \DOIprefix\doi{10.1088/0004-637X/777/2/127}.
\bibitem[{Pajola et~al.(2018)Pajola, Roush, Dalle~Ore, Marzo and
  Simioni}]{Pajola_2018}
\bibinfo{author}{Pajola, M.}, \bibinfo{author}{Roush, T.},
  \bibinfo{author}{Dalle~Ore, C.}, \bibinfo{author}{Marzo, G.A.},
  \bibinfo{author}{Simioni, E.}, \bibinfo{year}{2018}.
\newblock \bibinfo{title}{Phobos {MRO}/{CRISM} visible and near-infrared
  (0.5-2.5 µm) spectral modeling}.
\newblock \bibinfo{journal}{Planetary and Space Science} \bibinfo{volume}{154},
  \bibinfo{pages}{63--71}.
\bibitem[{Pang et~al.(1978)Pang, Pollack, Veverka, Lane and Ajello}]{Pang_1978}
\bibinfo{author}{Pang, K.D.}, \bibinfo{author}{Pollack, J.B.},
  \bibinfo{author}{Veverka, J.}, \bibinfo{author}{Lane, A.L.},
  \bibinfo{author}{Ajello, J.M.}, \bibinfo{year}{1978}.
\newblock \bibinfo{title}{The composition of {P}hobos: Evidence for
  carbonaceous chondrite surface from spectral analysis}.
\newblock \bibinfo{journal}{Science} \bibinfo{volume}{199},
  \bibinfo{pages}{64--66}.
\bibitem[{{P{\"a}tzold} et~al.(2016){P{\"a}tzold}, {Andert}, {Hahn}, {Asmar},
  {Barriot}, {Bird}, {H{\"a}usler}, {Peter}, {Tellmann}, {Gr{\"u}n},
  {Weissman}, {Sierks}, {Jorda}, {Gaskell}, {Preusker} and
  {Scholten}}]{Patzold_2016}
\bibinfo{author}{{P{\"a}tzold}, M.}, \bibinfo{author}{{Andert}, T.},
  \bibinfo{author}{{Hahn}, M.}, \bibinfo{author}{{Asmar}, S.W.},
  \bibinfo{author}{{Barriot}, J.P.}, \bibinfo{author}{{Bird}, M.K.},
  \bibinfo{author}{{H{\"a}usler}, B.}, \bibinfo{author}{{Peter}, K.},
  \bibinfo{author}{{Tellmann}, S.}, \bibinfo{author}{{Gr{\"u}n}, E.},
  \bibinfo{author}{{Weissman}, P.R.}, \bibinfo{author}{{Sierks}, H.},
  \bibinfo{author}{{Jorda}, L.}, \bibinfo{author}{{Gaskell}, R.},
  \bibinfo{author}{{Preusker}, F.}, \bibinfo{author}{{Scholten}, F.},
  \bibinfo{year}{2016}.
\newblock \bibinfo{title}{{A homogeneous nucleus for comet
  67P/Churyumov-Gerasimenko from its gravity field}}.
\newblock \bibinfo{journal}{Nature} \bibinfo{volume}{530},
  \bibinfo{pages}{63--65}.
\newblock \DOIprefix\doi{10.1038/nature16535}.
\bibitem[{{P{\"a}tzold} et~al.(2019){P{\"a}tzold}, {Andert}, {Hahn}, {Barriot},
  {Asmar}, {H{\"a}usler}, {Bird}, {Tellmann}, {Oschlisniok} and
  {Peter}}]{Patzold_2019}
\bibinfo{author}{{P{\"a}tzold}, M.}, \bibinfo{author}{{Andert}, T.P.},
  \bibinfo{author}{{Hahn}, M.}, \bibinfo{author}{{Barriot}, J.P.},
  \bibinfo{author}{{Asmar}, S.W.}, \bibinfo{author}{{H{\"a}usler}, B.},
  \bibinfo{author}{{Bird}, M.K.}, \bibinfo{author}{{Tellmann}, S.},
  \bibinfo{author}{{Oschlisniok}, J.}, \bibinfo{author}{{Peter}, K.},
  \bibinfo{year}{2019}.
\newblock \bibinfo{title}{{The Nucleus of comet 67P/Churyumov-Gerasimenko -
  Part I: The global view - nucleus mass, mass-loss, porosity, and
  implications}}.
\newblock \bibinfo{journal}{Monthly Notices of the Royal Astronomical Society}
  \bibinfo{volume}{483}, \bibinfo{pages}{2337--2346}.
\newblock \DOIprefix\doi{10.1093/mnras/sty3171}.
\bibitem[{{Pieters} and {Noble}(2016)}]{Pieters_2016}
\bibinfo{author}{{Pieters}, C.M.}, \bibinfo{author}{{Noble}, S.K.},
  \bibinfo{year}{2016}.
\newblock \bibinfo{title}{{Space weathering on airless bodies}}.
\newblock \bibinfo{journal}{Journal of Geophysical Research (Planets)}
  \bibinfo{volume}{121}, \bibinfo{pages}{1865--1884}.
\newblock \DOIprefix\doi{10.1002/2016JE005128}.
\bibitem[{{Poch} et~al.(2016a){Poch}, {Pommerol}, {Jost}, {Carrasco}, {Szopa}
  and {Thomas}}]{Poch_2016_2}
\bibinfo{author}{{Poch}, O.}, \bibinfo{author}{{Pommerol}, A.},
  \bibinfo{author}{{Jost}, B.}, \bibinfo{author}{{Carrasco}, N.},
  \bibinfo{author}{{Szopa}, C.}, \bibinfo{author}{{Thomas}, N.},
  \bibinfo{year}{2016}a.
\newblock \bibinfo{title}{{Sublimation of ice-tholins mixtures: A morphological
  and spectro-photometric study}}.
\newblock \bibinfo{journal}{Icarus} \bibinfo{volume}{266},
  \bibinfo{pages}{288--305}.
\newblock \DOIprefix\doi{10.1016/j.icarus.2015.11.006}.
\bibitem[{{Poch} et~al.(2016b){Poch}, {Pommerol}, {Jost}, {Carrasco}, {Szopa}
  and {Thomas}}]{Poch_2016}
\bibinfo{author}{{Poch}, O.}, \bibinfo{author}{{Pommerol}, A.},
  \bibinfo{author}{{Jost}, B.}, \bibinfo{author}{{Carrasco}, N.},
  \bibinfo{author}{{Szopa}, C.}, \bibinfo{author}{{Thomas}, N.},
  \bibinfo{year}{2016}b.
\newblock \bibinfo{title}{{Sublimation of water ice mixed with silicates and
  tholins: Evolution of surface texture and reflectance spectra, with
  implications for comets}}.
\newblock \bibinfo{journal}{Icarus} \bibinfo{volume}{267},
  \bibinfo{pages}{154--173}.
\newblock \DOIprefix\doi{10.1016/j.icarus.2015.12.017}.
\bibitem[{{Poggiali} et~al.(2024){Poggiali}, {Fossi, L.}, {Wargnier, A.},
  {Beccarelli, J.}, {Brucato, J. R.}, {Barucci, M. A.}, {Beck, P.}, {Matsuoka,
  M.}, {Nakamura, T.}, {Merlin, F.}, {Fornasier, S.}, {Pajola, M.},
  {Doressoundiram, A.}, {Gautier, T.} and {David, G.}}]{Poggiali_2024}
\bibinfo{author}{{Poggiali}, G.}, \bibinfo{author}{{Fossi, L.}},
  \bibinfo{author}{{Wargnier, A.}}, \bibinfo{author}{{Beccarelli, J.}},
  \bibinfo{author}{{Brucato, J. R.}}, \bibinfo{author}{{Barucci, M. A.}},
  \bibinfo{author}{{Beck, P.}}, \bibinfo{author}{{Matsuoka, M.}},
  \bibinfo{author}{{Nakamura, T.}}, \bibinfo{author}{{Merlin, F.}},
  \bibinfo{author}{{Fornasier, S.}}, \bibinfo{author}{{Pajola, M.}},
  \bibinfo{author}{{Doressoundiram, A.}}, \bibinfo{author}{{Gautier, T.}},
  \bibinfo{author}{{David, G.}}, \bibinfo{year}{2024}.
\newblock \bibinfo{title}{Grain size effects on the infrared spectrum of
  mineral mixtures with dark components: New laboratory experiments to
  interpret low-albedo rocky planetary surfaces}.
\newblock \bibinfo{journal}{Astronomy and Astrophysics} \bibinfo{volume}{685},
  \bibinfo{pages}{A14}.
\newblock \URLprefix \url{https://doi.org/10.1051/0004-6361/202347681},
  \DOIprefix\doi{10.1051/0004-6361/202347681}.
\bibitem[{{Pommerol} et~al.(2019){Pommerol}, {Jost}, {Poch}, {Yoldi}, {Brouet},
  {Gracia-Bern{\'a}}, {Cerubini}, {Galli}, {Wurz}, {Gundlach}, {Blum},
  {Carrasco}, {Szopa} and {Thomas}}]{Pommerol_2019}
\bibinfo{author}{{Pommerol}, A.}, \bibinfo{author}{{Jost}, B.},
  \bibinfo{author}{{Poch}, O.}, \bibinfo{author}{{Yoldi}, Z.},
  \bibinfo{author}{{Brouet}, Y.}, \bibinfo{author}{{Gracia-Bern{\'a}}, A.},
  \bibinfo{author}{{Cerubini}, R.}, \bibinfo{author}{{Galli}, A.},
  \bibinfo{author}{{Wurz}, P.}, \bibinfo{author}{{Gundlach}, B.},
  \bibinfo{author}{{Blum}, J.}, \bibinfo{author}{{Carrasco}, N.},
  \bibinfo{author}{{Szopa}, C.}, \bibinfo{author}{{Thomas}, N.},
  \bibinfo{year}{2019}.
\newblock \bibinfo{title}{{Experimenting with Mixtures of Water Ice and Dust as
  Analogues for Icy Planetary Material. Recipes from the Ice Laboratory at the
  University of Bern}}.
\newblock \bibinfo{journal}{Space Science Reviews} \bibinfo{volume}{215},
  \bibinfo{pages}{37}.
\newblock \DOIprefix\doi{10.1007/s11214-019-0603-0}.
\bibitem[{{Ryan} et~al.(2024){Ryan}, {Ballouz}, {Macke}, {Connolly} and
  {Lauretta}}]{Ryan_2024}
\bibinfo{author}{{Ryan}, A.J.}, \bibinfo{author}{{Ballouz}, R.L.},
  \bibinfo{author}{{Macke}, R.J.}, \bibinfo{author}{{Connolly}, H.C.},
  \bibinfo{author}{{Lauretta}, D.S.}, \bibinfo{year}{2024}.
\newblock \bibinfo{title}{{Physical and Thermal Properties of OSIRIS-REx
  Samples: Insight into the Evolution of Bennu and Its Regolith}}, in:
  \bibinfo{booktitle}{LPI Contributions}, p. \bibinfo{pages}{1594}.
\bibitem[{{Salisbury} et~al.(1991){Salisbury}, {D'Aria} and
  {Jarosewich}}]{Salisbury_1991}
\bibinfo{author}{{Salisbury}, J.W.}, \bibinfo{author}{{D'Aria}, D.M.},
  \bibinfo{author}{{Jarosewich}, E.}, \bibinfo{year}{1991}.
\newblock \bibinfo{title}{{Midinfrared (2.5-13.5 {\ensuremath{\mu}}m)
  reflectance spectra of powdered stony meteorites}}.
\newblock \bibinfo{journal}{Icarus} \bibinfo{volume}{92},
  \bibinfo{pages}{280--297}.
\newblock \DOIprefix\doi{10.1016/0019-1035(91)90052-U}.
\bibitem[{{Saunders} et~al.(1986){Saunders}, {Fanale}, {Parker}, {Stephens} and
  {Sutton}}]{Saunders_1986}
\bibinfo{author}{{Saunders}, R.S.}, \bibinfo{author}{{Fanale}, F.P.},
  \bibinfo{author}{{Parker}, T.J.}, \bibinfo{author}{{Stephens}, J.B.},
  \bibinfo{author}{{Sutton}, S.}, \bibinfo{year}{1986}.
\newblock \bibinfo{title}{{Properties of filamentary sublimation residues from
  dispersions of clay in ice}}.
\newblock \bibinfo{journal}{Icarus} \bibinfo{volume}{66},
  \bibinfo{pages}{94--104}.
\newblock \DOIprefix\doi{10.1016/0019-1035(86)90009-6}.
\bibitem[{{Schmidt} and {Bourguignon}(2019)}]{Schmidt_2019}
\bibinfo{author}{{Schmidt}, F.}, \bibinfo{author}{{Bourguignon}, S.},
  \bibinfo{year}{2019}.
\newblock \bibinfo{title}{{Efficiency of BRDF sampling and bias on the average
  photometric behavior}}.
\newblock \bibinfo{journal}{Icarus} \bibinfo{volume}{317},
  \bibinfo{pages}{10--26}.
\newblock \DOIprefix\doi{10.1016/j.icarus.2018.06.025},
  \href{http://arxiv.org/abs/1806.03898}{\tt arXiv:1806.03898}.
\bibitem[{{Schmidt} and {Fernando}(2015)}]{Schmidt_2015}
\bibinfo{author}{{Schmidt}, F.}, \bibinfo{author}{{Fernando}, J.},
  \bibinfo{year}{2015}.
\newblock \bibinfo{title}{{Realistic uncertainties on Hapke model parameters
  from photometric measurement}}.
\newblock \bibinfo{journal}{Icarus} \bibinfo{volume}{260},
  \bibinfo{pages}{73--93}.
\newblock \DOIprefix\doi{10.1016/j.icarus.2015.07.002},
  \href{http://arxiv.org/abs/1506.08089}{\tt arXiv:1506.08089}.
\bibitem[{{Schr{\"o}der} et~al.(2014){Schr{\"o}der}, {Grynko}, {Pommerol},
  {Keller}, {Thomas} and {Roush}}]{Schroder_2014}
\bibinfo{author}{{Schr{\"o}der}, S.E.}, \bibinfo{author}{{Grynko}, Y.},
  \bibinfo{author}{{Pommerol}, A.}, \bibinfo{author}{{Keller}, H.U.},
  \bibinfo{author}{{Thomas}, N.}, \bibinfo{author}{{Roush}, T.L.},
  \bibinfo{year}{2014}.
\newblock \bibinfo{title}{{Laboratory observations and simulations of phase
  reddening}}.
\newblock \bibinfo{journal}{Icarus} \bibinfo{volume}{239},
  \bibinfo{pages}{201--216}.
\newblock \DOIprefix\doi{10.1016/j.icarus.2014.06.010},
  \href{http://arxiv.org/abs/1701.05822}{\tt arXiv:1701.05822}.
\bibitem[{{Schr{\"o}der} et~al.(2021){Schr{\"o}der}, {Poch}, {Ferrari}, {De
  Angelis}, {Sultana}, {Potin}, {Beck}, {De Sanctis} and
  {Schmitt}}]{Schroder_2021}
\bibinfo{author}{{Schr{\"o}der}, S.E.}, \bibinfo{author}{{Poch}, O.},
  \bibinfo{author}{{Ferrari}, M.}, \bibinfo{author}{{De Angelis}, S.},
  \bibinfo{author}{{Sultana}, R.}, \bibinfo{author}{{Potin}, S.M.},
  \bibinfo{author}{{Beck}, P.}, \bibinfo{author}{{De Sanctis}, M.C.},
  \bibinfo{author}{{Schmitt}, B.}, \bibinfo{year}{2021}.
\newblock \bibinfo{title}{{Dwarf planet (1) Ceres surface bluing due to high
  porosity resulting from sublimation}}.
\newblock \bibinfo{journal}{Nature Communications} \bibinfo{volume}{12},
  \bibinfo{pages}{274}.
\newblock \DOIprefix\doi{10.1038/s41467-020-20494-5}.
\bibitem[{{Shepard} and {Cloutis}(2011)}]{Shepard_2011}
\bibinfo{author}{{Shepard}, M.K.}, \bibinfo{author}{{Cloutis}, E.},
  \bibinfo{year}{2011}.
\newblock \bibinfo{title}{{Laboratory Measurements of Band Depth Variation with
  Observation Geometry}}, in: \bibinfo{booktitle}{42nd Annual Lunar and
  Planetary Science Conference}, p. \bibinfo{pages}{1043}.
\bibitem[{{Shepard} and {Helfenstein}(2007)}]{Shepard_2007}
\bibinfo{author}{{Shepard}, M.K.}, \bibinfo{author}{{Helfenstein}, P.},
  \bibinfo{year}{2007}.
\newblock \bibinfo{title}{{A test of the Hapke photometric model}}.
\newblock \bibinfo{journal}{Journal of Geophysical Research (Planets)}
  \bibinfo{volume}{112}, \bibinfo{pages}{E03001}.
\newblock \DOIprefix\doi{10.1029/2005JE002625}.
\bibitem[{{Shkuratov} et~al.(2012){Shkuratov}, {Kaydash}, {Korokhin},
  {Velikodsky}, {Petrov}, {Zubko}, {Stankevich} and {Videen}}]{Shkuratov_2012}
\bibinfo{author}{{Shkuratov}, Y.}, \bibinfo{author}{{Kaydash}, V.},
  \bibinfo{author}{{Korokhin}, V.}, \bibinfo{author}{{Velikodsky}, Y.},
  \bibinfo{author}{{Petrov}, D.}, \bibinfo{author}{{Zubko}, E.},
  \bibinfo{author}{{Stankevich}, D.}, \bibinfo{author}{{Videen}, G.},
  \bibinfo{year}{2012}.
\newblock \bibinfo{title}{{A critical assessment of the Hapke photometric
  model}}.
\newblock \bibinfo{journal}{Journal of Quantitiative Spectroscopy and Radiative
  Transfer} \bibinfo{volume}{113}, \bibinfo{pages}{2431--2456}.
\newblock \DOIprefix\doi{10.1016/j.jqsrt.2012.04.010}.
\bibitem[{{Shkuratov} et~al.(1999){Shkuratov}, {Starukhina}, {Hoffmann} and
  {Arnold}}]{Shkuratov_1999}
\bibinfo{author}{{Shkuratov}, Y.}, \bibinfo{author}{{Starukhina}, L.},
  \bibinfo{author}{{Hoffmann}, H.}, \bibinfo{author}{{Arnold}, G.},
  \bibinfo{year}{1999}.
\newblock \bibinfo{title}{{A Model of Spectral Albedo of Particulate Surfaces:
  Implications for Optical Properties of the Moon}}.
\newblock \bibinfo{journal}{Icarus} \bibinfo{volume}{137},
  \bibinfo{pages}{235--246}.
\newblock \DOIprefix\doi{10.1006/icar.1998.6035}.
\bibitem[{{Simonelli} et~al.(1998){Simonelli}, {Wisz}, {Switala}, {Adinolfi},
  {Veverka}, {Thomas} and {Helfenstein}}]{Simonelli_1998}
\bibinfo{author}{{Simonelli}, D.P.}, \bibinfo{author}{{Wisz}, M.},
  \bibinfo{author}{{Switala}, A.}, \bibinfo{author}{{Adinolfi}, D.},
  \bibinfo{author}{{Veverka}, J.}, \bibinfo{author}{{Thomas}, P.C.},
  \bibinfo{author}{{Helfenstein}, P.}, \bibinfo{year}{1998}.
\newblock \bibinfo{title}{{Photometric Properties of PHOBOS Surface Materials
  From Viking Images}}.
\newblock \bibinfo{journal}{Icarus} \bibinfo{volume}{131},
  \bibinfo{pages}{52--77}.
\newblock \DOIprefix\doi{10.1006/icar.1997.5800}.
\bibitem[{{Skorov} et~al.(2011){Skorov}, {van Lieshout}, {Blum} and
  {Keller}}]{Skorov_2011}
\bibinfo{author}{{Skorov}, Y.V.}, \bibinfo{author}{{van Lieshout}, R.},
  \bibinfo{author}{{Blum}, J.}, \bibinfo{author}{{Keller}, H.U.},
  \bibinfo{year}{2011}.
\newblock \bibinfo{title}{{Activity of comets: Gas transport in the
  near-surface porous layers of a cometary nucleus}}.
\newblock \bibinfo{journal}{Icarus} \bibinfo{volume}{212},
  \bibinfo{pages}{867--876}.
\newblock \DOIprefix\doi{10.1016/j.icarus.2011.01.018},
  \href{http://arxiv.org/abs/1101.2525}{\tt arXiv:1101.2525}.
\bibitem[{{Souchon} et~al.(2011){Souchon}, {Pinet}, {Chevrel}, {Daydou},
  {Baratoux}, {Kurita}, {Shepard} and {Helfenstein}}]{Souchon_2011}
\bibinfo{author}{{Souchon}, A.L.}, \bibinfo{author}{{Pinet}, P.C.},
  \bibinfo{author}{{Chevrel}, S.D.}, \bibinfo{author}{{Daydou}, Y.H.},
  \bibinfo{author}{{Baratoux}, D.}, \bibinfo{author}{{Kurita}, K.},
  \bibinfo{author}{{Shepard}, M.K.}, \bibinfo{author}{{Helfenstein}, P.},
  \bibinfo{year}{2011}.
\newblock \bibinfo{title}{{An experimental study of Hapke's modeling of natural
  granular surface samples}}.
\newblock \bibinfo{journal}{Icarus} \bibinfo{volume}{215},
  \bibinfo{pages}{313--331}.
\newblock \DOIprefix\doi{10.1016/j.icarus.2011.06.023}.
\bibitem[{{Stephan} et~al.(2017){Stephan}, {Jaumann}, {Krohn}, {Schmedemann},
  {Zambon}, {Tosi}, {Carrozzo}, {McFadden}, {Otto}, {De Sanctis}, {Ammannito},
  {Matz}, {Roatsch}, {Preusker}, {Raymond} and {Russell}}]{Stephan_2017}
\bibinfo{author}{{Stephan}, K.}, \bibinfo{author}{{Jaumann}, R.},
  \bibinfo{author}{{Krohn}, K.}, \bibinfo{author}{{Schmedemann}, N.},
  \bibinfo{author}{{Zambon}, F.}, \bibinfo{author}{{Tosi}, F.},
  \bibinfo{author}{{Carrozzo}, F.G.}, \bibinfo{author}{{McFadden}, L.A.},
  \bibinfo{author}{{Otto}, K.}, \bibinfo{author}{{De Sanctis}, M.C.},
  \bibinfo{author}{{Ammannito}, E.}, \bibinfo{author}{{Matz}, K.D.},
  \bibinfo{author}{{Roatsch}, T.}, \bibinfo{author}{{Preusker}, F.},
  \bibinfo{author}{{Raymond}, C.A.}, \bibinfo{author}{{Russell}, C.T.},
  \bibinfo{year}{2017}.
\newblock \bibinfo{title}{{An investigation of the bluish material on Ceres}}.
\newblock \bibinfo{journal}{Geophysics Research Letters} \bibinfo{volume}{44},
  \bibinfo{pages}{1660--1668}.
\newblock \DOIprefix\doi{10.1002/2016GL071652}.
\bibitem[{Sugita et~al.(2019)Sugita, Honda, Morota, Kameda, Sawada, Tatsumi,
  Yamada, Honda, Yokota, Kouyama, Sakatani, Ogawa, Suzuki, Okada, Namiki,
  Tanaka, Iijima, Yoshioka, Hayakawa, Cho, Matsuoka, Hirata, Hirata, Miyamoto,
  Domingue, Hirabayashi, Nakamura, Hiroi, Michikami, Michel, Ballouz, Barnouin,
  Ernst, Schröder, Kikuchi, Hemmi, Komatsu, Fukuhara, Taguchi, Arai, Senshu,
  Demura, Ogawa, Shimaki, Sekiguchi, Müller, Hagermann, Mizuno, Noda,
  Matsumoto, Yamada, Ishihara, Ikeda, Araki, Yamamoto, Abe, Yoshida, Higuchi,
  Sasaki, Oshigami, Tsuruta, Asari, Tazawa, Shizugami, Kimura, Otsubo, Yabuta,
  Hasegawa, Ishiguro, Tachibana, Palmer, Gaskell, Corre, Jaumann, Otto,
  Schmitz, Abell, Barucci, Zolensky, Vilas, Thuillet, Sugimoto, Takaki, Suzuki,
  Kamiyoshihara, Okada, Nagata, Fujimoto, Yoshikawa, Yamamoto, Shirai, Noguchi,
  Ogawa, Terui, Kikuchi, Yamaguchi, Oki, Takao, Takeuchi, Ono, Mimasu,
  Yoshikawa, Takahashi, Takei, Fujii, Hirose, Nakazawa, Hosoda, Mori, Shimada,
  Soldini, Iwata, Abe, Yano, Tsukizaki, Ozaki, Nishiyama, Saiki, Watanabe and
  Tsuda}]{Sugita_2019}
\bibinfo{author}{Sugita, S.}, \bibinfo{author}{Honda, R.},
  \bibinfo{author}{Morota, T.}, \bibinfo{author}{Kameda, S.},
  \bibinfo{author}{Sawada, H.}, \bibinfo{author}{Tatsumi, E.},
  \bibinfo{author}{Yamada, M.}, \bibinfo{author}{Honda, C.},
  \bibinfo{author}{Yokota, Y.}, \bibinfo{author}{Kouyama, T.},
  \bibinfo{author}{Sakatani, N.}, \bibinfo{author}{Ogawa, K.},
  \bibinfo{author}{Suzuki, H.}, \bibinfo{author}{Okada, T.},
  \bibinfo{author}{Namiki, N.}, \bibinfo{author}{Tanaka, S.},
  \bibinfo{author}{Iijima, Y.}, \bibinfo{author}{Yoshioka, K.},
  \bibinfo{author}{Hayakawa, M.}, \bibinfo{author}{Cho, Y.},
  \bibinfo{author}{Matsuoka, M.}, \bibinfo{author}{Hirata, N.},
  \bibinfo{author}{Hirata, N.}, \bibinfo{author}{Miyamoto, H.},
  \bibinfo{author}{Domingue, D.}, \bibinfo{author}{Hirabayashi, M.},
  \bibinfo{author}{Nakamura, T.}, \bibinfo{author}{Hiroi, T.},
  \bibinfo{author}{Michikami, T.}, \bibinfo{author}{Michel, P.},
  \bibinfo{author}{Ballouz, R.L.}, \bibinfo{author}{Barnouin, O.S.},
  \bibinfo{author}{Ernst, C.M.}, \bibinfo{author}{Schröder, S.E.},
  \bibinfo{author}{Kikuchi, H.}, \bibinfo{author}{Hemmi, R.},
  \bibinfo{author}{Komatsu, G.}, \bibinfo{author}{Fukuhara, T.},
  \bibinfo{author}{Taguchi, M.}, \bibinfo{author}{Arai, T.},
  \bibinfo{author}{Senshu, H.}, \bibinfo{author}{Demura, H.},
  \bibinfo{author}{Ogawa, Y.}, \bibinfo{author}{Shimaki, Y.},
  \bibinfo{author}{Sekiguchi, T.}, \bibinfo{author}{Müller, T.G.},
  \bibinfo{author}{Hagermann, A.}, \bibinfo{author}{Mizuno, T.},
  \bibinfo{author}{Noda, H.}, \bibinfo{author}{Matsumoto, K.},
  \bibinfo{author}{Yamada, R.}, \bibinfo{author}{Ishihara, Y.},
  \bibinfo{author}{Ikeda, H.}, \bibinfo{author}{Araki, H.},
  \bibinfo{author}{Yamamoto, K.}, \bibinfo{author}{Abe, S.},
  \bibinfo{author}{Yoshida, F.}, \bibinfo{author}{Higuchi, A.},
  \bibinfo{author}{Sasaki, S.}, \bibinfo{author}{Oshigami, S.},
  \bibinfo{author}{Tsuruta, S.}, \bibinfo{author}{Asari, K.},
  \bibinfo{author}{Tazawa, S.}, \bibinfo{author}{Shizugami, M.},
  \bibinfo{author}{Kimura, J.}, \bibinfo{author}{Otsubo, T.},
  \bibinfo{author}{Yabuta, H.}, \bibinfo{author}{Hasegawa, S.},
  \bibinfo{author}{Ishiguro, M.}, \bibinfo{author}{Tachibana, S.},
  \bibinfo{author}{Palmer, E.}, \bibinfo{author}{Gaskell, R.},
  \bibinfo{author}{Corre, L.L.}, \bibinfo{author}{Jaumann, R.},
  \bibinfo{author}{Otto, K.}, \bibinfo{author}{Schmitz, N.},
  \bibinfo{author}{Abell, P.A.}, \bibinfo{author}{Barucci, M.A.},
  \bibinfo{author}{Zolensky, M.E.}, \bibinfo{author}{Vilas, F.},
  \bibinfo{author}{Thuillet, F.}, \bibinfo{author}{Sugimoto, C.},
  \bibinfo{author}{Takaki, N.}, \bibinfo{author}{Suzuki, Y.},
  \bibinfo{author}{Kamiyoshihara, H.}, \bibinfo{author}{Okada, M.},
  \bibinfo{author}{Nagata, K.}, \bibinfo{author}{Fujimoto, M.},
  \bibinfo{author}{Yoshikawa, M.}, \bibinfo{author}{Yamamoto, Y.},
  \bibinfo{author}{Shirai, K.}, \bibinfo{author}{Noguchi, R.},
  \bibinfo{author}{Ogawa, N.}, \bibinfo{author}{Terui, F.},
  \bibinfo{author}{Kikuchi, S.}, \bibinfo{author}{Yamaguchi, T.},
  \bibinfo{author}{Oki, Y.}, \bibinfo{author}{Takao, Y.},
  \bibinfo{author}{Takeuchi, H.}, \bibinfo{author}{Ono, G.},
  \bibinfo{author}{Mimasu, Y.}, \bibinfo{author}{Yoshikawa, K.},
  \bibinfo{author}{Takahashi, T.}, \bibinfo{author}{Takei, Y.},
  \bibinfo{author}{Fujii, A.}, \bibinfo{author}{Hirose, C.},
  \bibinfo{author}{Nakazawa, S.}, \bibinfo{author}{Hosoda, S.},
  \bibinfo{author}{Mori, O.}, \bibinfo{author}{Shimada, T.},
  \bibinfo{author}{Soldini, S.}, \bibinfo{author}{Iwata, T.},
  \bibinfo{author}{Abe, M.}, \bibinfo{author}{Yano, H.},
  \bibinfo{author}{Tsukizaki, R.}, \bibinfo{author}{Ozaki, M.},
  \bibinfo{author}{Nishiyama, K.}, \bibinfo{author}{Saiki, T.},
  \bibinfo{author}{Watanabe, S.}, \bibinfo{author}{Tsuda, Y.},
  \bibinfo{year}{2019}.
\newblock \bibinfo{title}{The geomorphology, color, and thermal properties of
  ryugu: Implications for parent-body processes}.
\newblock \bibinfo{journal}{Science} \bibinfo{volume}{364},
  \bibinfo{pages}{eaaw0422}.
\newblock \URLprefix
  \url{https://www.science.org/doi/abs/10.1126/science.aaw0422},
  \DOIprefix\doi{10.1126/science.aaw0422},
  \href{http://arxiv.org/abs/https://www.science.org/doi/pdf/10.1126/science.aaw0422}{\tt
  arXiv:https://www.science.org/doi/pdf/10.1126/science.aaw0422}.
\bibitem[{{Sullivan} et~al.(2007){Sullivan}, {Arvidson}, {Grotzinger}, {Knoll},
  {Golombek}, {Jolliff}, {Squyres} and {Weitz}}]{Sullivan_2007}
\bibinfo{author}{{Sullivan}, R.}, \bibinfo{author}{{Arvidson}, R.},
  \bibinfo{author}{{Grotzinger}, J.}, \bibinfo{author}{{Knoll}, A.},
  \bibinfo{author}{{Golombek}, M.}, \bibinfo{author}{{Jolliff}, B.},
  \bibinfo{author}{{Squyres}, S.}, \bibinfo{author}{{Weitz}, C.},
  \bibinfo{year}{2007}.
\newblock \bibinfo{title}{{Aeolian Geomorphology with MER Opportunity at
  Meridiani Planum, Mars}}, in: \bibinfo{booktitle}{38th Annual Lunar and
  Planetary Science Conference}, p. \bibinfo{pages}{2048}.
\bibitem[{{Sultana} et~al.(2021){Sultana}, {Poch}, {Beck}, {Schmitt} and
  {Quirico}}]{Sultana_2021}
\bibinfo{author}{{Sultana}, R.}, \bibinfo{author}{{Poch}, O.},
  \bibinfo{author}{{Beck}, P.}, \bibinfo{author}{{Schmitt}, B.},
  \bibinfo{author}{{Quirico}, E.}, \bibinfo{year}{2021}.
\newblock \bibinfo{title}{{Visible and near-infrared reflectance of hyperfine
  and hyperporous particulate surfaces}}.
\newblock \bibinfo{journal}{Icarus} \bibinfo{volume}{357},
  \bibinfo{pages}{114141}.
\newblock \DOIprefix\doi{10.1016/j.icarus.2020.114141},
  \href{http://arxiv.org/abs/2010.16136}{\tt arXiv:2010.16136}.
\bibitem[{{Sultana} et~al.(2023){Sultana}, {Poch}, {Beck}, {Schmitt},
  {Quirico}, {Spadaccia}, {Patty}, {Pommerol}, {Maturilli}, {Helbert} and
  {Alemanno}}]{Sultana_2023}
\bibinfo{author}{{Sultana}, R.}, \bibinfo{author}{{Poch}, O.},
  \bibinfo{author}{{Beck}, P.}, \bibinfo{author}{{Schmitt}, B.},
  \bibinfo{author}{{Quirico}, E.}, \bibinfo{author}{{Spadaccia}, S.},
  \bibinfo{author}{{Patty}, L.}, \bibinfo{author}{{Pommerol}, A.},
  \bibinfo{author}{{Maturilli}, A.}, \bibinfo{author}{{Helbert}, J.},
  \bibinfo{author}{{Alemanno}, G.}, \bibinfo{year}{2023}.
\newblock \bibinfo{title}{{Reflection, emission, and polarization properties of
  surfaces made of hyperfine grains, and implications for the nature of
  primitive small bodies}}.
\newblock \bibinfo{journal}{Icarus} \bibinfo{volume}{395},
  \bibinfo{pages}{115492}.
\newblock \DOIprefix\doi{10.1016/j.icarus.2023.115492},
  \href{http://arxiv.org/abs/2302.10111}{\tt arXiv:2302.10111}.
\bibitem[{{Szabo} et~al.(2022){Szabo}, {Poppe}, {Biber}, {Mutzke}, {Pichler},
  {J{\"a}ggi}, {Galli}, {Wurz} and {Aumayr}}]{Szabo_2022}
\bibinfo{author}{{Szabo}, P.S.}, \bibinfo{author}{{Poppe}, A.R.},
  \bibinfo{author}{{Biber}, H.}, \bibinfo{author}{{Mutzke}, A.},
  \bibinfo{author}{{Pichler}, J.}, \bibinfo{author}{{J{\"a}ggi}, N.},
  \bibinfo{author}{{Galli}, A.}, \bibinfo{author}{{Wurz}, P.},
  \bibinfo{author}{{Aumayr}, F.}, \bibinfo{year}{2022}.
\newblock \bibinfo{title}{{Deducing Lunar Regolith Porosity From Energetic
  Neutral Atom Emission}}.
\newblock \bibinfo{journal}{Geophysics Research Letters} \bibinfo{volume}{49},
  \bibinfo{pages}{e2022GL101232}.
\newblock \DOIprefix\doi{10.1029/2022GL101232}.
\bibitem[{{Tatsumi} et~al.(2018){Tatsumi}, {Domingue}, {Hirata}, {Kitazato},
  {Vilas}, {Lederer}, {Weissman}, {Lowry} and {Sugita}}]{Tatsumi_2018}
\bibinfo{author}{{Tatsumi}, E.}, \bibinfo{author}{{Domingue}, D.},
  \bibinfo{author}{{Hirata}, N.}, \bibinfo{author}{{Kitazato}, K.},
  \bibinfo{author}{{Vilas}, F.}, \bibinfo{author}{{Lederer}, S.},
  \bibinfo{author}{{Weissman}, P.R.}, \bibinfo{author}{{Lowry}, S.C.},
  \bibinfo{author}{{Sugita}, S.}, \bibinfo{year}{2018}.
\newblock \bibinfo{title}{{Vis-NIR disk-integrated photometry of asteroid 25143
  Itokawa around opposition by AMICA/Hayabusa}}.
\newblock \bibinfo{journal}{Icarus} \bibinfo{volume}{311},
  \bibinfo{pages}{175--196}.
\newblock \DOIprefix\doi{10.1016/j.icarus.2018.04.001}.
\bibitem[{{Thomas} et~al.(2008){Thomas}, {Alexander} and
  {Keller}}]{Thomas_2008}
\bibinfo{author}{{Thomas}, N.}, \bibinfo{author}{{Alexander}, C.},
  \bibinfo{author}{{Keller}, H.U.}, \bibinfo{year}{2008}.
\newblock \bibinfo{title}{{Loss of the Surface Layers of Comet Nuclei}}.
\newblock \bibinfo{journal}{Space Science Reviews} \bibinfo{volume}{138},
  \bibinfo{pages}{165--177}.
\newblock \DOIprefix\doi{10.1007/s11214-008-9332-5}.
\bibitem[{{Thomas} et~al.(2011){Thomas}, {Stelter}, {Ivanov}, {Bridges},
  {Herkenhoff} and {McEwen}}]{Thomas_2011}
\bibinfo{author}{{Thomas}, N.}, \bibinfo{author}{{Stelter}, R.},
  \bibinfo{author}{{Ivanov}, A.}, \bibinfo{author}{{Bridges}, N.T.},
  \bibinfo{author}{{Herkenhoff}, K.E.}, \bibinfo{author}{{McEwen}, A.S.},
  \bibinfo{year}{2011}.
\newblock \bibinfo{title}{{Spectral heterogeneity on Phobos and Deimos: HiRISE
  observations and comparisons to Mars Pathfinder results}}.
\newblock \bibinfo{journal}{Planetary and Space Science} \bibinfo{volume}{59},
  \bibinfo{pages}{1281--1292}.
\newblock \DOIprefix\doi{10.1016/j.pss.2010.04.018}.
\bibitem[{{Thomas} et~al.(1979){Thomas}, {Veverka}, {Bloom} and
  {Duxbury}}]{Thomas_1979}
\bibinfo{author}{{Thomas}, P.}, \bibinfo{author}{{Veverka}, J.},
  \bibinfo{author}{{Bloom}, A.}, \bibinfo{author}{{Duxbury}, T.},
  \bibinfo{year}{1979}.
\newblock \bibinfo{title}{{Grooves on Phobos: their distribution, morphology
  and possible origin.}}
\newblock \bibinfo{journal}{Journal of Geophysics Research}
  \bibinfo{volume}{84}, \bibinfo{pages}{8457--8477}.
\newblock \DOIprefix\doi{10.1029/JB084iB14p08457}.
\bibitem[{{Tsuchiyama} et~al.(2014){Tsuchiyama}, {Uesugi}, {Uesugi}, {Nakano},
  {Noguchi}, {Matsumoto}, {Matsuno}, {Nagano}, {Imai}, {Shimada}, {Takeuchi},
  {Suzuki}, {Nakamura}, {Noguchi}, {Abe}, {Yada} and
  {Fujimura}}]{Tsuchiyama_2014}
\bibinfo{author}{{Tsuchiyama}, A.}, \bibinfo{author}{{Uesugi}, M.},
  \bibinfo{author}{{Uesugi}, K.}, \bibinfo{author}{{Nakano}, T.},
  \bibinfo{author}{{Noguchi}, R.}, \bibinfo{author}{{Matsumoto}, T.},
  \bibinfo{author}{{Matsuno}, J.}, \bibinfo{author}{{Nagano}, T.},
  \bibinfo{author}{{Imai}, Y.}, \bibinfo{author}{{Shimada}, A.},
  \bibinfo{author}{{Takeuchi}, A.}, \bibinfo{author}{{Suzuki}, Y.},
  \bibinfo{author}{{Nakamura}, T.}, \bibinfo{author}{{Noguchi}, T.},
  \bibinfo{author}{{Abe}, M.}, \bibinfo{author}{{Yada}, T.},
  \bibinfo{author}{{Fujimura}, A.}, \bibinfo{year}{2014}.
\newblock \bibinfo{title}{{Three-dimensional microstructure of samples
  recovered from asteroid 25143 Itokawa: Comparison with LL5 and LL6 chondrite
  particles}}.
\newblock \bibinfo{journal}{Meteoritics \& Planetary Science}
  \bibinfo{volume}{49}, \bibinfo{pages}{172--187}.
\newblock \DOIprefix\doi{10.1111/maps.12177}.
\bibitem[{{Ulamec} et~al.(2023){Ulamec}, {Michel}, {Grott}, {B{\"o}ttger},
  {Schr{\"o}der}, {H{\"u}bers}, {Cho}, {Rull}, {Murdoch}, {Vernazza},
  {Prieto-Ballesteros}, {Biele}, {Tardivel}, {Arrat}, {Hagelschuer},
  {Knollenberg}, {Vivet}, {Sunday}, {Jorda}, {Groussin}, {Robin} and
  {Miyamoto}}]{Ulamec_2023}
\bibinfo{author}{{Ulamec}, S.}, \bibinfo{author}{{Michel}, P.},
  \bibinfo{author}{{Grott}, M.}, \bibinfo{author}{{B{\"o}ttger}, U.},
  \bibinfo{author}{{Schr{\"o}der}, S.}, \bibinfo{author}{{H{\"u}bers}, H.W.},
  \bibinfo{author}{{Cho}, Y.}, \bibinfo{author}{{Rull}, F.},
  \bibinfo{author}{{Murdoch}, N.}, \bibinfo{author}{{Vernazza}, P.},
  \bibinfo{author}{{Prieto-Ballesteros}, O.}, \bibinfo{author}{{Biele}, J.},
  \bibinfo{author}{{Tardivel}, S.}, \bibinfo{author}{{Arrat}, D.},
  \bibinfo{author}{{Hagelschuer}, T.}, \bibinfo{author}{{Knollenberg}, J.},
  \bibinfo{author}{{Vivet}, D.}, \bibinfo{author}{{Sunday}, C.},
  \bibinfo{author}{{Jorda}, L.}, \bibinfo{author}{{Groussin}, O.},
  \bibinfo{author}{{Robin}, C.}, \bibinfo{author}{{Miyamoto}, H.},
  \bibinfo{year}{2023}.
\newblock \bibinfo{title}{{Science objectives of the MMX rover}}.
\newblock \bibinfo{journal}{Acta Astronautica} \bibinfo{volume}{210},
  \bibinfo{pages}{95--101}.
\newblock \DOIprefix\doi{10.1016/j.actaastro.2023.05.012}.
\bibitem[{{Usui} et~al.(2020){Usui}, {Bajo}, {Fujiya}, {Furukawa}, {Koike},
  {Miura}, {Sugahara}, {Tachibana}, {Takano} and {Kuramoto}}]{Usui_2020}
\bibinfo{author}{{Usui}, T.}, \bibinfo{author}{{Bajo}, K.i.},
  \bibinfo{author}{{Fujiya}, W.}, \bibinfo{author}{{Furukawa}, Y.},
  \bibinfo{author}{{Koike}, M.}, \bibinfo{author}{{Miura}, Y.N.},
  \bibinfo{author}{{Sugahara}, H.}, \bibinfo{author}{{Tachibana}, S.},
  \bibinfo{author}{{Takano}, Y.}, \bibinfo{author}{{Kuramoto}, K.},
  \bibinfo{year}{2020}.
\newblock \bibinfo{title}{{The Importance of Phobos Sample Return for
  Understanding the Mars-Moon System}}.
\newblock \bibinfo{journal}{Space Science Reviews} \bibinfo{volume}{216},
  \bibinfo{pages}{49}.
\newblock \DOIprefix\doi{10.1007/s11214-020-00668-9}.
\bibitem[{{Vernazza} et~al.(2012){Vernazza}, {Delbo}, {King}, {Izawa},
  {Olofsson}, {Lamy}, {Cipriani}, {Binzel}, {Marchis}, {Mer{\'\i}n} and
  {Tamanai}}]{Vernazza_2012}
\bibinfo{author}{{Vernazza}, P.}, \bibinfo{author}{{Delbo}, M.},
  \bibinfo{author}{{King}, P.L.}, \bibinfo{author}{{Izawa}, M.R.M.},
  \bibinfo{author}{{Olofsson}, J.}, \bibinfo{author}{{Lamy}, P.},
  \bibinfo{author}{{Cipriani}, F.}, \bibinfo{author}{{Binzel}, R.P.},
  \bibinfo{author}{{Marchis}, F.}, \bibinfo{author}{{Mer{\'\i}n}, B.},
  \bibinfo{author}{{Tamanai}, A.}, \bibinfo{year}{2012}.
\newblock \bibinfo{title}{{High surface porosity as the origin of emissivity
  features in asteroid spectra}}.
\newblock \bibinfo{journal}{Icarus} \bibinfo{volume}{221},
  \bibinfo{pages}{1162--1172}.
\newblock \DOIprefix\doi{10.1016/j.icarus.2012.04.003}.
\bibitem[{{Vernazza} et~al.(2013){Vernazza}, {Fulvio}, {Brunetto}, {Emery},
  {Dukes}, {Cipriani}, {Witasse}, {Schaible}, {Zanda}, {Strazzulla} and
  {Baragiola}}]{Vernazza_2013}
\bibinfo{author}{{Vernazza}, P.}, \bibinfo{author}{{Fulvio}, D.},
  \bibinfo{author}{{Brunetto}, R.}, \bibinfo{author}{{Emery}, J.P.},
  \bibinfo{author}{{Dukes}, C.A.}, \bibinfo{author}{{Cipriani}, F.},
  \bibinfo{author}{{Witasse}, O.}, \bibinfo{author}{{Schaible}, M.J.},
  \bibinfo{author}{{Zanda}, B.}, \bibinfo{author}{{Strazzulla}, G.},
  \bibinfo{author}{{Baragiola}, R.A.}, \bibinfo{year}{2013}.
\newblock \bibinfo{title}{{Paucity of Tagish Lake-like parent bodies in the
  Asteroid Belt and among Jupiter Trojans}}.
\newblock \bibinfo{journal}{Icarus} \bibinfo{volume}{225},
  \bibinfo{pages}{517--525}.
\newblock \DOIprefix\doi{10.1016/j.icarus.2013.04.019}.
\bibitem[{Wargnier et~al.(2024)Wargnier, Gautier, Doressoundiram, Poggiali,
  Beck, Poch, Quirico, Nakamura, Miyamoto, Kameda, Hasselmann, Ruscassier,
  Buch, Fornasier and Barucci}]{Wargnier_2024a}
\bibinfo{author}{Wargnier, A.}, \bibinfo{author}{Gautier, T.},
  \bibinfo{author}{Doressoundiram, A.}, \bibinfo{author}{Poggiali, G.},
  \bibinfo{author}{Beck, P.}, \bibinfo{author}{Poch, O.},
  \bibinfo{author}{Quirico, E.}, \bibinfo{author}{Nakamura, T.},
  \bibinfo{author}{Miyamoto, H.}, \bibinfo{author}{Kameda, S.},
  \bibinfo{author}{Hasselmann, P.H.}, \bibinfo{author}{Ruscassier, N.},
  \bibinfo{author}{Buch, A.}, \bibinfo{author}{Fornasier, S.},
  \bibinfo{author}{Barucci, M.A.}, \bibinfo{year}{2024}.
\newblock \bibinfo{title}{Spectro-photometry of phobos simulants: I.
  detectability of hydrated minerals and organic bands}.
\newblock \bibinfo{journal}{Icarus} \bibinfo{volume}{421},
  \bibinfo{pages}{116216}.
\newblock \URLprefix
  \url{https://www.sciencedirect.com/science/article/pii/S0019103524002768},
  \DOIprefix\doi{https://doi.org/10.1016/j.icarus.2024.116216}.
\bibitem[{{Warner} et~al.(2017){Warner}, {Golombek}, {Sweeney}, {Fergason},
  {Kirk} and {Schwartz}}]{Warner_2017}
\bibinfo{author}{{Warner}, N.H.}, \bibinfo{author}{{Golombek}, M.P.},
  \bibinfo{author}{{Sweeney}, J.}, \bibinfo{author}{{Fergason}, R.},
  \bibinfo{author}{{Kirk}, R.}, \bibinfo{author}{{Schwartz}, C.},
  \bibinfo{year}{2017}.
\newblock \bibinfo{title}{{Near Surface Stratigraphy and Regolith Production in
  Southwestern Elysium Planitia, Mars: Implications for Hesperian-Amazonian
  Terrains and the InSight Lander Mission}}.
\newblock \bibinfo{journal}{Space Science Reviews} \bibinfo{volume}{211},
  \bibinfo{pages}{147--190}.
\newblock \DOIprefix\doi{10.1007/s11214-017-0352-x}.
\bibitem[{{Yada} et~al.(2021){Yada}, {Abe}, {Okada}, {Nakato}, {Yogata},
  {Miyazaki}, {Hatakeda}, {Kumagai}, {Nishimura}, {Hitomi}, {Soejima},
  {Yoshitake}, {Iwamae}, {Furuya}, {Uesugi}, {Karouji}, {Usui}, {Hayashi},
  {Yamamoto}, {Fukai}, {Sugita}, {Cho}, {Yumoto}, {Yabe}, {Bibring},
  {Pilorget}, {Hamm}, {Brunetto}, {Riu}, {Lourit}, {Loizeau}, {Lequertier},
  {Moussi-Soffys}, {Tachibana}, {Sawada}, {Okazaki}, {Takano}, {Sakamoto},
  {Miura}, {Yano}, {Ireland}, {Yamada}, {Fujimoto}, {Kitazato}, {Namiki},
  {Arakawa}, {Hirata}, {Yurimoto}, {Nakamura}, {Noguchi}, {Yabuta}, {Naraoka},
  {Ito}, {Nakamura}, {Uesugi}, {Kobayashi}, {Michikami}, {Kikuchi}, {Hirata},
  {Ishihara}, {Matsumoto}, {Noda}, {Noguchi}, {Shimaki}, {Shirai}, {Ogawa},
  {Wada}, {Senshu}, {Yamamoto}, {Morota}, {Honda}, {Honda}, {Yokota},
  {Matsuoka}, {Sakatani}, {Tatsumi}, {Miura}, {Yamada}, {Fujii}, {Hirose},
  {Hosoda}, {Ikeda}, {Iwata}, {Kikuchi}, {Mimasu}, {Mori}, {Ogawa}, {Ono},
  {Shimada}, {Soldini}, {Takahashi}, {Takei}, {Takeuchi}, {Tsukizaki},
  {Yoshikawa}, {Terui}, {Nakazawa}, {Tanaka}, {Saiki}, {Yoshikawa}, {Watanabe}
  and {Tsuda}}]{Yada_2022}
\bibinfo{author}{{Yada}, T.}, \bibinfo{author}{{Abe}, M.},
  \bibinfo{author}{{Okada}, T.}, \bibinfo{author}{{Nakato}, A.},
  \bibinfo{author}{{Yogata}, K.}, \bibinfo{author}{{Miyazaki}, A.},
  \bibinfo{author}{{Hatakeda}, K.}, \bibinfo{author}{{Kumagai}, K.},
  \bibinfo{author}{{Nishimura}, M.}, \bibinfo{author}{{Hitomi}, Y.},
  \bibinfo{author}{{Soejima}, H.}, \bibinfo{author}{{Yoshitake}, M.},
  \bibinfo{author}{{Iwamae}, A.}, \bibinfo{author}{{Furuya}, S.},
  \bibinfo{author}{{Uesugi}, M.}, \bibinfo{author}{{Karouji}, Y.},
  \bibinfo{author}{{Usui}, T.}, \bibinfo{author}{{Hayashi}, T.},
  \bibinfo{author}{{Yamamoto}, D.}, \bibinfo{author}{{Fukai}, R.},
  \bibinfo{author}{{Sugita}, S.}, \bibinfo{author}{{Cho}, Y.},
  \bibinfo{author}{{Yumoto}, K.}, \bibinfo{author}{{Yabe}, Y.},
  \bibinfo{author}{{Bibring}, J.P.}, \bibinfo{author}{{Pilorget}, C.},
  \bibinfo{author}{{Hamm}, V.}, \bibinfo{author}{{Brunetto}, R.},
  \bibinfo{author}{{Riu}, L.}, \bibinfo{author}{{Lourit}, L.},
  \bibinfo{author}{{Loizeau}, D.}, \bibinfo{author}{{Lequertier}, G.},
  \bibinfo{author}{{Moussi-Soffys}, A.}, \bibinfo{author}{{Tachibana}, S.},
  \bibinfo{author}{{Sawada}, H.}, \bibinfo{author}{{Okazaki}, R.},
  \bibinfo{author}{{Takano}, Y.}, \bibinfo{author}{{Sakamoto}, K.},
  \bibinfo{author}{{Miura}, Y.N.}, \bibinfo{author}{{Yano}, H.},
  \bibinfo{author}{{Ireland}, T.R.}, \bibinfo{author}{{Yamada}, T.},
  \bibinfo{author}{{Fujimoto}, M.}, \bibinfo{author}{{Kitazato}, K.},
  \bibinfo{author}{{Namiki}, N.}, \bibinfo{author}{{Arakawa}, M.},
  \bibinfo{author}{{Hirata}, N.}, \bibinfo{author}{{Yurimoto}, H.},
  \bibinfo{author}{{Nakamura}, T.}, \bibinfo{author}{{Noguchi}, T.},
  \bibinfo{author}{{Yabuta}, H.}, \bibinfo{author}{{Naraoka}, H.},
  \bibinfo{author}{{Ito}, M.}, \bibinfo{author}{{Nakamura}, E.},
  \bibinfo{author}{{Uesugi}, K.}, \bibinfo{author}{{Kobayashi}, K.},
  \bibinfo{author}{{Michikami}, T.}, \bibinfo{author}{{Kikuchi}, H.},
  \bibinfo{author}{{Hirata}, N.}, \bibinfo{author}{{Ishihara}, Y.},
  \bibinfo{author}{{Matsumoto}, K.}, \bibinfo{author}{{Noda}, H.},
  \bibinfo{author}{{Noguchi}, R.}, \bibinfo{author}{{Shimaki}, Y.},
  \bibinfo{author}{{Shirai}, K.}, \bibinfo{author}{{Ogawa}, K.},
  \bibinfo{author}{{Wada}, K.}, \bibinfo{author}{{Senshu}, H.},
  \bibinfo{author}{{Yamamoto}, Y.}, \bibinfo{author}{{Morota}, T.},
  \bibinfo{author}{{Honda}, R.}, \bibinfo{author}{{Honda}, C.},
  \bibinfo{author}{{Yokota}, Y.}, \bibinfo{author}{{Matsuoka}, M.},
  \bibinfo{author}{{Sakatani}, N.}, \bibinfo{author}{{Tatsumi}, E.},
  \bibinfo{author}{{Miura}, A.}, \bibinfo{author}{{Yamada}, M.},
  \bibinfo{author}{{Fujii}, A.}, \bibinfo{author}{{Hirose}, C.},
  \bibinfo{author}{{Hosoda}, S.}, \bibinfo{author}{{Ikeda}, H.},
  \bibinfo{author}{{Iwata}, T.}, \bibinfo{author}{{Kikuchi}, S.},
  \bibinfo{author}{{Mimasu}, Y.}, \bibinfo{author}{{Mori}, O.},
  \bibinfo{author}{{Ogawa}, N.}, \bibinfo{author}{{Ono}, G.},
  \bibinfo{author}{{Shimada}, T.}, \bibinfo{author}{{Soldini}, S.},
  \bibinfo{author}{{Takahashi}, T.}, \bibinfo{author}{{Takei}, Y.},
  \bibinfo{author}{{Takeuchi}, H.}, \bibinfo{author}{{Tsukizaki}, R.},
  \bibinfo{author}{{Yoshikawa}, K.}, \bibinfo{author}{{Terui}, F.},
  \bibinfo{author}{{Nakazawa}, S.}, \bibinfo{author}{{Tanaka}, S.},
  \bibinfo{author}{{Saiki}, T.}, \bibinfo{author}{{Yoshikawa}, M.},
  \bibinfo{author}{{Watanabe}, S.i.}, \bibinfo{author}{{Tsuda}, Y.},
  \bibinfo{year}{2021}.
\newblock \bibinfo{title}{{Preliminary analysis of the Hayabusa2 samples
  returned from C-type asteroid Ryugu}}.
\newblock \bibinfo{journal}{Nature Astronomy} \bibinfo{volume}{6},
  \bibinfo{pages}{214--220}.
\newblock \DOIprefix\doi{10.1038/s41550-021-01550-6}.
\bibitem[{{Young} et~al.(2019){Young}, {Poston}, {Wray}, {Hand} and
  {Carlson}}]{Young_2019}
\bibinfo{author}{{Young}, C.L.}, \bibinfo{author}{{Poston}, M.J.},
  \bibinfo{author}{{Wray}, J.J.}, \bibinfo{author}{{Hand}, K.P.},
  \bibinfo{author}{{Carlson}, R.W.}, \bibinfo{year}{2019}.
\newblock \bibinfo{title}{{The mid-IR spectral effects of darkening agents and
  porosity on the silicate surface features of airless bodies}}.
\newblock \bibinfo{journal}{Icarus} \bibinfo{volume}{321},
  \bibinfo{pages}{71--81}.
\newblock \DOIprefix\doi{10.1016/j.icarus.2018.10.032}.
\bibitem[{{Zolensky} et~al.(2002){Zolensky}, {Nakamura}, {Gounelle},
  {Mikouchi}, {Kasama}, {Tachikawa} and {Tonui}}]{Zolensky_2002}
\bibinfo{author}{{Zolensky}, M.E.}, \bibinfo{author}{{Nakamura}, K.},
  \bibinfo{author}{{Gounelle}, M.}, \bibinfo{author}{{Mikouchi}, T.},
  \bibinfo{author}{{Kasama}, T.}, \bibinfo{author}{{Tachikawa}, O.},
  \bibinfo{author}{{Tonui}, E.}, \bibinfo{year}{2002}.
\newblock \bibinfo{title}{{Mineralogy of Tagish Lake: An ungrouped type 2
  carbonaceous chondrite}}.
\newblock \bibinfo{journal}{Meteoritics \& Planetary Science}
  \bibinfo{volume}{37}, \bibinfo{pages}{737--761}.
\newblock \DOIprefix\doi{10.1111/j.1945-5100.2002.tb00852.x}.

\end{thebibliography}

\end{document}